\documentclass[letters,a4paper,fleqn,usenatbib]{mnras}

\usepackage{newtxtext,newtxmath}

\usepackage[T1]{fontenc}
\usepackage{ae,aecompl}

\usepackage{graphicx}	
\usepackage{amsmath}	
\usepackage{hyperref}
\usepackage{pdflscape}
\usepackage{wasysym}
\usepackage{siunitx}
\usepackage{mathtools}
\usepackage{color}
\usepackage{gensymb}

\newcommand{\appx}[1]{\color{blue}#1\mbox{ }\color{black}}

\title[Taurid meteoroids complex]{A Proposed Alternative Dynamical History for 2P/Encke which Explains the Taurid Meteoroid Complex}

\author[A. Egal et al.]{
A. Egal,$^{1,2,3}$\thanks{E-mail: aegal@uwo.ca (AE)}
P. Wiegert,$^{1,2}$
P. G. Brown$^{1,2}$
\\
$^{1}$Department of Physics and Astronomy, The University of Western Ontario, London, Ontario N6A 3K7, Canada\\
$^{2}$ Institute for Earth and Space Exploration (IESX), The University of Western Ontario, London, Ontario N6A 3K7, Canada\\
$^{3}$IMCCE, Observatoire de Paris, PSL Research University, CNRS, Sorbonne Universit\'{e}s, UPMC Univ. Paris 06, Univ. Lille, France\\
}

\date{Accepted 2022 June 28; Received 2022 May 31; in original form XYZ}

\pubyear{2022}

\begin{document}
\label{firstpage}
\pagerange{\pageref{firstpage}--\pageref{lastpage}}
\maketitle

\begin{abstract}
The Taurid Meteoroid Complex (TMC) is a broad stream of meteoroids that produces several annual meteor showers on Earth. If the linkage between these showers and 2P/Encke is at the centre of most TMC models, the small size and low activity of the comet suggest that 2P/Encke is not the unique parent body of the Taurids.
Here we simulate the formation of the TMC from 2P/Encke and several NEAs. In total, we explored more than a hundred stream formation scenarios using clones of 2P/Encke. Each modelled stream was integrated and compared with present-day Taurid observations. As previously reported, we find that even slight variations of 2P/Encke's orbit modifies considerably the characteristics of the simulated showers. Most of the comet's clones, including the nominal one, appear to reproduce the radiant structure of the Taurid meteors but do not match the observed time and duration of the showers. However, the radiants and timing of most Taurid showers are well reproduced by a particular clone of the comet. Our analysis thus suggest that with this specific dynamical history, 2P/Encke is the sole parent of the four major TMC showers which have ages from 7-21 ka. Our modelling also predicts that the 2022 Taurid Resonant Swarm return will be comparable in strength to the 1998, 2005 and 2015 returns.  While purely dynamical models of Encke’s orbit -limited by chaos- may fail to reveal the comet's origin, its meteor showers may provide the trail of breadcrumbs needed to backtrack our way out of the labyrinth.
\end{abstract}




\section{Introduction}

Comet 2P/Encke is one of the most puzzling comets discovered.  It is a comparatively large and bright comet, with a nuclear diameter of 2.4 km \citep{Fernandez2000,Lisse2004} and has one of the shortest orbital periods measured ($\simeq$3.3 years). Given its large size, it is surprising that it was only first observed in the 18th century. Contrasting with the historic assumption that the comet is particularly "gassy" \citep{AHearn1985}, Encke was found to have among the most massive of known cometary dust-trails, extended along its orbit with a noticeable asymmetry \citep{Sykes1992}. 

Adding to the oddities associated with the comet is the possible association of several Near-Earth Objects (NEOs) on similar orbits which may have a genetic relationship with Encke (cf. Section \ref{sec:NEAs_analysis}), though the dynamical significance of this linkage has been questioned \citep{Valsecchi1995}. The current orbit of Encke is decoupled from Jupiter and therefore very difficult for a Jupiter Family Comet (JFC) to evolve toward \citep{Valsecchi1995}.

Past work to address this question has focused on numerical integration of JFCs with and without non-gravitational forces to attempt to quantify dynamical pathways likely to place Encke near its present orbit \citep{Levison2006}. However, purely dynamical models of Encke are limited by the effects of chaos, and its past orbital history remains a major unsolved question.

One component of the Encke complex to be explored in connection with its past orbital evolution is the Taurid meteoroid stream complex (TMC). The TMC comprises a series of at least four significant showers, namely the nighttime Northern and Southern Taurids (\#017 NTA and \#002 STA), and the daytime $\beta$ Taurids and $\zeta$ Perseids (\#173 BTA and \#172 ZPE). Presuming these showers originate primarily from Encke or a proto-Encke object on a similar orbit, the strength, duration of activity, timing of peaks and radiants of these showers provide a series of potential constraints on possible past dynamical evolution and dust production of the comet.

Early models of Taurid stream formation (which assumed that most meteoroids were injected into the stream from the comet) were motivated in part by the similarity in the longitude of perihelion of the Taurids and Encke, as well as the spread in Taurid aphelia \citep{Whipple1952}. Models of the Taurids based solely on cometary activity from Encke (or an object on a similar orbit) resulted in ages for the complex ranging from 5 ka - 100 ka \citep{Whipple1952, Jones1986,Babadzhanov1990,Steel1996}, and all struggled to explain the observed wide dispersion of the stream's orbital elements.

More recently, it has been proposed that Encke is simply the largest member of a broader Taurid complex, produced by the breakup of a larger progenitor comet which entered the inner solar system 20 ka ago \citep{Steel1996a,Napier2010}. \cite{Steel1991} were the first to numerically model the Taurid streams as just one part of the dust complex resulting from this proto-Encke comet. This hypothetical parent object would have been originally in the 7:2 mean motion resonance with Jupiter, but otherwise had the current orbital elements of Encke (which is not in this resonance currently). 

The authors assumed material was ejected from a parent object with speeds up to 2 km/s to replicate the large spread in semi-major axis of the Taurid stream. Ejecting meteoroids at different epochs back to 15 ka before present, \cite{Steel1991} found the best match to the current N. and S. Taurid orbits resulted from ejections near perihelion within the last 10 ka, but with ongoing subsequent fragmentation of the original body since that time. They also proposed a splitting of the original object, with collisions ejecting material from daughter fragments; these were needed primarily to better fit the large spread in observed Taurid semi-major axes. Similarly, \cite{Babadzhanov1990} used the spread in longitudes of perihelion for the Taurid showers to estimate an age of 8 ka - 18 ka for the complex. This was based on differences in the precession rate as a function of semi-major axis.

Most recently, \cite{Tomko2019} have modelled the Taurid streams by ejecting particles from the nominal backward integrated orbit of Encke. They released 10 000 particles at Encke's perihelion passages 1, 2, 4, 8 and 16 ka in the past. For each set of models a range of eight $\beta$ values was also chosen to capture the role of radiation forces on subsequent particle evolution, representing a total of 400 000 test particles. These were released with speeds from 70-92 m/s in isotropic directions at perihelion from 2P. They then chose test particles at the current epoch which had Minimum Orbital Intersection Distances (MOIDs) < 0.05 AU with the Earth's orbit as representative of the presently observed showers. 

\cite{Tomko2019} also compared model orbits with measured orbits of all meteor showers in the IAU Meteor Data Center (MDC) to identify probable linked minor Taurid showers (which they term {\it filaments}). The authors confirmed the link between Encke and the NTA, STA, BTA and ZPE meteor showers. In total their model predicted some 21 presently observable Taurid showers (once duplicates were removed) of which they identify 16 as being in the IAU MDC. However, the veracity of the model and connection with these streams is entirely based on orbital similarity with a subjectively chosen D-criterion cut of 0.1; it does not compare predicted times or intensity of shower activity. This is particularly problematic for the Taurids as their radiants are embedded in the highly populated helion and anti-helion sporadic meteor sources, making chance associations more likely.

Most of the preceding works focused on reproducing the large dispersion in orbital elements associated with the Taurid stream or matching radiants in a qualitative sense. However, the N and S Taurids have broad radiants and are long-lived; it is therefore difficult to establish stream membership \citep[e.g][]{Stohl1992} using orbital elements or radiants alone. The timing of the activity of the individual showers is also an important measurement which may provide clues to the formation of the Taurid complex, but has not been used in any modelling efforts to date. 

Here we examine the question as to whether or not all four of the major Taurid streams (namely, the NTA, STA, BTA and ZPE) could be produced from material ejected by Encke (or a proto-Encke object on a similar orbit) alone, using ejection speeds typical of cometary gas-drag meteoroid production \citep{Jones1995a}. We use observations of shower duration, strength and spread in radiants/orbital elements of the major Taurid streams as model constraints. Given that Encke's orbit is chaotic, we explore the formation of the Taurid streams from a range of potential Encke clones, that is, particles on orbits very close to that of Encke currently but with differing evolution under backward integration. In principle, if we are able to find a clone which produces the radiants, timing and activity levels of the main four showers, we may provisionally interpret this as a potential past dynamical pathway for Encke most consistent with meteoroid streams observed at Earth today.

The paper is divided into 8 main parts as follows. In section \ref{sec:Encke}, we examine the physical properties, dust production and orbital stability of comet 2P/Encke. The general characteristics of the NTA, STA, BTA and ZPE, analyzed in a precursor work to this study \citep{Egal2022}, are summarized in section \ref{sec:meteor_observations}. Section \ref{sec:methods} details the stream models, initial conditions and calibration process used to simulate the meteor showers.

In section \ref{sec:nominal_ejection}, we perform a detailed analysis of the meteoroid swarm produced by 2P/Encke. We find the model to reproduce the radiants structure of the four major Taurid showers, but to fail in predicting the peak time and duration of most showers. This led us to explore in section \ref{sec:NEAs_ejection} if the meteoroids ejected from several NEOs, released during the breakup of a larger parent body, provided predictions in better agreement with meteor observations. The inclusion of meteoroids ejecta from additional parent bodies proved to improve the modelling of the showers' activity, but is still insufficient to reproduce the general TMC characteristics. 

In Section \ref{sec:clones_ejection}, we explored more than a hundred alternate evolutions of comet Encke, and identified one specific clone that reproduce the main features of the TMC. The implications and limitations of this specific orbital history of Encke on the TMC formation are discussed in Sections \ref{sec:clones_ejection} and \ref{sec:discussion}.

\section{Comet 2P/Encke} \label{sec:Encke}

Comet 2P/Encke is a short-period comet discovered in 1786 by the astronomer Pierre M\'echain. The comet is named after Johann Franz Encke, who recognized the periodic character of the comet in 1819 and successfully predicted its return in 1822. 

Because of the peculiarity of its orbit, Encke is the sole member of its own class called Encke-type comets. The particular characteristics of Encke are two fold - it possesses the shortest orbital period of all known comets (today close to 3.3 years), and evolves on an orbit dynamically decoupled from the influence of Jupiter because of gravitational interactions with the terrestrial planets \citep{Whipple1940}. Despite the lack of close encounters between Encke and Jupiter, the gravitational effect of the giant planet is still the main mechanism causing the secular variations of the comet's orbit as described in detail in \cite{Whipple1940} or \cite{Valsecchi1995}. In this section, we summarize the main characteristics of Encke's nucleus, activity and orbital evolution. \\[0.75cm]

\subsection{Nucleus and activity}

\subsubsection{Encke's nucleus and dust trail} \label{sec:nucleus}

The nucleus of Encke is dark, with a geometric albedo of 0.046 $\pm$ 0.023 and an effective radius of 2.4 $\pm$ 0.3 km \citep{Fernandez2000,Lisse2004}. Encke's spectrum is compatible with an X-class object of type Xe in the Bus-Demeo taxonomy \citep{DeMeo2009}. Between 400 and 800 nm, the nucleus' spectrum is essentially featureless but displays a moderate reddening \citep{Tubiana2015}. Particles ejected from the comet have also been found to be redder than the Sun and the zodiacal light, suggesting the presence of organic-based compounds within the dust \citep{Ishiguro2007}. The meteoroids appear to have a low albedo of 0.04 to 0.06 \citep{Lisse2004}. 

The Infrared Astronomical Satellite (IRAS) detected an extended dust trail along Encke's orbit, formed by the accumulation of material ejected during several orbital periods \citep{Sykes1986,Sykes1992}. The dust emitted by the nucleus formed one of the most extensive trails detected by the satellite, partly because the comet's orbit is shielded from strong gravitational interactions with Jupiter. Observations of the comet's dust trail in the mid-infrared and optical revealed that the core of trail is more than 20 000 km wide and composed of particles in the range 1 mm to 10 cm  \citep{Reach2000,Epifani2001,Lowry2003,Ishiguro2007}. Estimates of the particle's size distribution within the trail, based on the power law fit of the time-averaged size distribution, ranged from about 2.8 to 3.6 \citep{Lisse2004,Epifani2001,Fulle1990}.

The discovery of Encke's dust trail and the elevated dust-to-gas mass ratios measured by \cite{Reach2000}, \cite{Lisse2004} and \cite{Ishiguro2007} contrasted with the previous assumption that the comet is particularly "gassy" \citep{AHearn1985}. Most of the dust ejection occurs in a short arc around perihelion, though some dust emission is detected at heliocentric distances of 1.2 AU \citep{Lisse2004} or 2.6 AU \citep{Epifani2001}.  The current mass loss rate of the nucleus is estimated to be close to $10^{12}$ - $10^{13}$ g per orbit
\citep{Sykes1992,Reach2000,Lisse2004,Ishiguro2007}. 

\subsubsection{Af$\rho$ profile}

To model a meteoroid stream, it is necessary to estimate the variation of a comet's dust production with heliocentric distance $r_h$. This is used to establish a correlation between the number of particles simulated and the number of meteoroids ejected from the nucleus at $r_h$ \citep{Egal2020a}. The quantity of dust ejected by the nucleus is often described by the Af$\rho$ parameter defined by \cite{AHearn1984}. This parameter depends on the nucleus' Bond albedo, phase angle, the filling factor of the dust within the field of view and the aperture radius of the telescope used for the observation. Raw Af$\rho$ estimates need therefore to be corrected for phase angle effects and the observations' aperture radius before comparison with other measurements. 

Despite the short period of Encke which tends to increase the frequency of observations of the comet, there are only a few published estimates of the nucleus' Af$\rho$ along its orbit. Most of the available Af$\rho$ measurements are provided by \cite{Osip1992}, who measured the dust emission of Encke during several returns in 1977, 1980, 1984 and 1990. Unfortunately, Af$\rho$ estimates were determined for heliocentric distances of 0.53 to 1.23 AU, but not close to perihelion ($q\simeq0.34$ AU) where the comet is the most active. 

\begin{figure}
	\centering
	\includegraphics[width=.49\textwidth]{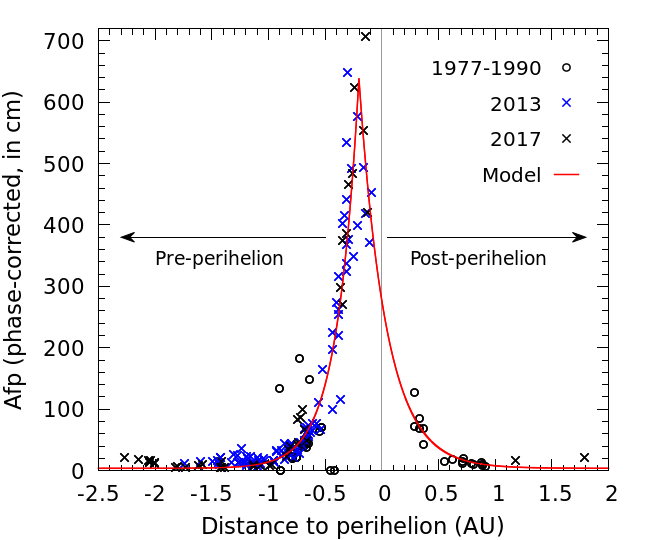}
	\caption{\label{fig:Afp} Phase-corrected Af$\rho$ measurements of comet Encke during the 1977, 1980, 1984 and 1990 apparitions \protect\citep[black circles][]{Osip1992} and 2013 and 2017 observations provided by the Cometas Obs network (blue and black crosses). The abscissa (X) represents the comet's heliocentric distance to perihelion, arbitrarily negative for pre-perihelion measurements and positive for post-perihelion times ($X=(r_h-q)\frac{t-t(q)}{|t-t(q)|}$, with $t(q)$ the time at perihelion).}
\end{figure}

In order to determine Encke's Af$\rho$ profile with the heliocentric distance, we corrected the published estimates of \cite{Osip1992} for the phase effect. For each date of observation, we computed the comet's phase angle and modified the original Af$\rho$ value using the Schleicher-Marcus composite dust phase function described in \cite{Egal2020c}. The resulting Af$\rho$ as function of the distance to perihelion is presented in Figure \ref{fig:Afp}. The figure include additional Af$\rho$ measurements kindly provided by the Cometas Obs amateur network\footnote{\url{http://www.astrosurf.com/cometas-obs/}}, which determined the comet's dust production during its 2013 and 2017 apparitions. These estimates, once corrected for the phase angle, fit well with those of \cite{Osip1992} and provide an estimate of the likely dust emission from Encke close to perihelion. 

We modelled the evolution of Af$\rho$ with $r_h$ using a double-exponential function of the form:

\begin{equation}\label{eq:afrho}
\begin{aligned}
& Af\rho(X)=K_1+Af\rho(X_\text{max})\times10^{-\gamma |X-X_\text{max}|}, \\
& \gamma=\gamma_1 \text{ if } X\le X_\text{max} \text{ and } \gamma=\gamma_2 \text{ if } X\ge X_\text{max}, \\
& X=(r_h-q)\frac{t-t(q)}{|t-t(q)|}\\
\end{aligned}
\end{equation}

where $r_h$ is the comet's heliocentric distance at time $t$, $q$ is the perihelion distance, $t(q)$ the time of perihelion passage, and $\gamma_{1,2}$, $X_\text{max}$ and $K_1$ parameters which are to be determined by the fitting process. The best fit solution of the Af$\rho$ profile was obtained for $\gamma_1 = 2.24$, $\gamma_2 = 1.82$, $X_\text{max}=-0.2$ AU, Af$\rho(X_\text{max})\simeq635$ cm and $K_1=3.74$ cm. The modelled profile is represented by the red line in Figure \ref{fig:Afp}. We find the comet's dust production to peak pre-perihelion, with a slightly steeper pre-perihelion branch compared to post-perihelion. 

Our result differs somewhat from that of \cite{AHearn1995}, who found the comet's dust production to be asymmetric with a $r_h$-dependence of -0.81 pre-perihelion and -2.99 post-perihelion. However, since no measurement of Encke's dust production below 0.53 AU was available to those authors, we consider the model of Figure \ref{fig:Afp} to provide a more accurate description of the current Af$\rho$ variations of the comet.   

 \subsection{Orbital evolution} \label{sec:orbital_evolution_2P}
 
2P/Encke has been observed at each perihelion return since 1818 with the exception of 1944. Despite the numerous observations of the comet, no simple dynamical model of the comet's motion allows all its apparitions to be linked with accuracy \citep{Usanin2017}. The combination of planetary gravitational perturbations and variable non-gravitational forces acting upon the nucleus likely explains the divergence between the models and the observations. 
 
 \subsubsection{The reproducibility of Encke's past orbit} 
  
 As the primary goal of the present work is to determine whether or not Encke alone is capable of forming the four main Taurid streams, its past orbital evolution is central to our study. We use as our starting conditions the orbital solution provided by the JPL\footnote{https://ssd.jpl.nasa.gov/sbdb.cgi, accessed in June 2021}, considering either 1) constant or 2) no non-gravitational acceleration parameters (solutions 1 \& 2 respectively in Table \ref{tab:IC}).
 For each case, a thousand clones of the comet's orbit were created according to the multivariate normal distribution defined by the solution's covariance matrix. Each clone was then integrated until 30 000 BCE, as described in Section \ref{sec:integration}. The time evolution of the clones' orbital elements is presented in Appendix \appx{A,} either with \appx{(A1)} or without \appx{(A2)} non-gravitational forces. 
 
 \begin{table*}
 \resizebox{\textwidth}{!}{
 	\begin{tabular}{crccccccccccc}
 		Solution & Body & JD & e & a & q & i & $\Omega$ & $\omega$ & $m$ & $A_1$ & $A_2$ & $A_3$ \\
 		& & & & (AU) & (AU) & ($\degree$) & ($\degree$) & ($\degree$) & ($\degree$) & (AU/d$^2$) & (AU/d$^2$) & (AU/d$^2$) \\[0.1cm]
 		1 & 2P/Encke & 2457097.50000 &  0.8483 & 2.2151 & 0.3360 & 11.7818 & 334.5678 & 186.5437 & 143.2720 & -2.49e$^{-11}$ & -2.69e$^{-12}$ & 3.86e$^{-9}$ \\
 		2 & 2P/Encke & 2457259.50000 &  0.8483 & 2.2151 & 0.3360 & 11.7814 & 334.5678 & 186.5468 & 143.2720 & 0.0 & 0.0 & 0.0 \\
 		3 & 2P/Encke & 2456618.24851 & 0.8486 & 2.2151 & 0.3354 & 11.7815 & 334.5683 & 186.5430 & 0.0000 & -- & --  & -- \\
 	    4 & 2004 TG10 & 2459000.50000 & 0.8618 & 2.2334 & 0.3086 & 4.1811 & 205.0870 & 317.3676 & 213.3681 & -- & --  & -- \\
 	    5 & 2005 TF50 & 2459000.50000 & 0.8687 & 2.2733 & 0.2984 & 10.6970 & 0.6430 & 159.9071 & 74.4271 & -- & --  & -- \\
 	    6 & 2005 UR   & 2453668.50000 & 0.8817 & 2.2478 & 0.2660 & 6.9330 & 20.0281 & 140.4776 & 346.6137 & -- & --  & -- \\
 	    7 & 2015 TX24 & 2459000.50000 & 0.8721 & 2.2660 & 0.2900 & 6.0426 & 32.9362 & 127.0567 & 112.6679 & -- & --  & -- \\
 	\end{tabular}}
 	\caption{\label{tab:IC} Orbital elements of 2P/Encke and NEOs used as a starting point for the numerical integrations in this work.}
 \end{table*}

 Apart from a slight but sudden dispersion of the clone swarm around 600 BCE, we observe in Figure \appx{A2} an almost linear increase in the dispersion of the clones' orbital elements backward in time. Such increases were found to be caused by distant encounters of the clones with Earth. We see a general decrease in perihelion distance over the last 20-30 ka, but overall the orbital elements of the clones remain in a narrow range, underscoring the stability of Encke's orbit and its lack of close approaches to Jupiter. Using the current non-gravitational parameters for the comet (though we recognize that these certainly must have varied over this time frame), we see in Figure \appx{A1} a very similar evolution. At least for fixed non-gravitational parameters, the changes in the orbital elements are minor, in agreement with \cite{Tomko2019}. More importantly, the overall evolution remains confined to a restricted region of phase space, meaning that there is a high probability that our backward integrations are at least somewhat representative of the true evolution of Encke's orbit.
 
 The secular evolution of the comet's angular elements is the key to its stability \citep{Whipple1940}. When the perihelion argument favours a possible close encounter with Jupiter ($\omega$ of 0$\degree$ or 180$\degree$), the orbit's inclination is at maximum which increases the average distance of the comet from the ecliptic plane. When the inclination is close to zero, we observe that the perihelion argument adopts the values of 90$\degree$ or 270$\degree$, which moves its aphelion (and perihelion) further away from Jupiter.
 
  \begin{figure}
 	\centering
 	\includegraphics[trim={0 1.8cm 0 1.8cm},clip,width=.49\textwidth]{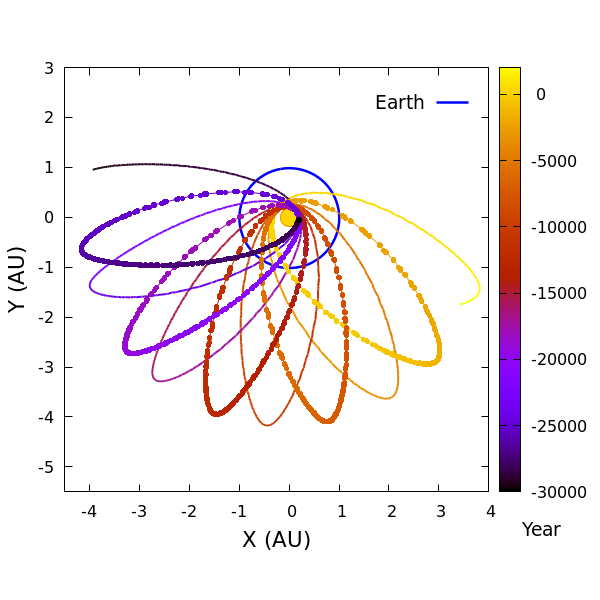}
 	\caption{The position of 2P/Encke's ascending node (solid lines) and descending node (dotted lines) between 30 000 BCE and 2021. The position of the comet was integrated from the orbital solution provided in Table \ref{tab:IC}, without non-gravitational forces (solution 2). The blue solid line represents the orbit of Earth in 2021. The nodal locations are colour-coded by date.}
 	\label{fig:2P_nodes}
 \end{figure}
  
 As a first approximation, we consider the past dynamical evolution of Encke integrated from solution 2 in Table \ref{tab:IC} (without non-gravitational forces) to be a representative model for the comet's evolution. We will call this model {\it nominal Encke}.  Figure \ref{fig:2P_nodes} presents nodal locations (ascending and descending nodes) of Encke's nominal orbit between 30 000 BCE and 2021. Because of the angular precession of its orbit, Encke's node has crossed the Earth's orbit multiple times during the past millennia. In one precession cycle of the comet's perihelion argument - which lasts about 5000-6000 years \citep{Egal2021} - meteoroids ejected from the comet may create four meteor showers visible on Earth, pre/post-perihelion and with a northern and southern component \citep{Babadzhanov2001}.
 
\subsection{On the origin of comet Encke} \label{sec:origin}

Explaining how Encke arrived onto a short-period orbit decoupled from the influence of Jupiter poses a challenge to current dynamical models. Backward and forward integrations of JFCs and Encke-type objects have highlighted the existence of dynamical pathways between these two orbital groups \citep[e.g.,][]{Valsecchi1995,Steel1996,Levison2006}. It was found that the gravitational perturbation exerted by terrestrial planets alone have the potential to decouple JFC to Encke-type orbits. However, the time required to perform such a transition exceeds the physical lifetime of JFCs \citep{Levison2006}. 

The inclusion of strong nongravitational forces (NGF) to the model can reduce the transition time between JFCs and Encke-type orbits \citep{Steel1996,Pittich2004}. However, the validity of such an approach has been questioned because of the unrealistically high NGFs used in these studies, and because of the difficulty of modelling highly variable NGF over time \citep{Levison2006}. 

Another hypothesis to explain the creation of Encke is that the comet was dormant during most of its transition from a JFC-type orbit and became active tens to hundreds of thousands of apparitions later because of a significant change in its orbit. A scenario in which the comet suddenly became active at the end of the 18th century could explain why Encke was undetected before 1786, given that its large size and present orbit should have made it easily visible at earlier epochs.

However, backward integrations of the comet's motion do not reveal any significant close encounters with any planets that could have modified Encke's orbit at that epoch \citep[in this work or in][]{Usanin2017}. A possible alternative explanation may be the formation of Encke after the breakup of a larger comet in the inner solar system \citep{Clube1984,Steel1991}. It has been suggested that Encke will become dormant in the coming decades \citep{Sekanina1969,Usanin2017} before eventually colliding with the Sun \citep{Valsecchi1995}.

Purely dynamical models of Encke's orbit are plagued by the effects of chaos. Though stable in a broad sense, planetary encounters cause small scale changes in the comet's motion which make backwards integration of its current orbit keenly sensitive to small differences in starting conditions. However,  meteor observations provide a potentially powerful lever against this problem, since the true past orbital evolution of Encke must be consistent with the Taurid meteoroid complex as whole, presuming it is the sole parent of the TMC. As we will show, a careful comparison of the observed meteoroid distribution within the present TMC, when coupled to formation of the stream using a range of different possible past dynamical histories of Encke itself, allows interesting constraints to be placed on the comet's prior evolution. 
 
 \section{Taurid meteor showers} \label{sec:meteor_observations}
 
Meteoroid stream simulation is fundamentally an inversion problem, though historically it has been done through forward modelling \citep{Egal2020a}. For cases where the parent is clearly linked to a given stream and in an orbit which does not suffer significant perturbations, it is possible to numerically invert the observed activity profile, timing and radiant of a stream to fit dust production parameters of a parent comet (with various assumptions). This procedure has recently been refined in a series of works by \cite{Egal2019, Egal2020b} and applied to the October Draconids and Halleyid streams. It has been shown to successfully reproduce past shower activity and make future predictions of shower intensity and timing which have been validated \citep[e.g.,][]{Egal2018}. 

Such an inversion process requires calibration in the form of stream activity as a function of time. The model produces predictions of radiant distribution and orbital elements which can then be checked against observations once inversion is completed. For this reason, it is essential to establish calibrated (at least in a relative sense) observed profiles for stream complexes for model inversions.  \\[1cm]

\subsection{General observational characteristics of the four major Taurid showers}  

In \cite{Egal2022}, we analyzed more than two decades of visual, optical and radar meteor measurements to derive the observational properties of the four major Taurid showers (NTA, STA, BTA and ZPE). This work provided quantitative measurements of the showers' activity, radiant and orbital elements variation as a function of solar longitude. These data form the modelling constraints which any successful dynamical simulations of the Taurid Meteoroid Complex must reproduce. The general characteristics of the major Taurid showers as determined in \cite{Egal2022} were found to be:

\begin{enumerate}
	\item The NTA meteors are generally visible between 197$\degree$ and 255$\degree$ solar longitude (SL), with maximum rates of 5-6 meteors per hour forming a plateau between 220$\degree$ and 232$\degree$.
	\item The STA can be observed between 170$\degree$ and 245$\degree$ SL, and display two wide peaks of activity. The first one occurs around SL 197-198$\degree$, and is twice as pronounced in radar data than in optical observations (reaching about 10 to 11 meteors per hour). The STA are therefore notably enriched in small particles at the beginning of their activity. A second peak is detected around SL 220$\degree$ in radar and optical data, reaching an average of 5 to 6 meteors per hour. 
	\item The daytime BTA and ZPE are mainly observed using radar instruments. From measurements by the Canadian Meteor Orbit Radar (CMOR), the BTA are detected between 84$\degree$ and 106$\degree$ SL, with maximum meteor rates of about 9 meteors per hour reached around 91-95$\degree$ SL. 
	\item The ZPE are active between 55$\degree$ and 94$\degree$ SL, and present a broad maximum of 12-14 meteors per hour between 76.5$\degree$ and 81.5$\degree$ SL. Observations of the shower are contaminated by the nearby daytime Arietids around 69-73$\degree$ SL. 
	\item The mass distribution index of the NTA and STA is estimated to be close to 1.9 from visual and optical observations. Radar measurements indicate a mass index of 1.81 $\pm$ 0.05 for the ZPE and 1.87 $\pm$ 0.05 for the BTA. Physical properties of the meteoroids are compatible with a cometary nature, with an anti-correlation of strength with particle size. 
	\item The variations in time and intensity of the showers' main peak of activity suggest a filamentary structure of the Taurid meteoroid stream. However, the small number of meteors detected at each apparition of these showers may account for the observed variability.
	\item The northern meteoroid branch (ZPE) is the dominant of the two daytime showers, while the southern branch (STA) is the most active of the two nighttime showers. Differences in strength may indicate a distinct evolution history of the ZPE/STA compared with the NTA/BTA.
	\item The years showing stronger activity associated with the Taurid Swarm Complex (TSC), composed of meteoroids trapped into the 7:2 Mean Motion Resonance (MMR) with Jupiter, produce enhanced meteor activity only for the STA. In particular, the swarm years are marked by a higher proportion of fireballs. Enhanced meteor rates were not identified for the NTA, BTA or ZPE during potential swarm year returns. 
	\item Several meter-sized meteoroids have been identified within the TSC \citep{Spurny2017}. However, we find no evidence of the TSC existence in CMOR data, suggesting that the swarm of meteoroids lacks sub-mm particles. This suggests that smaller meteoroids are removed from the 7:2 MMR much faster than fireball-producing meteoroids. 
	\item The average orbital elements of the NTA and STA change significantly with solar longitude, while variations of the core of the BTA and ZPE showers are less pronounced.
	\item A positive correlation in (a,e) is observed for the four showers in optical and radar data. However, the core of the BTA and ZPE in CMOR data show an opposite trend, which may reflect a difference in age or evolution history between the core and the wings of the showers (or between sub-mm and larger particles). 
	\item The dispersion of the showers' perihelion longitudes suggest that the NTA/BTA branch was formed over a shorter timescale than the STA/ZPE branch. The precession timescales associated with the core of each shower require a minimum age of 6 ka to form the four major showers from a single, unique parent body. 
\end{enumerate}

\subsection{NEOs of the Taurid complex} \label{sec:NEAs_analysis}

The link between the four major Taurid showers and the peculiar comet Encke was originally established by \cite{Whipple1940} and \cite{Whipple1952}. With time, multiple minor showers with radiants located in the constellations of Taurus, Aries, Cetus and Pisces (among others) have been proposed as being associated with the Taurids and several Near-Earth Asteroids (NEAs) were suggested as additional parent bodies of the TMC \citep[e.g.,][]{OS1988,Steel1991,Babadzhanov1990,Stohl1990,Babadzhanov2001,Porubcan2002,Porubcan2006,Babadzhanov2008,Madiedo2013,Bucek2014,Jenniskens2016}. In particular, several Apollo-type asteroids have been identified as potentially larger members  of the Taurid Swarm Complex \citep{Spurny2017,Olech2017,Devillepoix2021}. 

After analyzing the observational properties of the Taurid showers, \cite{Clube1984} expanded on the original hypothesis of \cite{Whipple1940} and \cite{Whipple1952} that the Taurid complex was created by the breakup of a large comet a few millennia ago. Sublimation-driven meteoroid ejection and successive fragmentation of the nucleus were proposed to explain the formation of the TMC, the TSC, and large NEAs within the complex. 
\cite{Asher1993b}, \cite{Asher1994b} and \cite{Steel1996a} first proposed a list of NEAs associated with the complex. Subsequent studies have greatly expanded this potential NEO list \citep[e.g.][]{Babadzhanov2001,Porubcan2006,Babadzhanov2008,Brown2010,Jopek2011,Popescu2014,Olech2016,Dumitru2017}. 

In \cite{Egal2021}, we explored dynamical linkages among 51 NEAs associated with the Taurid complex, as well as with comet Encke. We found that only twelve pairs of bodies in our sample approached each other with both a low Minimum Orbit Intersection Distance (MOID < 0.05 AU) and small relative velocity (<500 m/s) in the past 20 ka. In particular, four NEAs were found to approach each other and comet Encke at a similar epoch around 3200 BCE (and more unlikely around 4700 BCE as well). These asteroids, namely 2004 TG10, 2005 TF50, 2005 UR and 2015 TX24 were previously suggested by others as associated with or producing fireballs at Earth \citep[e.g.][]{Olech2017,Spurny2017,Devillepoix2021}. Our earlier study confirmed a potential past dynamical linkage was possible solely among these four NEAs and 2P/Encke - we found no other similarly significant linkages among the remaining 47 proposed Taurid complex NEOs.

\cite{Egal2021} showed that most of the 15 TSC fireballs detected by the European Fireball Network in 2015 also approached Encke between 3000 BCE and 3500 BCE. While a common origin for these fireballs, the four NEAs and Encke is still an unanswered question, no alternative scenario investigated in \cite{Egal2021} explained satisfactorily the orbital convergence of all these bodies around a common epoch. While this dynamical analysis is suggestive of a genetic link, a spectral analysis of 2004 TG10, 2005 TF50, 2005 UR and 2015 TX24 is required to prove or reject this hypothesis. 

\section{Methods} \label{sec:methods}

Meteoroid stream simulations performed in this work follow the methodology described in \cite{Egal2019}. Using an ephemeris of the comet's past orbital behaviour as its basis, millions of test particles are ejected from the comet nucleus and numerically integrated forward in time. The characteristics of  Earth-impacting particles, as well as the dynamical evolution of the meteoroid stream, are recorded at discrete time steps. In order to calibrate the model with meteor observations, our primary simulation outputs are: 

\begin{enumerate}
	\item the date and duration of potential meteor activity on Earth
	\item the characteristics (age, size, etc.) of the trails involved
	\item the stream's structure close to Earth 
	\item the radiants of Earth-intercepting meteoroids 
	\item the approximate shower intensity
\end{enumerate}

In this work, we have conducted a series of independent simulations of the formation of the Taurid meteoroid complex. Our analysis includes the development of meteoroid trails using different possible orbital histories of Encke, as well as from other potential parent bodies. The simulation sets were all created following the same methodology, but some differ in the number of particles integrated, the age of the trails or the frequency of particle ejection. The parameters of each simulation set considered in this work are presented in Table \ref{tab:simulation_parameters} and in Appendix E. The purpose of this section is to summarize the main model parameters chosen for our meteoroid stream simulations. For additional information about the model, the reader is referred to \cite{Egal2019} and \cite{Egal2020c}.

\begin{table*}
	\begin{tabular}{lrrrrrrrrl}
		ID & Body  & $N_\text{app}/\text{sz1}$ & $N_\text{app}/\text{sz2}$ & $N_\text{app}/\text{sz3}$ & $Np_\text{/app}$ & 
		$Np_\text{tot}$ & $f_{app}$ & Y$_{beg}$ & Solution \\ 
		Run 0 & 2P/Encke  & 312 & 0  & 0  & 100  & 3.12$\times10^4$ & 33 & 30 000 BCE  & 2 \\
		Run 1     & 2P/Encke  & 92  & 92 & 91 & 1000 & 2.75$\times10^5$ & 100 & 30 000 BCE  & 2 \\  
		Run 2     & 2P/Encke  & 369 & 4278 & 329 &  300 & 1.49$\times10^6$ &  1-2* & 30 000 BCE  & 2 \\[0.1cm]
		\hline 
		\hline
		TG1 & 2004 TG10 & 1 & 1 & 1 & 1000 & 3.00$\times10^3$ & once & 3200 BCE & 4\\
		TF1 & 2005 TF50 & 1 & 1 & 1 & 1000 & 3.00$\times10^3$ & once & 3200 BCE & 5\\ 
		UR1 & 2005 UR & 1 & 1 & 1 & 1000 & 3.00$\times10^3$ & once & 3200 BCE & 6\\
		TX1 & 2015 TX24 & 1 & 1 & 1 & 1000 & 3.00$\times10^3$ & once & 3200 BCE & 7\\
		2P1 & 2P/Encke & 1 & 1 & 1 & 1000 & 3.00$\times10^3$ & once & 3200 BCE & 2 \\
		\hline
		\hline 
		TG2 & 2004 TG10 & 0 & 52 & 0 & 300 & 1.56$\times10^4$ & 30 & 3200 BCE & 4 \\
		TF2 & 2005 TF50 & 0 & 52 & 0 & 300 & 1.56$\times10^4$ & 30 & 3200 BCE & 5 \\
		UR2 & 2005 UR & 0 & 52 & 0 & 300 & 1.56$\times10^4$ & 30 & 3200 BCE & 6 \\
		TX2 & 2015 TX24 & 0 & 52 & 0 & 300 & 1.56$\times10^4$ & 30 & 3200 BCE & 7 \\
		2P2 & 2P/Encke & 0 & 52 & 0 & 300 & 1.56$\times10^4$ & 30 & 3200 BCE & 2 \\
		\hline
	\end{tabular}
	\caption{\label{tab:simulation_parameters} Summary of the simulation  parameters used to generate synthetic meteoroid streams from comet 2P/Encke and several NEOs. The first and second columns indicate the name of the simulation set and the parent body selected. Meteoroid trails were generated at every $f_{app}$ apparition of the comet since year Y$_{beg}$. 
	Most models involve the simulation of three synthetic trails at the selected perihelion passages, covering the size ranges sz1 (0.1-1 mm), sz2 (1-10 mm) and sz3 (1-10 cm). Since the value of $f_{app}$ may differ depending on the size range considered, parameters $N_\text{app}/\text{sz1,2,3}$ represent the number of apparitions of the parent body at which meteoroids were ejected in the size ranges sz1,2,3. $Np_\text{tot}$ indicates the total number of particles simulated for each data set. Finally, the column Solution represents which orbital solution for 2P/Encke as summarized in Table \ref{tab:IC} was used as the parent body ephemeris when ejecting meteoroids.}
\end{table*}
 
 \subsection{Stream modelling} 
 
 \subsubsection{Particle ejection}\label{sec:ejection}
 
 In this work, we wish to simulate the meteoroids ejected from comet Encke going back to 30 000 BCE, a reasonable upper limit to the expected age of the complex given prior work \citep[e.g.][]{Steel1991, Babadzhanov1990, Tomko2019}. Our model assumes ejection speeds consistent with gas drag release; we model the comet as a spherical nucleus of 2.4 km in radius, with a density of 1 g~cm$^{-3}$, a geometric albedo of 0.046 and assume that 10\% of the nucleus' surface is active (cf., Section \ref{sec:nucleus}). Meteoroids are taken to have a density and albedo similar to the comet. 
 
 As described in \cite{Egal2019}, particles are ejected at each time step where the heliocentric distance $r_h$ is below 0.5 AU. Though most of Encke's dust emission is confined to heliocentric distances below 1.3 AU (cf. Figure \ref{fig:Afp}), we release particles only around perihelion since a sensitivity analysis revealed that the ejection location of the meteoroids do not significantly alter the stream structure around the Earth at the present epoch (cf. Appendix B).
 
 Simulated meteoroids are ejected each day from the sunlit hemisphere of the nucleus, following the ejection velocity model of \cite{Crifo1997}. This model is restricted to the sublimation of frozen water, responsible for $\sim$90\% of the volatiles flowing out of a comet below 3 AU \citep{Combi2004}, and provides ejection velocities from nearly zero to 180 m/s in our simulations. 

 Given the considerable time span considered in this work ($\simeq$32 ka), particles were not ejected at each apparition of the comet. Meteoroid trails were simulated every $f_\text{app}$ of the comet, with $f_\text{app}$ constant for each individual run. The particles created are divided among the following three size/mass/magnitude bins:
 
 \begin{enumerate}
 	\item sz1: $[10^{-4},10^{-3}]$ m, $[10^{-9},10^{-6}]$ kg, [+10 to +3] mag
 	\item sz2: $[10^{-3},10^{-2}]$ m, $[10^{-6},10^{-3}]$ kg, [+3 to -4] mag
 	\item sz3: $[10^{-2},10^{-1}]$ m, $[10^{-3},1]$ kg, [-4 to -10] mag
 \end{enumerate}

 These bins approximately correspond to meteors detectable by patrol radars (bin 1), visual/video means (bin 2) and as fireballs (bin 3). Details of the simulations parameters (size bin considered, number of particles ejected and frequency of apparition of the comet $f_{app}$) are presented in Table \ref{tab:simulation_parameters}.
 
 \subsubsection{Alternative parent bodies}
 
 In addition to comet Encke, we have investigated the potential contribution of NEAs 2004 TG10, 2005 TF50, 2005 UR and 2015 TX24 to the Taurid meteor showers. In \cite{Egal2021}, we analyzed the consequences to TMC formation of a hypothetical breakup event resulting in the separation of these four asteroids around 3200 BCE. Using simple forward models mimicking both a collisional fragmentation and a gentler separation of cometary fragments 5 to 6 millennia ago, we concluded that such breakup event is insufficient to reproduce the full extent of the TMC but may contribute to the core of the Taurid showers. 
 
 In Section \ref{sec:NEAs_ejection}, we explore the possibility of additional meteoroid ejection from 2004 TG10, 2005 TF50, 2005 UR and 2015 TX24. No conclusive proof of cometary-type activity for these NEAs has been found; however, given the lack of available observations we cannot strongly exclude such activity. To complement \cite{Egal2021}'s analysis, we simulate the ejection of meteoroid streams by cometary processes from each of these asteroids beginning in 3200 BCE (and 4900 BCE), corresponding to the time of the possible fragmentation event found in \cite{Egal2021}. For each NEA, we assume a nucleus density of 1g~cm$^{-3}$ and a diameter of 1.32 km for 2004 TG10, 300 m for 2005 TF50, 170 m for 2005 UR and 250 m for 2015 TX24 \citep{Lamy2004,Nugent2015,Masiero2017}. The bodies' Bond albedos were taken as 0.017 for 2004 TG10, 0.04 for 2005 TF50 and 0.07 for 2005 UR and 2015 TX24 \citep{Nugent2015,Masiero2017}. The parameters of the meteoroid stream simulations which we performed from various TMC parents are summarized in Table \ref{tab:simulation_parameters}. 
 
 In Section \ref{sec:clones_ejection}, we analyze alternative ephemerides for comet Encke over the past 32 ka. It has been proposed that larger NGFs may have acted on Encke in the past \citep{Steel1996}, raising doubts about the nominal solution ("nominal Encke") analyzed in Section \ref{sec:orbital_evolution_2P}. Using variations of the A1, A2, and A3 NGF coefficients \citep{Marsden1973} found for Encke as a proxy, we explored possible orbital histories of the comet and compared the meteoroid streams modelled from these alternative ephemerides with nominal Encke. In total, we have generated a full synthetic Taurid stream for 112 potential past orbital histories of Encke as described in Section \ref{sec:ejection}. 
 
 \subsubsection{Integration} \label{sec:integration}
 
 The comet and meteoroids in our simulations are all integrated with a 15$^\text{th}$ order RADAU algorithm \citep{Everhart1985}, with a precision control parameter LL of 12 and an external time step of one day. The integration is performed considering the gravitational influence of the Sun, the Moon, and the eight planets of the solar system. Planet positions are obtained using the INPOP13 planetary solution \citep{Fienga2014}. General relativistic corrections and non-gravitational forces are also taken into account. The meteoroids are integrated as test particles in the stream, under the influence of solar radiation pressure and Poynting-Robertson drag. The Yarkovsky-Radzievskii effect is neglected since the size of our simulated particles does not exceed 10 cm \citep{Vokrouhlicky2000}.  The integrations are performed in a Sun-centered, ecliptic J2000 coordinate system. 
 
 \subsection{Data analysis}
 
 \subsubsection{Terminology}
 
 Meteoroids that approach Earth at the current epoch are selected as potential meteors and compared with Taurid observations. In order to avoid confusion, we adopt in this paper the terminology defined in \cite{Egal2020c}. The term "nodal crossing location" is applied to the position of the particle when physically crossing the ecliptic plane, while the words "nodes", "node location" or "nodal footprint" indicate the ecliptic position of the particle's ascending and descending nodes (as determined by its osculating orbit at the epoch of interest) even if the particle itself is far from the ecliptic plane. 
 
 The terms "impacting particles" and "impactors" are applied to simulated meteoroids whose actual position in space best represent meteoroids which may collide with the Earth. In this work, we retain as possible impactors meteoroids that have a Minimum Orbital Intersecting Distance (MOID) below $DX=0.01$ AU (Sections \ref{sec:nominal_ejection} and \ref{sec:NEAs_ejection}) or $DX=0.05$ AU (Section \ref{sec:clones_ejection}) with Earth's orbit at the present epoch. Because typical Taurid orbits have low inclinations ($i$ < 10$\degree$), we find that selection based on the MOID and not the nodal crossing distance is most realistic in reproducing particles that  approach the Earth closely, though their nodes may be far from the Earth's orbit. 
 
 To complement the MOID-based selection, we also examine the relative time between the Earth and the particle's arrival at the orbital location where the MOID is reached. We use this parameter, denoted $\Delta DT$, to remove or decrease the contribution of particles that approach the MOID long before (or after) the Earth has reached this location. In our simulations, meteoroids that would arrive more than 30 days (Sections \ref{sec:nominal_ejection} and \ref{sec:NEAs_ejection}) or 100 days (Section \ref{sec:clones_ejection}) from the Earth's passage of the MOID longitude are excluded from our analysis.
 
 \subsubsection{Weighting scheme} \label{section:weights}
 
 Simulated particles selected with the $DX$ and $DT$ criteria are used to model the meteor showers' activity profiles and radiant structures. To bridge the gap between the small number of particles simulated and the real number of meteoroids ejected by the parent body, each particle is assigned a weight representing the actual number of meteoroids released by the comet under similar ejection circumstances \citep{Egal2019}. The weighted number of particles approaching Earth at a given date is then measured to estimate the model predicted meteoroid flux $\mathcal{F}$. This may then be converted into a zenithal hourly rate \citep[ZHR,][]{Koschack1990} for ease of comparison with observations. 
 
 The weighting scheme adopted in this work follows the general methodology presented in \cite{Egal2020b}. The weight of each particle depends on:
 
 \begin{itemize}
 
 \item The initial number of particles ejected at a given epoch and with a given size ($W_0$)
 \item The dust production of the parent body at that heliocentric distance ($W_{r_h}$, measured via Af$\rho$)
 \item The differential  size frequency distribution of the particle radii after ejection ($W_u$)
 \end{itemize}
 
 Among these three weights, the value of the size distribution index $u$ in $W_u$ has the largest influence on the shape of the modelled activity profiles. 
 To improve agreement between the modelled and observed activity, additional empirical weights can be applied to the simulation. However, these only have secondary effects on the shape of the modelled profiles. Such weights can involve the age of the simulated trails, or the distance of the particle to the planet at the time of the prediction \citep[cf.][]{Egal2020c}. However, given the complexity of reproducing four distinct meteor showers with one unique model, we choose to restrict the number of tunable parameters to a minimum. 

In this work, we therefore keep only one variable parameter, to be determined from our calibration fits, within our weighting scheme: the size distribution index $u$ (involved in the computation of $W_u$). The values of $W_0$ and $W_{r_h}$, which depend on the number of particles simulated and the heliocentric distance of the particle at its ejection respectively, are determined automatically during the processing and do not require any specific calibration. In addition, a fixed normalization coefficient $K$ is used to scale all the ZHR predictions to the average activity level observed for the Taurid showers as described in \cite{Egal2019}.

 \subsubsection{Validation}
 
 To validate our stream models, we calibrate our simulations on more than two decades of observations of the NTA, STA, BTA and ZPE. The radiant structure, average intensity and variability of each shower as measured with visual, optical and radar systems and presented in \cite{Egal2022} form the basis of this constraining data. The major results of that earlier study have been summarized in Section \ref{sec:meteor_observations}. 
 
 In this work, we compare the orbital elements and the radiants of our modelled Taurid meteoroids with the observations of the Canadian Meteor Orbital Radar (CMOR, \cite{Webster2004}; wavelet analysis and individual meteors), the photographic meteors selected by \cite{Svoren2011}, the nighttime Taurids sub-components identified by \cite{Jenniskens2016} and any meteor referenced as a NTA or STA in the optical CAMS database between 2010 and 2016 (v3.0). Details about each dataset are provided in \cite{Egal2022}. 
 
 As a consequence of the low meteor rates of the Taurid showers, the activity profiles of the NTA, STA, BTA and ZPE display significant variability from year to year. The number, location and strength of the activity peaks were measured to change over time and vary between optical and radar observations. For this reason, we do not seek to reproduce the characteristics of each shower during every specific year, but rather focus our analysis on modelling the average duration, peak time and intensity of the four showers. 
 
 To accomplish this, we calibrate our modelled streams on the average activity profile of the NTA, STA, BTA and ZPE presented in \cite{Egal2022} (see their Figures 1 and 2). Since the CMOR radar provides  consistent, long-term coverage of both day and nighttime Taurid showers for almost 20 years,  we use the average activity measured by CMOR between 2002 and 2021 as the primary baseline of our comparison. We note, however, that the shape of CMOR activity  profiles, after being scaled to match the shower intensity measured from visual observations, are in good agreement with optical data. The main difference between the optical and radar activity profiles is the first peak of the STA around 197$\degree$ SL, which is at least twice as strong in the radar profile than in visual data. 
 
 \section{Meteoroid Ejection from the Nominal Orbit of 2P/Encke} \label{sec:nominal_ejection}
 
To provide a baseline for comparison of stream activity produced by the ensemble of clones of Encke, in this section we explore in detail the evolution and current activity of the meteoroids released by our nominal clone of comet Encke, integrated without NGFs (i.e., using solution 2 of Table \ref{tab:IC}). Since this orbital solution is a representative model for the comet's evolution within observational uncertainties (cf. section \ref{sec:orbital_evolution_2P}), we eject a large number of meteoroids from the comet beginning in 30 000 BCE to produce good number statistics. However, as we will see, this nominal orbital solution does not provide a good match to the overall characteristics of the Taurid complex. To illustrate the resulting mismatch and motivate the search for alternative clones, we describe the results of this model in some detail. 
 
 Using this nominal orbital evolution for Encke, we performed three sets of meteoroids ejection to investigate different aspects of the stream's orbital evolution. These simulations are referred to as "Run 0, 1, 2"  in Table \ref{tab:simulation_parameters}. In Run 0, particles of size 0.1 to 1 mm were ejected once every 33 apparitions of the comet (i.e., roughly every 100 years) to monitor the dispersion of small particles more sensitive to radiative forces. Run 1 encompassed particles of all sizes and was meant to compare the evolution of particles across the different size bins, with close to 100 particles in each size range ejected every 300 years from the comet's nominal orbit. Run 2 focused on increasing the ejection frequency of particles that would produce visual and optical meteors, with meteoroids of sizes of 1 mm to 1 cm being simulated for almost every apparition of the comet. Taken together, this "nominal Encke" dataset comprises about 1.8 million particles generated from the comet's nominal ephemeris (integrated without NGF) from 30 000 BCE to the current time. 
 
 \subsection{Nodal crossing locations}
 
 \begin{figure}
  \includegraphics[trim={0.1cm 0.65cm 0.1cm 1.5cm},clip,width=.5\textwidth]{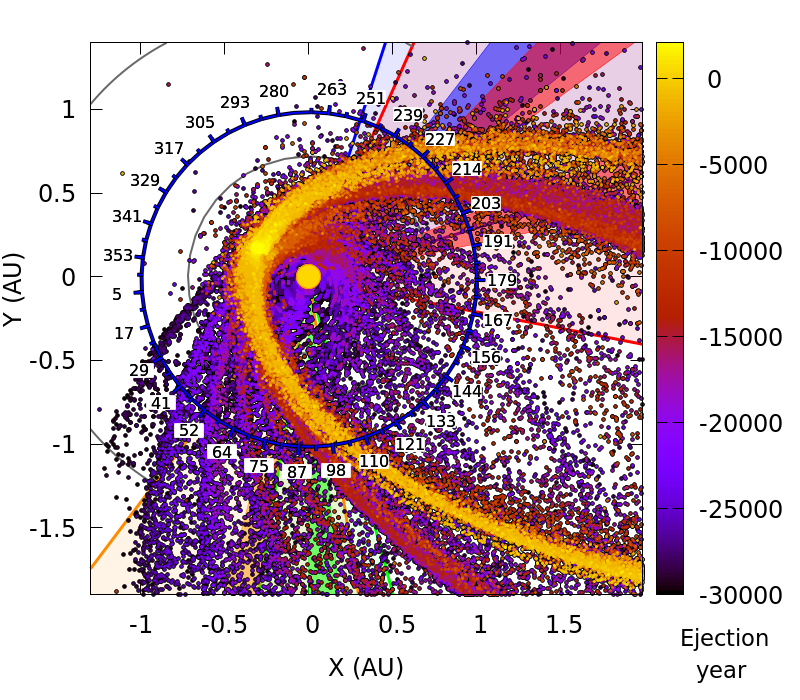}
  \caption{\label{fig:nominal_nodes} Nodal crossing locations of the meteoroids ejected from 2P/Encke's nominal clone (solution 2), shown in 2021. The orbit of the Earth is represented in blue, with the corresponding solar longitude indicated along the orbit. The nodal crossing locations are color-coded as a function of the particle's ejection epoch. The shaded areas represent the duration and maximum activity times of the NTA (in blue), STA (in red), BTA (in green) and ZPE (in orange). }
\end{figure}

 The nodal crossing locations of the meteoroid stream produced from solution 2 circa 2021 is shown in Figure \ref{fig:nominal_nodes}. An animation representing the meteoroids orbital motion since 30 000 BCE using the Run 1 data set is available in supplementary material (cf. Figure C1.1). After ejection, differential precession of the meteoroid orbits causes the nodal footprint of the stream to extend along ellipsoid-shapes in the ecliptic plane (hereafter called "streamlets"). With time, these streamlets also evolve to intersect the Earth's orbit at larger solar longitudes, creating the multiple branches observed in Figure \ref{fig:nominal_nodes}. 
 
Figure \ref{fig:nominal_nodes} also shows four zones of different colors corresponding to the average time and duration of the major Taurid showers determined in \cite{Egal2022}. The time of each shower is indicated by the semi-transparent areas, while epochs of maximum activity are represented in darker colors. Blue, red, green and orange areas are related to the NTA, STA, BTA and ZPE meteor showers respectively. 
 
Examining Figure \ref{fig:nominal_nodes}, we see that meteoroids ejected from nominal Encke over the last 30 ka now cover a wide range of solar longitude, ranging from 20 to 260$\degree$ SL. The highest concentration of meteoroids around Earth's orbit is found between SL 90-110$\degree$ and 200-235$\degree$, overlapping the observed maximum activity times of the BTA and NTA. However, the number of meteoroids approaching Earth's orbit is much smaller outside these longitudes ranges, and no strong density enhancement is noted at the time of maximum activity of the ZPE (around 77$\degree$ SL) or at the first STA peak of activity (at 197$\degree$ SL). 
 
It may seem in Figure \ref{fig:nominal_nodes} that there is a gradient in the meteoroids age as function of the solar longitude. However, meteoroids ejected a few tens of thousand years ago also cross the Earth's orbit where the younger trails are located; the nodal footprint of these old trails is simply more dispersed than material ejected in the past millennia. Differential precession acting upon long timescales explains why meteoroids ejected 20 to 30 ka ago cover a solar longitude range of about 240$\degree$; in our simulation, only material of that age can contribute to the early activity of the STA and the ZPE.    
 
 \begin{figure}
     \centering
     \includegraphics[width=.5\textwidth]{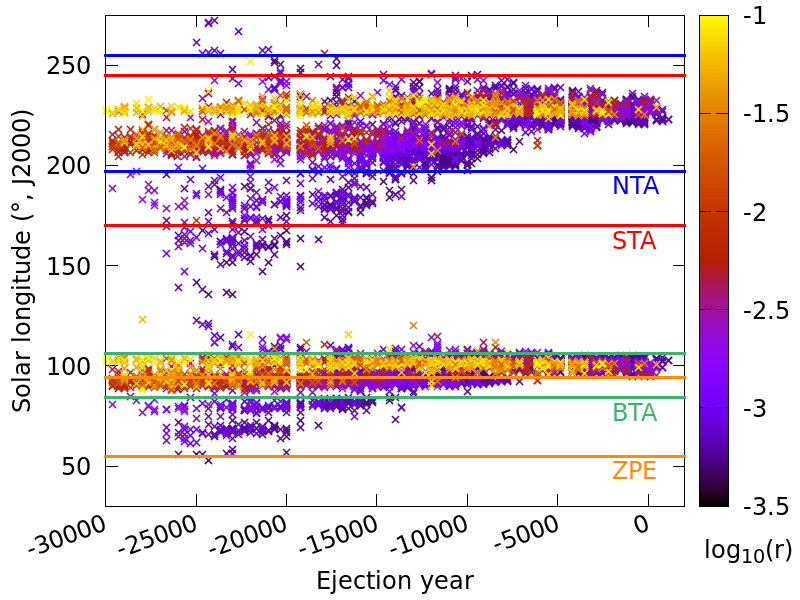}
     \caption{Solar longitudes of the particles approaching Earth's orbit circa 2021, as a function of the ejection epoch for solution 2. Particles are color coded by size, ranging from 0.5 mm to 10 cm in radii. The blue, red, green and orange lines indicate the beginning and end of the NTA, STA, BTA and ZPE activity respectively as reported by CMOR. 
     \label{fig:nominal_age}}
 \end{figure}
 
 The lack of meteoroids during the beginning of the STA and ZPE activity is visible in Figure \ref{fig:nominal_age}, which represents the solar longitude of the meteoroids approaching Earth in 2021 as function of the ejection epoch for solution 2. Only meteoroids with radii greater than 0.5 mm (corresponding to the CMOR detection limit) and approaching Earth's orbit with a MOID below 0.05 AU are presented. The average activity period of the NTA (197-255$\degree$), STA (170-245$\degree$), BTA (84-106$\degree$) and ZPE (55-94$\degree$) is indicated with red, blue, green and orange lines respectively. 

As shown in Figure \ref{fig:nominal_nodes}, most of the simulated meteoroids are concentrated around 94-106$\degree$ and 220-230$\degree$ SL. The SL dispersion of the meteoroids in 2021 increases for old trails, allowing a lower limit to be set for the age of the modelled showers. For this simulation set, trails ejected prior to 10 000 BCE and 17 000 BCE are necessary to reproduce the durations of the NTA and STA respectively. We observe that the BTA can be constructed with trails ejected after 13 000 BCE, while the ZPE require the contribution of meteoroids ejected prior to 20 000 BCE. The preliminary analysis of this stream model indicates that the NTA/BTA branch may be younger than the STA/ZPE, which is consistent with the different perihelion longitudes dispersion reported for each shower (cf. Section \ref{sec:meteor_observations}). 
 
 \subsection{Activity profile} \label{sec:nominal_profile}
 
The showers' model activity profile based on solution 2 is constructed by retaining particles having a MOID with Earth below 0.01 AU; we consider these as potential impactors. In a second step, particles that are more than 30 days of approaching or leaving the location of the Earth at the longitude of the MOID are excluded from the impactor population. This second selection stage permits examination of only those particles that are relatively close to the MOID when the Earth is physically near the MOID as well. 
 After selection, the particles are assigned a weight as described in Section \ref{section:weights}, and the modelled ZHR profile computed as detailed in \cite{Egal2019}.
 
  To compare with observations, we split the meteoroids into four branches depending on each particle's ecliptic radiant latitude (above or below the ecliptic plane) and solar longitude at impact (before or after 140$\degree$ SL). We thus obtain four modelled profiles at each year, that are compared with the average profiles measured by the CMOR radar \citep[cf. analysis in][]{Egal2022}. As little variation is seen in the activity and radiant structure modelled each year, we use year 2021 as a representative example of the results obtained with the nominal Encke dataset. For comparison, the average activity profile and radiant obtained from this model between 2002 and 2021 is provided in Appendix C2.1. 
  
\begin{figure}
     \centering
     \includegraphics[width=.5\textwidth]{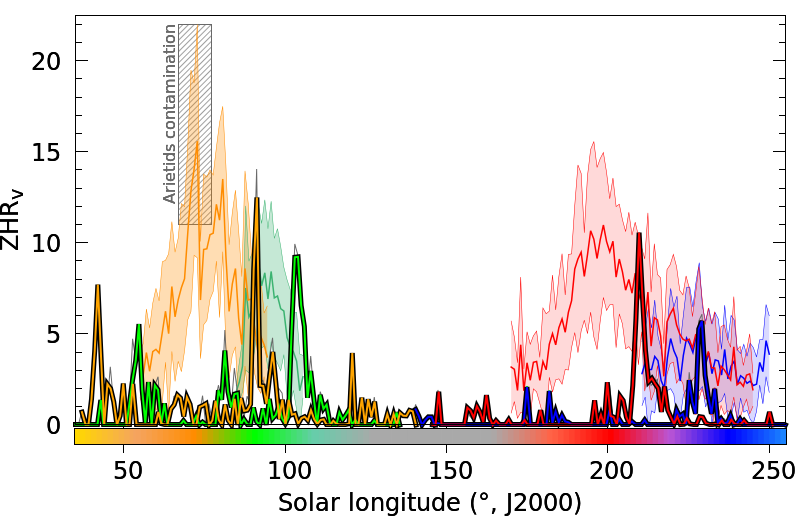}
     \caption{Simulated ZHR$_v$ profile of the NTA (blue), STA (red), BTA (green) and ZPE (orange) as thick solid lines as generated from meteoroids ejected using solution 2 for comet Encke. The observations of the average activity profile measured by CMOR as given in \protect\cite{Egal2022} are shown as thin lines with uncertainty bounds using the same colors.}
     \label{fig:nominal_profile}
 \end{figure}

  The four activity profiles modelled for year 2021 are shown in Figure \ref{fig:nominal_profile}. The modelled profiles were scaled to the average activity levels measured in the visual range, denoted by ZHR$_v$.
 As expected from the nodal footprint of the stream presented in Figure \ref{fig:nominal_nodes}, we see that our simulated profiles do not generally reproduce the time of maximum or duration of the four major Taurid showers. Though the maximum modelled NTA activity matches the reported peak time, the peak dates of the three other showers are discordant with the observations by several degrees in solar longitude. In particular, little activity is detected from the ZPE prior to 87$\degree$ SL, or from the STA before 205$\degree$ SL, in contradiction with observations.
 
 The modelled profiles in Figure \ref{fig:nominal_profile} were obtained using a unique size distribution index $u$ of 3.2 at the meteoroids ejection (i.e., a mass index $s$ of 1.75). We examined profiles obtained with different values of $u$ and found they were in no better agreement with the showers' observations. The profiles obtained by averaging our simulated activity over several years (cf. Appendix C2) do not resolve the large discrepancy between our model predictions and the observations. 
 
 \subsection{Radiants}

 \begin{figure}
     \centering
     \includegraphics[width=.5\textwidth]{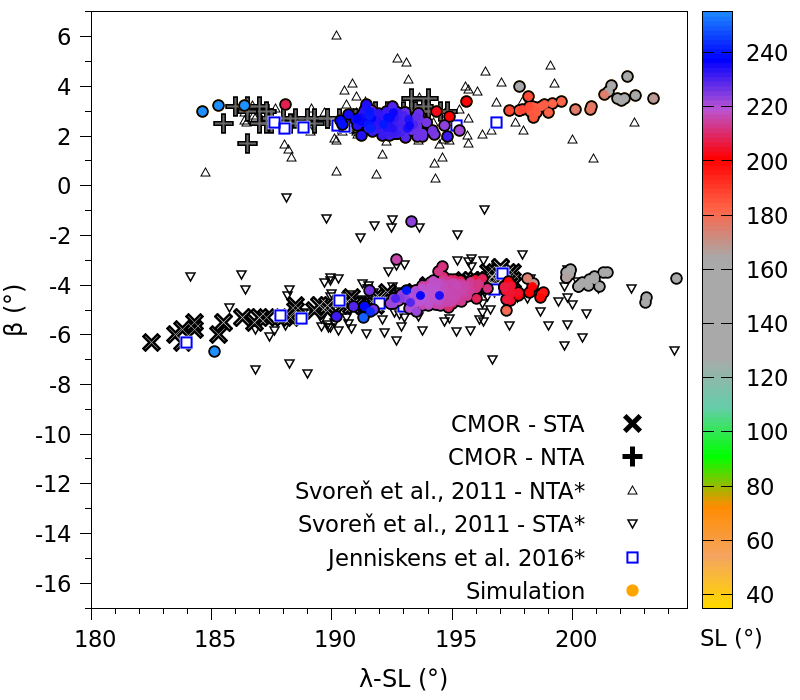}
     \includegraphics[width=.5\textwidth]{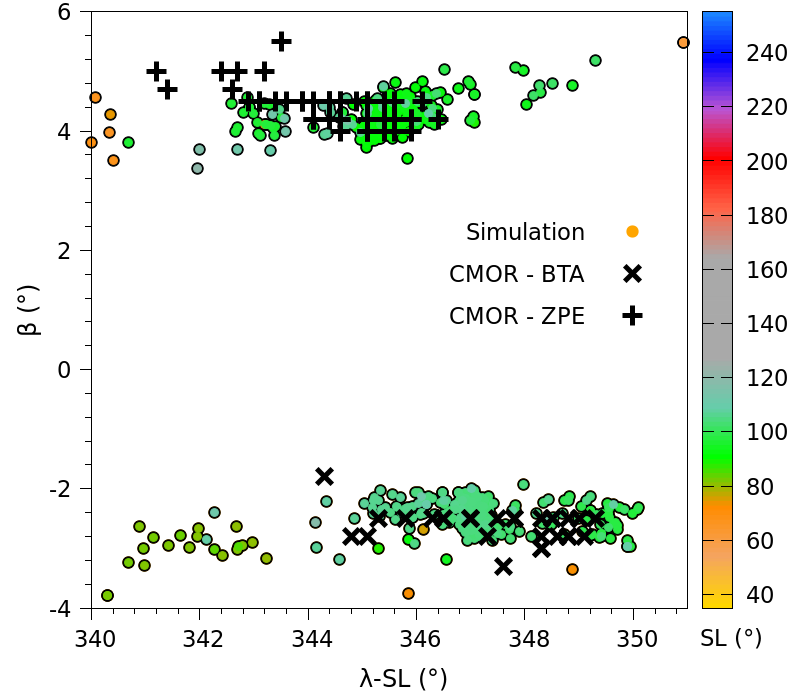}
     \caption{Ecliptic sun-centered longitude and latitude of the NTA, STA (top panel) and BTA, ZPE (bottom panel) simulated radiants as function of solar longitude (color coded). The simulated radiants are compared with optical observations presented in \protect\cite{Svoren2011} and \protect\cite{Jenniskens2016}, as well as with the average shower radiant location (in 1$\degree$ SL) determined from the wavelet analysis of CMOR 29 MHz data.}
     \label{fig:nominal_radiant}
 \end{figure}
 
 The ecliptic radiants of the impacting meteoroids from our nominal Encke simulation, color-coded as function of the solar longitude, are shown in Figure \ref{fig:nominal_radiant}. The simulated radiants in 2021 are compared with selected NTA and STA meteors from the IAU MDC \citep{Svoren2011} and the CAMS database \citep{Jenniskens2016}, as well as with the core of the shower observed with the CMOR radar (based on a wavelet analysis). The average location of the simulated radiants between 2002 and 2021, computed at each degree in solar longitude, is provided in Appendix C2.
 
Overall this is a good match between the simulated and observed radiants. Our model reproduces the radiant location and dispersion of the core of the four major showers, except for the early STA radiants (for $\lambda - SL<190\degree$). This missing portion of the radiant's structure is probably related to the lack of STA meteoroids in the simulations particularly at SL less than 205$\degree$, since a few meteors were found to populate this radiant area in different years (cf., Figure C1).
 
We note that the modelled ZPE radiants are in good agreement with the average radiant observed by CMOR, despite the fact that the arrival time of our simulated ZPE differs significantly from observations (see Figure \ref{fig:nominal_radiant}). This underscores the need for stream models to consider both radiant and activity profiles to improve fidelity.
 
 In agreement with previous work \citep[e.g.,][]{Tomko2019}, the comparison of our modelled radiants with the observations suggests a genetic linkage between comet Encke and the four major Taurid showers. However, the timing of model predicted activity and duration of these showers is very different from observations. From these comparisons, we conclude that we cannot reproduce these showers from our nominal version of comet Encke alone.  The divergence between our modelled stream and the Taurid meteor showers may be explained by two main hypotheses: either the selected orbital solution considered for Encke does not represent the past trajectory of the comet, or different parent bodies are necessary to reproduce the extent of the Taurid stream complex. We examine the second of these possibilities first.

 
 \section{Forming the Taurids using NEAs} \label{sec:NEAs_ejection}
 
The long activity of the major Taurid showers has posed a challenge to most models of the TMC to date. In order to explain the wide distribution of orbits in the complex, scenarios combining sublimation-driven meteoroid ejection and successive fragmentations of a larger, progenitor cometary nucleus have been proposed \citep{Clube1984,Steel1991,Asher1991}. In line with this hypothesis, a long list of  NEAs associated with the TMC has also been proposed and some NEAs directly linked to minor Taurid showers and fireballs (cf. Section \ref{sec:NEAs_analysis}). 
Most previous work has considered such NEAs as generally not the source of the Taurid meteoroids, but rather as the largest remnants of the fragmentation/splitting of a precursor parent body which released the meteoroids into space. Taurid complex NEAs may therefore help identify specific dust ejection events from past millennia that contribute to the present Taurid activity. 

In \cite{Egal2021}, we identified NEAs 2004 TG10, 2005 TF50, 2005 UR and 2015 TX24 as being possibly dynamically linked to comet Encke and several STA meteoroids. We refer to these five bodies, including the four NEAs and comet Encke, as the G5 group. We found that all of the G5 members were at a small MOID and reduced relative velocity with respect to each other around a common epoch circa 3200 BCE. Though we were not able to entirely exclude the possibility that this orbital convergence is due to a coincidental combination of several dynamical effects, this phenomenon could also be the marker of the fragmentation of a large parent body about 5 to 6 millennia ago \citep{Egal2021}. 

Though such a breakup event may have resulted in substantial dust production, the simulation of meteoroids ejected during this event was found insufficient to explain the full extent of the TMC and all the features of the TSC observed in 2015 \citep{Egal2021}. In this section, we perform a more detailed analysis to investigate the expected contribution of material released during such a splitting event to the four major Taurid showers. 

\subsection{breakup with small velocity ejection}

We begin by simulating a low-speed breakup scenario, corresponding to the separation of cometary fragments with a velocity of a few meters per second \citep{Boehnhardt2004}. Since the orbit of the hypothetical progenitor is unknown, we eject 3000 particles from each NEA and Encke in 3200 BCE  \citep[the time corresponding to the possible breakup found in][]{Egal2021}, for a total of 15 000 simulated meteoroids. The particles were ejected with velocities of 2 m/s, and integrated until 2021 as described in Section \ref{sec:integration}. 

\begin{figure}
    \centering
    \includegraphics[width=0.49\textwidth]{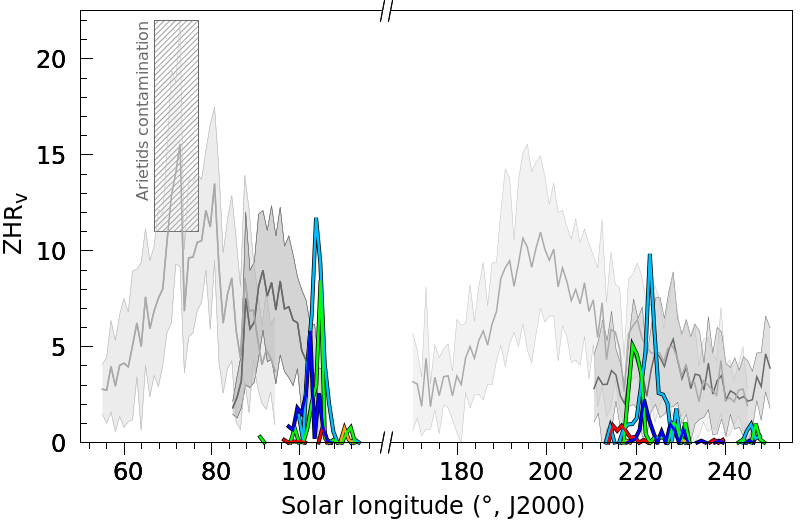}
    \includegraphics[width=0.49\textwidth]{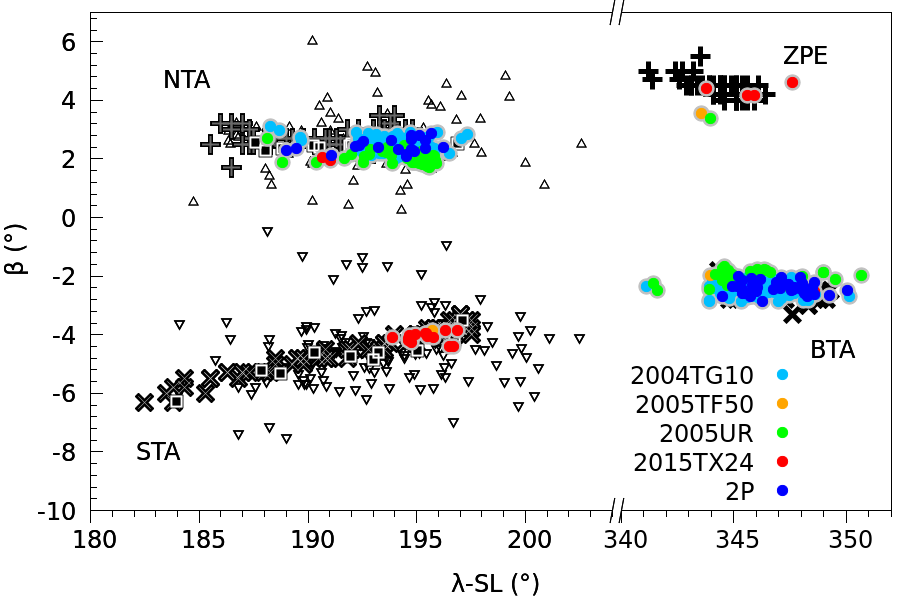}
    \caption{Simulated activity profiles (top) and sun-centered ecliptic radiants (bottom) of meteoroids approaching Earth's orbit in 2021. Particles were released from NEAs 2004 TG10 (cyan), 2005 TF50 (orange), 2005 UR (green), 2015 TX24 (red) and comet Encke (dark blue) in 3200 BCE, with an ejection velocity of 2 m/s. The average activity profile of the NTA, STA, BTA and ZPE determined from CMOR measurements is shown in grey.}
    \label{fig:NEAs_ms}
\end{figure}

The activity profiles and the radiants of particles approaching Earth's orbit with a MOID below 0.01 AU in 2021 are presented in Figure \ref{fig:NEAs_ms}. The nodal distribution of the simulated trails in 2021, very similar to the 2015 distribution provided in \cite{Egal2021}, is not presented here. As found by these authors, meteoroids ejected using the model of \cite{Crifo1997}, leading to ejection velocities of 1 to 125 m/s, produced identical results for stream activity today as the model using ejection speeds of 2 m/s. Only the profiles and radiants of the stream model ejected at 2 m/s from the nucleus are therefore discussed in this section.

Figure \ref{fig:NEAs_ms} shows that meteoroids ejected from each of these five objects would all be expected to produce meteor activity at similar solar longitudes, concentrated around 96-110$\degree$ SL and 212-232$\degree$ SL. NEAs 2004 TG10, 2005 UR and comet Encke are the main contributors to the activity, with radiants concentrated in the NTA and BTA areas. We observe almost no activity related to asteroid 2005 TF50, and only a moderate contribution from NEA 2015 TX24 with radiants located in the STA and ZPE regions. 

As noted previously, low-speed ejecta from the G5 members in 3200 BCE do not reproduce the timing of maximum activity or the duration of the four major Taurid showers. Meteoroids ejected five thousand years ago from each object, including comet Encke, have not precessed sufficiently to produce the four radiant clusters observed for the Taurids. 

Though no conclusive cometary-like activity has been observed for any NEA of the G5 group, could subsequent (post 3200 BCE) cometary-type ejection from these objects increase the width of the profiles presented in Figure \ref{fig:NEAs_ms}? Assuming that the G5 NEAs are genetically linked to Encke (and thus possess a cometary nature), we simulate the ejection of additional meteoroids trails from these objects using the model of \cite{Crifo1997}. About 15 600 particles were released at every 30 apparitions of each G5 member since 3200 BCE (cf. Table \ref{tab:simulation_parameters}). The resulting radiants and activity profiles in 2021 are presented in Appendix D, Figure D1.

We observe that ongoing meteoroid production from the G5 does not significantly alter the radiant distribution or the solar longitude timing of activity in 2021. The main difference between ongoing ejection and a single, low-speed ejection event at 3200 BCE concerns the relative contribution of the different G5 members. We note that with the ejection of more recent trails, Encke becomes primarily responsible for the activity in 2021 (especially during the spring around 100$\degree$ SL); there is also a significant increase in the number of particles predicted to impact Earth from 2015 TX24 in the STA branch around 216$\degree$ SL. In contrast, activity from ejecta originating with NEA 2005 UR becomes negligible compared with Encke, 2004 TG10 and 2015 TX24. From the poor match between the simulated activity profiles and radiant distributions for this scenario, we conclude that the slow fragmentation of a large parent body about 5 000 years ago can not explain by itself the long duration and the radiant structure of the major Taurid showers, even assuming ongoing cometary-like activity from each of the residual G5 fragments.  

\subsection{breakup with high velocity ejection} \label{sec:highV_scenario}

\begin{figure}
    \centering
    \includegraphics[width=0.49\textwidth]{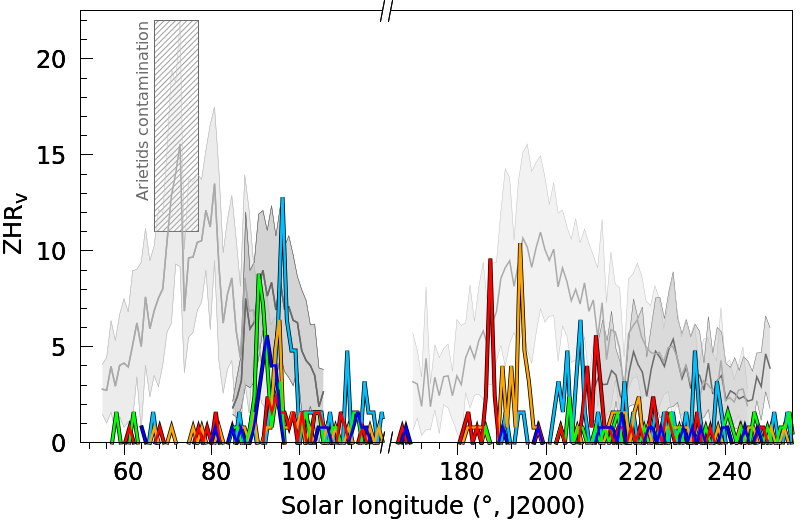}
    \includegraphics[width=0.49\textwidth]{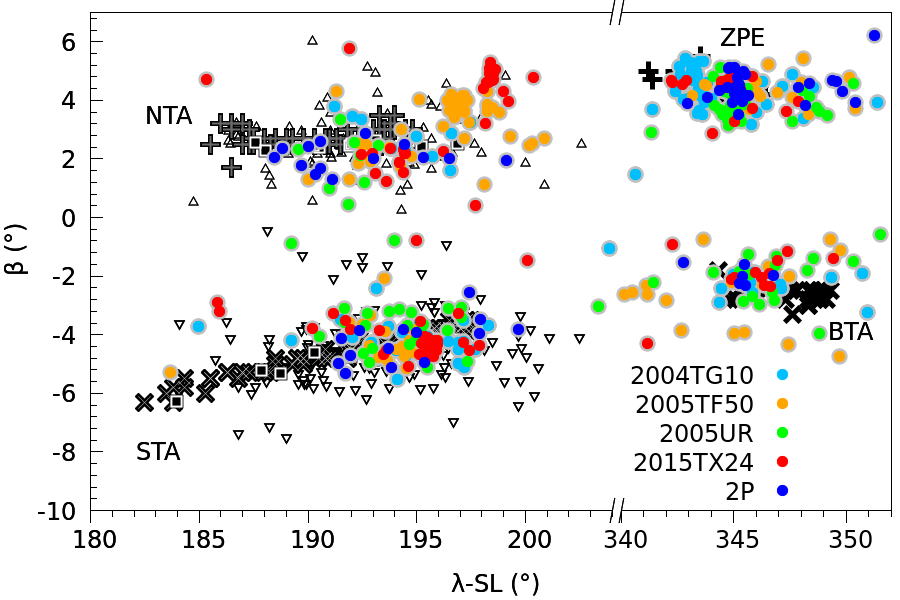}
    \caption{Simulated activity profiles (top) and sun-centered ecliptic radiants (bottom) of meteoroids approaching Earth's orbit in 2021. Particles were released from NEAs 2004 TG10 (cyan), 2005 TF50 (orange), 2005 UR (green), 2015 TX24 (red) and comet 2P/Encke (dark blue) in 3200 BCE, with an ejection velocity of 1 km/s. The average activity profile of the NTA, STA, BTA and ZPE determined from CMOR measurements is represented in grey. }
    \label{fig:NEAs_kms}
\end{figure}

We now examine the hypothesis of a high-velocity breakup of the G5 progenitor, caused by a collision with another body. The velocity dispersion of meteoroids ejected during such event could exceed a few hundred meters per second \citep{Hyodo2020}. To explore this scenario, we simulate the ejection of meteoroids from the G5 in 3200 BCE as previously, but considering this time ejection velocities of 1 km/s. The model activity profiles and radiants of meteoroids approaching Earth's orbit in 2021 with a MOID below 0.01 AU are presented in Figure \ref{fig:NEAs_kms}.

As expected from the meteoroids' nodal crossing distribution \citep[cf. Figure 15 in][]{Egal2021}, the activity profiles of the simulated showers cover a large range in solar longitude, from 56$\degree$ to more than 250$\degree$ SL. 2004 TG10 is found to produce significant activity at the time of the BTA, STA and late NTA. Encke and 2005 UR deliver meteoroids around 88-96$\degree$ SL, but do not significantly contribute to the autumn showers. In contrast, we find NEAs 2005 TF50 and 2015 TX24 now produce notable activity in spring (SL 92-98$\degree$) and during the early STA (SL 185-200$\degree$). Meteoroids ejected from asteroid 2015 TX24 also reach the Earth around 208-216$\degree$ SL. 

In the bottom panel of Figure \ref{fig:NEAs_kms}, we see that meteoroids ejected at high velocity from the G5 bodies cover most of the observed area of Taurid radiants. However, a careful check of the timing of each radiant (provided in Appendix D, Figures D2 and D3) reveals a systematic  difference between the simulated and the observed time of the radiants.

For example, the simulated peaks of activity around SL 88-98$\degree$, matching the time of the BTA meteor shower, are produced by meteoroids with radiants located above the ecliptic plane (i.e., compatible with the radiant of the ZPE). Similarly, the peak in activity produced by 2005 TF50 during the STA ($\simeq$196$\degree$ SL) arrives at Earth from radiants located in the NTA branch. The same conclusion is drawn when analyzing the autumn  peaks related to 2004 TG10 and 2015 TX24; that is this scenario would produce an "inversion" of the time and radiants of the modelled meteoroids compared with observations.

In principle, one could examine the evolution of high-speed meteoroids ejected from the G5 at different times, with greater ejection velocities or from different parent bodies in an effort to reduce the timing/radiant discrepancy. However, an unlimited range of scenarios becomes possible (and increasingly complex/unlikely). Hence we restrict ourselves to the conclusions presented in \cite{Egal2021}, which are that: 
\begin{enumerate}
    \item only the G5 members showed a plausible orbital convergence with each other in the past, among a sample of 52 bodies
    \item If real, this convergence occurred between 3000 BCE and 4900 BCE (and more convincingly around 3200 BCE) 
\end{enumerate}
 
Scenarios beyond these are outside the scope of this work. However, we also investigated the hypothesis of a high-velocity breakup of the G5 progenitor around 4700 BCE, corresponding to the other (and less likely) epoch of orbital convergence between the NEAs and Encke identified in \cite{Egal2021}. We generated about 3000 particles from each NEA around 4700 BCE, ejected with a velocity of 1 km/s. The particles were integrated until the present epoch and processed as described in Section \ref{sec:highV_scenario}. However, this model produced very similar results to the breakup at high-velocity simulated around 3200 BCE and was not further investigated.

In summary, our simple model of a high-speed progenitor breakup (violent nucleus fragmentation or collisional disruption) circa 3200 BCE fails to simultaneously explain both the radiant structure and the apparition time of the major Taurid showers. High-speed ejecta from Encke, combined with meteoroids released from four NEAs potentially linked with the comet, do not provide a satisfactory match with observations. In the following section, we explore a hybrid scenario combining the continuous ejection of meteoroids from an object located on Encke's orbit and the hypothetical fragmentation of the G5 progenitor five thousand years ago. 

\subsection{Hybrid Formation Scenario}

\cite{Steel1991} and \cite{Asher1991} interpreted the positive correlation of the Taurids' semi-major axis $a$ and eccentricity $e$ as a sign of continuous ejection of meteoroids from the parent body over several millennia. Based on the range of observed Taurid longitudes of ascending node $\Omega$, the authors estimated a possible age of $\sim$10 ka for the TMC. However, the numerical simulations of cometary ejection, even at high velocity (from 0.25 to 2 km/s), failed to reproduce the scatter observed for Taurid orbits from the data then available. 

These authors proposed that a hybrid model combining ejection at perihelion from the parent body and successive disruptions of cometary fragments in the asteroid belt (that originally separated from the progenitor around perihelion) could increase the scatter of the modelled Taurid orbits. It was suggested that distinct fragmentations of a progenitor cometary nuclei in the main asteroid belt could also explain the clustering in the ($q$,$\Omega$) plot observed for the NTA.  

\begin{figure}
     \centering
     \includegraphics[width=.5\textwidth]{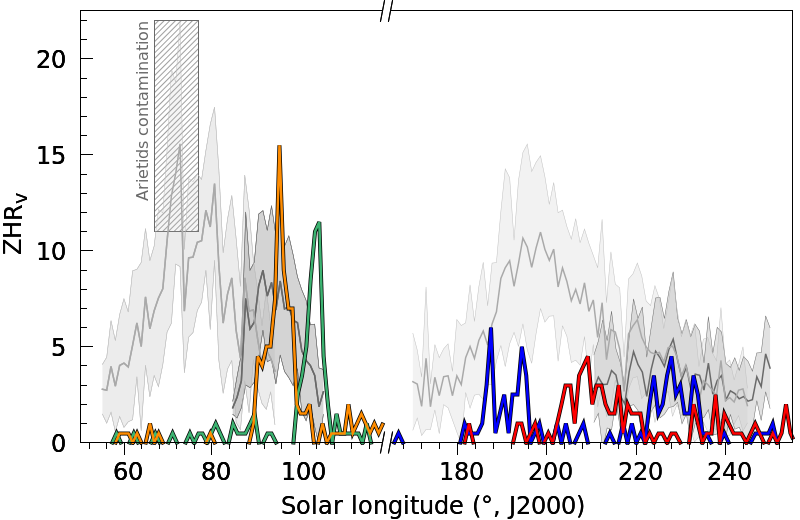}\\
     \includegraphics[width=.5\textwidth]{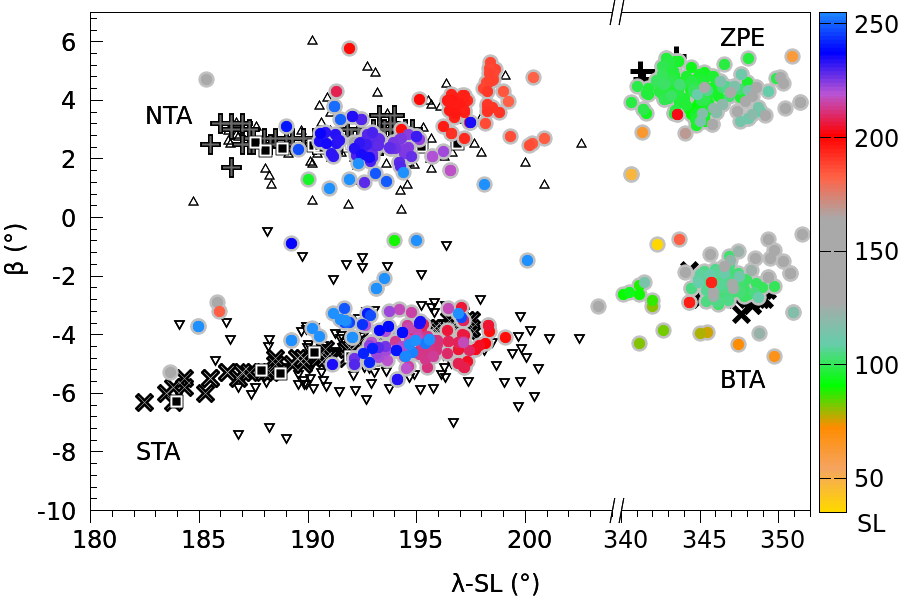}
     \caption{Simulated activity profile (top panel) and radiants (bottom panel) obtained when combining the continuous ejection of meteoroids from Encke since 30 000 BCE and the high-speed ejecta released by the comet and four NEAs around 3200 BCE (see text for details). The simulated profiles are coloured as a function of the epoch of apparition and the radiant location. Autumn showers with a radiant located above or below the ecliptic plane are represented in blue and red respectively. Spring showers with northern or southern radiants are plotted in orange and green. The sun-centered ecliptic radiants are color-coded as a function of the solar longitude. }
     \label{fig:NEAs_hybrid}
 \end{figure}

As a variation on this original hybrid scenario, we analyze the timing and radiant distributions obtained when combining the regular outgassing of comet Encke around perihelion since 30 000 BCE (nominal Encke) and the meteoroids ejected at 1 km/s from the G5 group around 3200 BCE. The modelled profiles and radiants computed from particles approaching Earth's orbit in 2021 (with a MOID $<$ 0.01 AU) are presented in Figure \ref{fig:NEAs_hybrid}. 
The activity profiles are color-coded as a function of the apparition date and location of the radiant: meteors located in the northern branch are represented in blue in autumn and orange in spring, while the southern branch is drawn in red in autumn and green in spring. The radiants are color-coded as a function of solar longitude at Earth intersection (bottom plot).

We find this hybrid model better reproduces the profile and radiants of the NTA and late STA ($>$200$\degree$ SL) than the previous scenarios. The meteoroids also populate the southern and northern branches of the stream intersecting Earth's orbit in spring; however, the modelled BTA peak occur about 10 days after the observed maximum activity of the shower ($\simeq$91-95$\degree$ SL) and almost no activity is recorded at the time of the ZPE. The model produces significant activity during the early STA (185-196$\degree$ SL), but with radiants located in the northern sky. 

The perihelion distance ($q$), semi-major axis ($a$), eccentricity ($e$) and longitude of the ascending node ($\Omega$) of the simulated meteoroids is presented in Appendix D, Figure D4. In that plot the orbital elements of particles retained in 2021 for each ejection model (low-speed splitting, high-speed breakup circa 3200 BCE and hybrid scenario) are compared with the orbits of the NTA, STA, BTA and ZPE measured by CAMS and CMOR (wavelet analysis). 

We observe that most simulated meteoroids of the STA and NTA branches occur within the orbital range reported by CAMS. The simulations are also in good agreement with the location of the BTA core detected by CMOR (wavelet analysis). However, as expected from our previous profiles and radiant plots, none of the explored scenarios reproduce the observed ZPE orbits. The dispersion of meteoroids ejected with a velocity of 1 km/s matches or exceeds the scatter observed for the NTA, STA and BTA streams. However, most simulated meteoroids are concentrated into much smaller orbital ranges, of sizes more comparable to the core of the streams observed by CAMS and CMOR \citep[cf. Appendix C2 in][]{Egal2022}. 

We find our hybrid model simulation to be compatible with the results of \cite{Steel1991} and \cite{Asher1991}. The modelled particles reproduce the general trends in ($a$, $e$) and ($\Omega$, $q$) summarized in \cite{Egal2022}. The model does not produce a clear trend in ($a$, $e$), hence we cannot reproduce the observed negative correlation between the meteoroids semi-major axis and eccentricity reported for the STA core. On the other hand, we note that ejection from the G5 members at high velocity (green or blue model in Figure D4) produce two clusters of NTA meteoroids in the ($\Omega$, $q$) plot (separated by the vertical line in the first panel of in Appendix D, Figure D4).

These clusters are located respectively before and after the limit $\Omega=205\degree$ identified by \cite{Steel1991} from photographic meteor observations. Though our simulated particles do not entirely reproduce the two NTA overlapping structures observed by CAMS between $\Omega=130-210\degree$ and $\Omega=170-290\degree$, the hypothesis of the collisional breakup of a large progenitor a few thousand years ago is consistent with the observed double slope in ($\Omega$, $q$) measured for optical NTA meteors \citep{Egal2022}. 

Though the hybrid scenario improves the overall agreement between our simulations and optical/radar observations compared to ejection from nominal Encke alone, the match is still far from perfect. The three sets of simulations show a systematic time offset (i.e., a shift in $\Omega$) for the NTA and STA compared with the wavelet analysis of CMOR data. There is also a moderate difference between the semi-major axis and eccentricity of the modelled meteor showers and CMOR measurements, though this is reduced when comparing the simulations with streams' cores as observed by CAMS \citep[for the NTA and STA, cf.][]{Egal2022}. The simulated streams do not entirely fill the orbital range covered by CAMS, and we are unable to reproduce the ZPE branch.

Our model combining a long-term dust emission from the nominal orbit of Encke and material ejected during the breakup of the G5 progenitor five thousand years ago therefore is able to reproduce some, but not all, of the major observed characteristics of the TMC. This suggests additional fragmentation events may be necessary to reproduce the full extent of the TMC, or that the orbital history of Encke diverges from the nominal scenario explored in Section \ref{sec:nominal_ejection}. 

Next we explore this second possibility for formation of the TMC, namely ejection from a single 2P/Encke-like object but on an orbit different from nominal Encke. 

\section{Exploring Taurid formation from clones of 2P/Encke} \label{sec:clones_ejection}

In this section, we examine the influence of alternative ephemerides for Encke on TMC formation. In particular, we wish to establish if an alternate history of Encke can explain the main features of the TMC today without recourse to any fragmentation events. Despite the apparent orbital stability of the comet (cf. Section \ref{sec:orbital_evolution_2P}), its orbit is chaotic, and the origin and the dynamical history of Encke remain uncertain. 

Previous models of the comet's evolution have hinted at the possibility of strong nongravitational forces acting on its nucleus in the past (see Section \ref{sec:origin}). This could result in a  different ephemeris than the nominal solution used as basis of our simulations in Section \ref{sec:nominal_ejection}. However, observational constraints on the direction or the magnitude of these past NGFs exist for the past two centuries at best \citep{Sekanina2021}, a period much shorter than the timescale of the TMC formation. We will see that additional non-trivial constraints on Encke's orbit are provided by present-day observations of the Taurid showers (at least under the assumption that Encke is their parent), as well as clues to the origin of this unusual comet.

\subsection{Stream models}

To examine the influence of different NGFs acting on Encke over several millennia, we created a thousand clones of the comet's nominal solution, and integrated these backward to 30 000 BCE. All the clones of the comet started on Encke's nominal orbit, but were given different (fixed) NGF coefficients. The clones' A$_1$, A$_2$ and A$_3$ parameters were randomly selected to lie between 0 and 100 times Encke's default values stated in Table \ref{tab:IC} (solution 1), and assumed to remain constant during the integration. We recognize that such an assumption is far from realistic given the known variability of cometary NGFs \citep{Sekanina1993}. At this stage, we therefore do not pretend to provide an accurate model for Encke; we rather use the NGFs as a simple tool to explore how alternative trajectories within a reasonable range of orbital phase space of the comet's nominal orbit may affect our stream simulations. 

Among the thousand clones generated for Encke, we selected 17 bodies that covered the full range in semi-major axis occupied by the swarm of particles during the integration. The initial conditions of these selected clones, as well as the evolution of their orbital elements, are provided in Table E1.1 and Figure E2.1. For each of these clones, we generated a full synthetic meteoroid swarm, with 100 particles ejected since 30 000 BCE at every 100 returns to perihelion of the body, as described in Section \ref{sec:integration}. In total, about 10 000 particles of sizes between 1 and 10 mm were simulated from each clone and integrated to the present epoch. More particles were generated for one particularly promising clone of our sample, which will be detailed in Section \ref{section:A4}. We refer to this collection of 17 stream models with varying orbital histories of Encke as sample "A".

To complete our investigation, we also performed additional stream simulations, covering an even larger range of possible clones of comet Encke. A first run was performed by considering 20 extreme values of NGF (clones B09 to B30 in Table E1.2), ranging from 0 to 10$^{-8}$ AU.d$^{-2}$ in most cases. Another simulation set, restricted to more realistic NGF estimates, explored 75 alternative scenarios of the comet's evolution (clones B2-04 to B5-24 in Table E1.3 and E1.4). For these data sets, 5 particles were ejected at each perihelion return of the body, with sizes ranging from 0.1 mm to 10 cm. The evolution of all 95 additional clones of Encke, referred to as sample "B", is represented in Figure E2.1. 

Combining these simulations sample sets, we examined the structure, activity and radiant of all resulting 112 hypothetical meteoroid streams released from Encke's clones since 30 000 BCE. Taken as a whole, our stream models comprised about 4.9 million particles integrated over 30 000 years. After a careful selection and weighting of the particles, we compared for each data set the resulting modelled stream to the observed characteristics of the major Taurid showers.

\begin{figure*}
 \includegraphics[width=.98\textwidth]{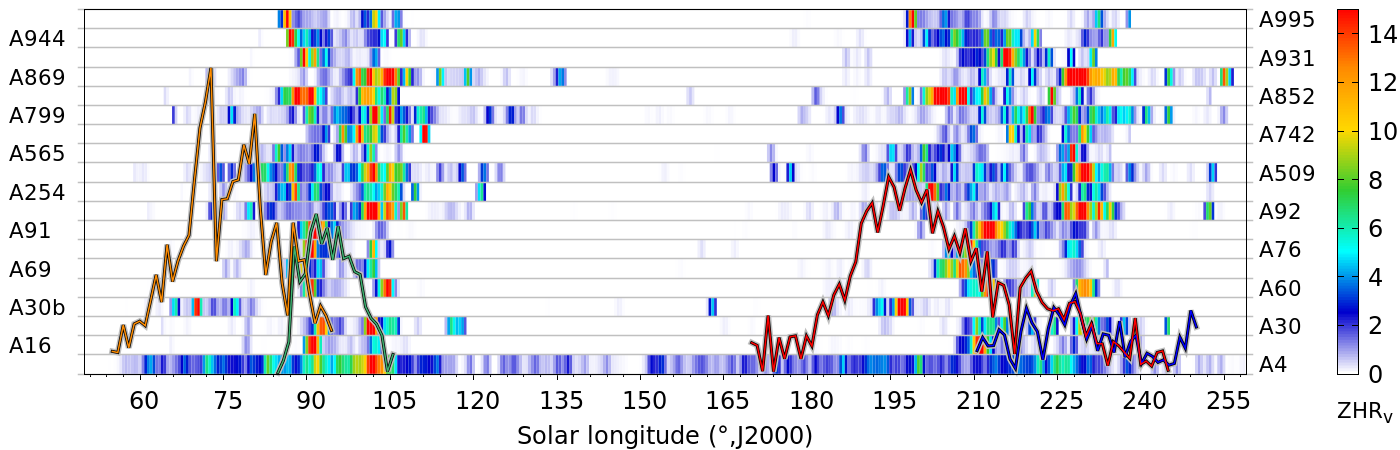}
 \caption{\label{fig:clonesA_profiles} Synthetic meteor activity produced in 2021 by meteoroid streams modelled from different ephemerides of comet Encke. The ZHR$_v$ profile as a function of solar longitude of the stream as seen at the Earth produced by each clone is represented by a horizontal band color-coded by the expected meteor rates. Maximum activity regions are represented by yellow to red areas, while moderate rates are illustrated in blue. The average activity profile of the NTA, STA, BTA and ZPE as measured by CMOR is presented for comparison with blue, red, green and orange lines respectively. The sample A labelling represents the number of the original random clone orbit from the original ensemble of 1000 Encke clones. }
\end{figure*}

\subsection{Selecting the most promising clones}

The activity profiles produced by the 17 clones of sample A are shown in Figure \ref{fig:clonesA_profiles}. For each model, the profile was constructed using meteoroids approaching Earth's orbit with a MOID below 0.01 AU in 2021. The modelled meteor rates are presented in the figure as a horizontal band, color-coded as a function of the model ZHR$_v$. This mode of representation allows quick comparison of the duration and location of the activity peaks (red and yellow areas) predicted for each clone with the average ZHR$_v$ measured by CMOR. Similar plots obtained for the 95 clones of sample "B" are provided in Figures E2.2 and E2.3 together with their radiant distribution (Figures E2.4 and E2.5). 

From Figure \ref{fig:clonesA_profiles}, we see that even a small variation of Encke's ephemeris produces significant modification of the shape, duration and timing of maximum activity of the modelled meteor showers. In contrast to the nominal model, we find some clones produce strong meteor activity at the time of the STA (e.g., A509, A565, A852, A995, B12, B21 or B30) and/or the ZPE (e.g., B23, B24,  B30, B3-04 to B3-06, B5-14, B5-22, where the dashed numbering is for internal record keeping, but still represents unique clones). We also see that numerous clones generate ZHR$_v$ peaks at the time of the NTA and BTA (e.g., A30, A869, B09, B16, etc.). An initial comparison suggests that none of the models considered shows a good match with the overall observed radar profiles of all four showers simultaneously in Figures \ref{fig:clonesA_profiles}, E2.2 or E2.3.

However, one of our simulated clones (A4) stands out from the others as particularly promising. Indeed, we note that only a few bodies in samples A and B generate meteor activity that approximate (or exceed) the long duration observed for the four Taurid showers (A4, B23, B24, B30, B4-04 and B3-04 to B3-07). However, most of these clones produce significant rates before 55$\degree$ SL or between 110$\degree$ and 160$\degree$ SL, where no strong Taurid activity is presently detected. After excluding these bodies, we find that only clone A4 and B3-07 are able to reproduce all the measured durations of the NTA, STA, BTA and ZPE showers simultaneously without predicting significant activity where none is observed. 

When comparing the time of the activity peaks predicted for each model, the superiority of clone A4 over B3-07 becomes clear. We observe that clone A4 produces significant activity at the time of the BTA (between 83$\degree$ and 106$\degree$ SL) and during the NTA maximum (between 221$\degree$ and 228$\degree$ SL). We also observe a secondary maximum of activity around 197$\degree$, matching the time reported by CMOR for the first STA peak, and around 72$\degree$ SL during the ZPE. In contrast, most of the activity produced by clone B3-07 is confined between 96-100$\degree$ SL and 231-237$\degree$ SL only, which is a several days after the reported BTA and NTA peaks. 

\subsection{Examining Clone A4} \label{section:A4}

\subsubsection{Profile and radiant}

To explore whether clone A4 of Encke alone can successfully reproduce the major observed features for all four Taurid showers,  we performed additional stream simulations with this particular ephemeris of Encke. We ejected $\simeq$60 particles at each perihelion passage of clone A4 beginning in 30 000 BCE, constructing a stream of about 600 000 particles. The meteoroids were simulated over the size ranges 0.1-1 mm, 1-10 mm and 10-100 mm, that roughly corresponds to the radar, optical and visual detector ranges. Particles approaching Earth with a MOID below 0.05 AU and $\Delta T < $ 100 days at the current epoch were retained as potential meteor-producers. 

Given the small number of meteoroids simulated (compared with the nominal scenario of Section \ref{sec:nominal_ejection}), it is not surprising that the model activity profiles display moderate year-to-year variability in peak magnitude (but not location) due to small number statistics. For this reason,  we average the contribution of the simulated showers between 2002 and 2021, and compare the results with the mean profiles measured by CMOR over the same time period. 

\begin{figure}
    \centering
    \includegraphics[width=0.5\textwidth]{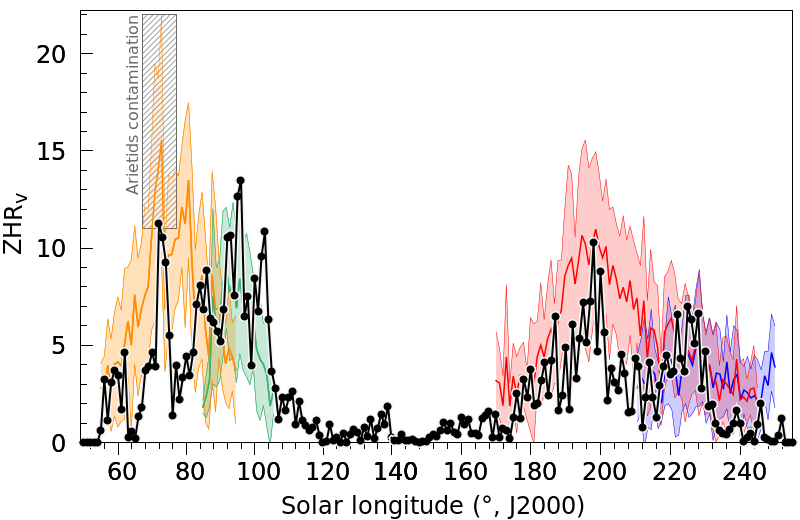}
    \includegraphics[width=0.5\textwidth]{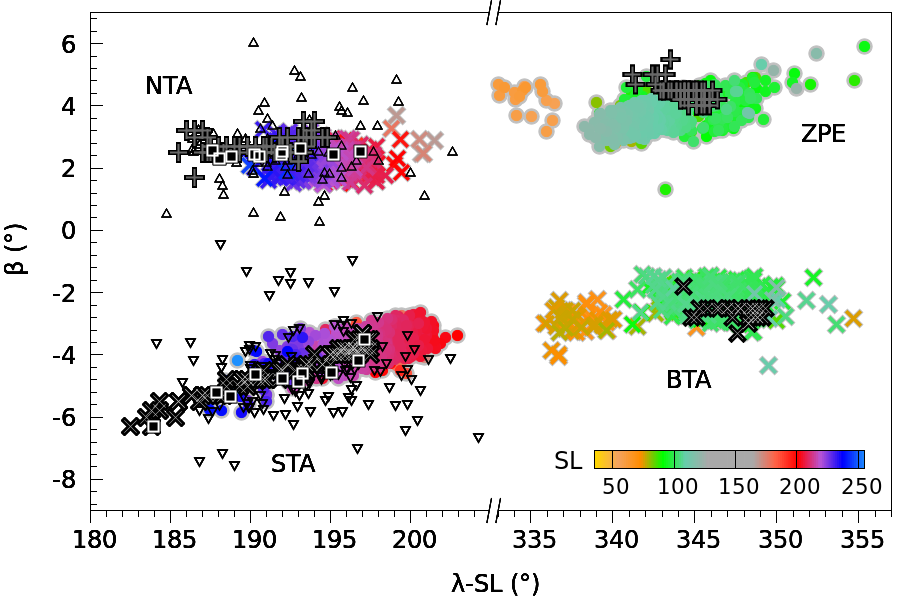}
    \caption{ \label{fig:A4} Simulated activity profile (top panel) and radiants (bottom panel) of the meteoroid stream generated from comet 2P/Encke's clone A4. The simulated profile (solid black line and dots) corresponds to the average activity modeled between 2002 and 2021, and compared with the mean activity recorded by CMOR (shaded regions) over the same time period. NTA, STA, BTA and ZPE profiles are represented in blue, red, green and orange respectively. The modelled radiants are color-coded as a function of the solar longitude, with BTA and NTA meteors plotted with crosses and ZPE or STA radiants with dots.}
\end{figure}

Figure \ref{fig:A4} shows the average meteor activity and radiants produced by clone A4. The characteristics of the showers simulated from clone A4 are in much better agreement with Taurid observations than any of our prior simulations. Most of the average modelled activity is confined between SL 55-107$\degree$ in spring and summer, and between SL 176-232$\degree$ in autumn. This model therefore reproduces the long duration of each Taurid shower (except for the late NTA after SL 232$\degree$), without predicting significant meteor rates at different times of the year when little to no Taurid activity is reported. 

In autumn, the simulated ZHR$_v$ profile peaks around 197-198$\degree$ SL, at the time of the first STA peak ($\simeq$197$\degree$), and between 222-228$\degree$ SL during the plateau of maximum NTA activity. In spring, the model predicts a maximum BTA intensity around 93-96$\degree$ SL, in agreement with CMOR observations. However, we note that our model does not fully reproduce the main peak observed for the ZPE, at 77$\degree$ SL; only the first peak of the ZPE, that shows contamination from the Arietids meteor shower in radar observations, is visible in the simulated profile around SL 74-75$\degree$. 

The radiants obtained with model A4 are also in agreement with the general structure of the Taurids (cf. Figure \ref{fig:A4}, bottom panel). In contrast to the fragmentation scenario of Section \ref{sec:NEAs_ejection}, most of the meteoroids ejected from body A4 have their radiants located in the northern hemisphere at the time of the NTA and ZPE, and in the southern hemisphere during the STA and BTA, consistent with observations. The model also reproduces the location and dispersion of the core of the four major showers, except for the early STA and NTA radiants (for $\lambda-SL<187\degree$). Additional simulations to increase the number of meteoroids retained for SL $>$ 231$\degree$, could potentially help fill in the small portion of missing NTA radiants. 

As expected from the simulated profile, we note that meteoroids produced from A4 do not fully reproduce both the timing and radiants of the ZPE. Though the model is in agreement with the observed ZPE radiants, the timing diverges by a few days from the reported date of maximum meteor activity. In contrast, some meteoroids occurring at the time of the ZPE have their radiants located in the BTA area, producing a small cluster of orange radiants visible in Figure \ref{fig:A4} for $\lambda-SL<340\degree$.

The orbital elements of the meteoroids ejected from clone A4 (and from the nominal clone), are compared with observations of the stream in Appendix F, Figures F1.1 to F1.4. As expected from the analysis of the showers' profiles and radiants, our simulations show  generally good agreement with the NTA, STA and BTA measured orbits. The simulated meteoroids are somewhat shifted in eccentricity and semi-major axis relative to the observations, but remain within the dispersion limits of the orbital elements. We find clone A4 to better reproduce the core of these showers than the nominal model, in particular for the BTA. Both models do not match well the time-evolution of the ZPE orbital elements.

Despite these limitations, we find that a single, unique weighting scheme applied to the stream formed by meteoroids ejected from body A4 since 30 000 BCE can reproduce the duration, timing and magnitude of the average NTA, STA, BTA and part of the ZPE meteor showers. We thus conclude that the four major Taurid showers can be reproduced via typical meteoroid ejection from a parent comet similar to Encke, but with a slightly different ephemeris than the nominal solution examined in Section \ref{sec:nominal_ejection}. This does not mean that the Taurids necessarily were produced this way, but the fact that this model produces a match to the four major Taurid streams that is significantly better in timing, duration, strength and radiant location compared to streams from more than 100 other Encke clones we used to generate potential showers is significant.  Thus the A4 ephemeris might provide insights into the history of 2P/Encke (whose origin and past orbital history is a mystery in itself) lead us to examine clone A4 in more detail.

\subsubsection{Meteoroid size distribution and age} \label{section:A4_distribution}

Having found a viable single orbital clone of Encke which simultaneously explains most of the properties of the four Taurid showers, we explore the implications of the model fit on the dust production and age of Encke.  In order to reproduce the relative intensity level of the four simulated showers, a higher size distribution index of the meteoroids at ejection was needed for the calibration of model A4. The best agreement between the model and CMOR observations (presented in Figure \ref{fig:A4}) was obtained using a size distribution index $u$ of 4.2, corresponding to a mass index of 2.06. This value exceeds the estimates of $u=$ 2.8 to $u=$ 3.6 derived from infrared and optical observations of Encke's dust particles in the range 1 mm to 10 cm (cf. Section  \ref{sec:nucleus}). 

\begin{figure}
  \includegraphics[width=.49\textwidth]{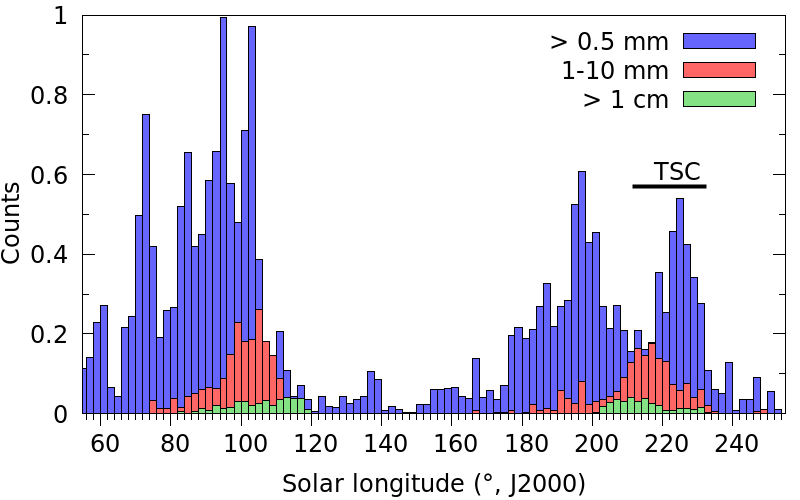}\\
  \includegraphics[width=.49\textwidth]{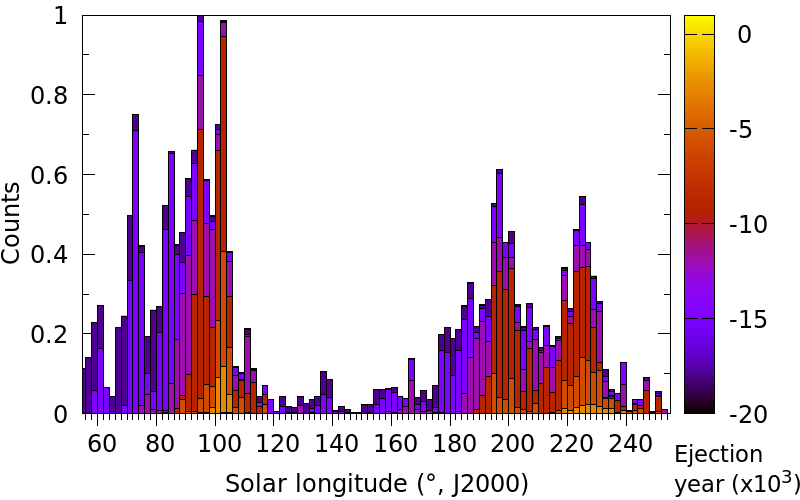}
  \caption{\label{fig:A4_distribution} Top: relative contribution of particles of different radii (from 0.5 mm to 10 cm) to the modelled average profile between 2002 and 2021 from meteoroids ejected by clone A4 of Encke. The solid line shows the observed time interval of fireballs associated with the Taurid Swarm Complex (TSC). Bottom: relative contribution of trails of different ages to the average activity profile modelled between 2002 and 2021 from A4. }
\end{figure}

The selection of a large $u$ value in our calibration, which increases the significance of small particles in our simulations, can be understood when examining the size distribution of the impacting particles as a function of the solar longitude. In Figure \ref{fig:A4_distribution} (top panel), we show the normalized distribution of particles impacting Earth for different size ranges, corresponding to meteoroids that could be detected with CMOR (blue) or optical instruments (red and green). The distribution of particles greater than 1 cm in radius, corresponding to meteors with magnitude brighter than -2 is represented in green.

We observe that in our model, most of the ZPE activity is produced by radar-sized meteoroids that would not be detected with optical instruments. Most of the early STA activity is caused by small particles as well, though mm-sized meteoroids are also present after SL 182$\degree$. This is consistent with the observation that the first STA peak around SL 197$\degree$ is twice as pronounced in radar data than in optical observations, implying that the STA are enriched in small particles at the beginning of their activity. Larger particles ($>$ 1 mm in radius) ejected from the clone A4 are found to contribute more significantly to the NTA, BTA and late STA activity. 

In our model, fireball-producing meteoroids only approach Earth between SL 200 and 232$\degree$ and between SL 86 and 120$\degree$, suggesting that some overdense radar echoes could be detected at the time of the BTA. This is also the time interval when a large swarm of impacts were detected on the moon in 1975 \citep{Duennebier1976}. 

Simulated cm-sized meteoroids are predicted from ejecta of A4 to encounter the Earth during the solar longitude interval where the returns of the Taurid resonant swarm (TSC) have been recorded (indicated with a solid black line in Figure \ref{fig:A4_distribution}). We note that the timing of most of our modelled fireballs from A4 in autumn precede by a few days the observed apparition of the resonant swarm. However, a full comparison between model A4's ejecta and the TSC observations requires identification in our simulations of those cm-sized meteoroids that are currently trapped into the 7:2 MMR with Jupiter. The detection of the TSC in our simulations is discussed in Section \ref{sec:discussion}.

We remark that the simulated profile, restricted to particles that would be detected by optical instruments (red and green distributions), produces enhanced meteor rates at the time of the first STA peak but does not match well the NTA activity. This suggests that the selected weighting scheme is not optimal for large particles, and that a better agreement with visual observations could be reached by tuning the calibration parameters on optical data alone. In particular, including a weight favouring the importance of young and dense trails \citep[similar to the $fM$ parameter of][]{Asher1999} would increase the average NTA and BTA levels and better reproduce the time and shape of the NTA $ZHR_v$ profile (see below). 

Since CMOR provides the only long-term, consistent source of measurements of all four Taurid showers, we prefer to calibrate our simulations on these radar observations. In addition, we aim to maintain the number of tunable parameters in our model as small as possible. However, we recognize that the adopted weighting solution is not unique, and that including additional weights could significantly improve the agreement between our model and optical data (cf. Section \ref{section:weights}). 

The bottom panel of Figure \ref{fig:A4_distribution} presents the contribution of trails of different ages to the average modelled profile. As observed for meteoroids ejected from the nominal orbit of Encke (cf. animation C1.1 available as a supplementary material), the differential precession of the simulated orbits over long timescales disperses the older trails over a wide range of solar longitudes. In Figure \ref{fig:A4_distribution}, the youngest trails only produce activity during the NTA and BTA meteor showers, while the older streams contribute to all four Taurid showers. The more recent trails involved in the NTA, STA and BTA showers are all concentrated around the reported times of maximum activity. 

Ejecting meteoroids from model A4, only material released prior to $\simeq$2000 BCE approaches the Earth at the present epoch; however, only trails ejected prior to $\simeq$5000 BCE produce significant meteor activity. The core of the NTA, STA and BTA showers can be modelled with trails ejected after 12 000 BCE, but the inclusion of older material is necessary to explain the ZPE and early STA activity. In our A4 simulations, trails generated prior to 19 000 BCE do not contribute to the present Taurid activity. 

The stream models generated from the nominal orbit of Encke or clone A4 both suggest that the NTA/BTA branch contains a higher proportion of young material than the STA/ZPE. This feature is produced primarily by the particle's precession rates rather than the comet's orbit, and is in agreement with the reported dispersion of the showers' perihelion longitudes. From Figure \ref{fig:A4_distribution}, we see that trails 7 to 14 ka old are probably sufficient to account for the formation of the NTA and BTA shower, while an age of 9 to 20 ka is more probable for the STA. Model A4 also suggests that the ZPE is the oldest of the four Taurid showers, with most meteors produced more than 16 ka ago. However, since model A4 does not reproduce accurately the characteristics of this shower, this estimate may be a lower limit.

Although clone A4 does not perfectly reproduce the activity and orbital structure of all aspects of the four major Taurid showers, the synthetic stream ejected from A4 is in much better agreement with observations than the nominal solution examined in Section \ref{sec:nominal_ejection} or any of the streams produced by the more than 100 clones of Encke we examined. Our simulations suggest that the general characteristics of the TMC can be modelled with classic cometary ejection from an object similar in size and activity to Encke between 5000 BCE and 19 000 BCE, but with a slightly different ephemeris than the nominal backward integrated solution for 2P/Encke.
		
\section{Discussion} \label{sec:discussion}

\begin{figure}
    \centering
    \includegraphics[width=0.5\textwidth]{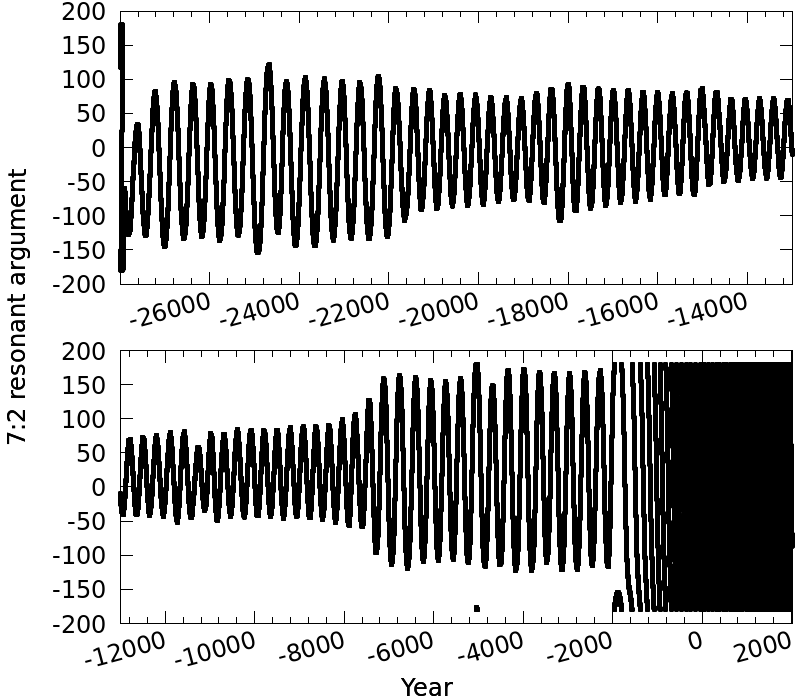}
    \caption{7:2 resonant argument for 2P/Encke's clone A4. A libration of the resonant angle $\sigma$ around 0$\degree$ indicates epochs when the body evolves inside the resonance. In particular, low-amplitude oscillations indicate that the body is strongly trapped into the 7:2 resonance. }
    \label{fig:72MMR_A4}
\end{figure}

Since clone A4 is able to best reproduce the Taurid meteoroid complex, we  focus in this section on the implications for the TMC formation assuming this specific orbital history of comet 2P/Encke. Because the 7:2 MMR plays an important role in the TMC evolution, we first examine if clone A4 spent significant time within the resonance in the past. To this end, we computed the value of the resonant argument $\sigma_{ 7:2}$ as described in \cite{Asher1991}: 
\begin{equation}
    \sigma_{ 7:2} = 7\lambda_J-2\lambda - 5\overline{\omega}
\end{equation}
where $\lambda_J$ is the mean longitude or Jupiter, and $\lambda$ and $\overline{\omega}$ the mean longitude and longitude of perihelion of the clone. The evolution of $\sigma_{ 7:2}$ during 30 ka is presented in Figure \ref{fig:72MMR_A4}. 

The librations of $\sigma_{ 7:2}$ around 0$\degree$ between 2000 BCE and 28 000 BCE clearly indicate that clone A4 has been trapped in the 7:2 MMR during most its evolution. In particular, we suspect clone A4 to have been strongly embedded into the resonance between 8000 BCE and 22 000 BCE, when the amplitude of the $\sigma_{ 7:2}$ oscillations decreases. 
We thus conclude that most of the meteoroids modelled from clone A4 that could contribute to the present-day Taurids (with ages ranging from 7 ka to 21 ka) were ejected in the 7:2 MMR. Material ejected after A4 left the resonance (around 2000 BCE) is not expected to produce any meteor activity at the current epoch (cf. Section \ref{section:A4_distribution}). 

\begin{figure*}
    \centering
    \includegraphics[width=\textwidth]{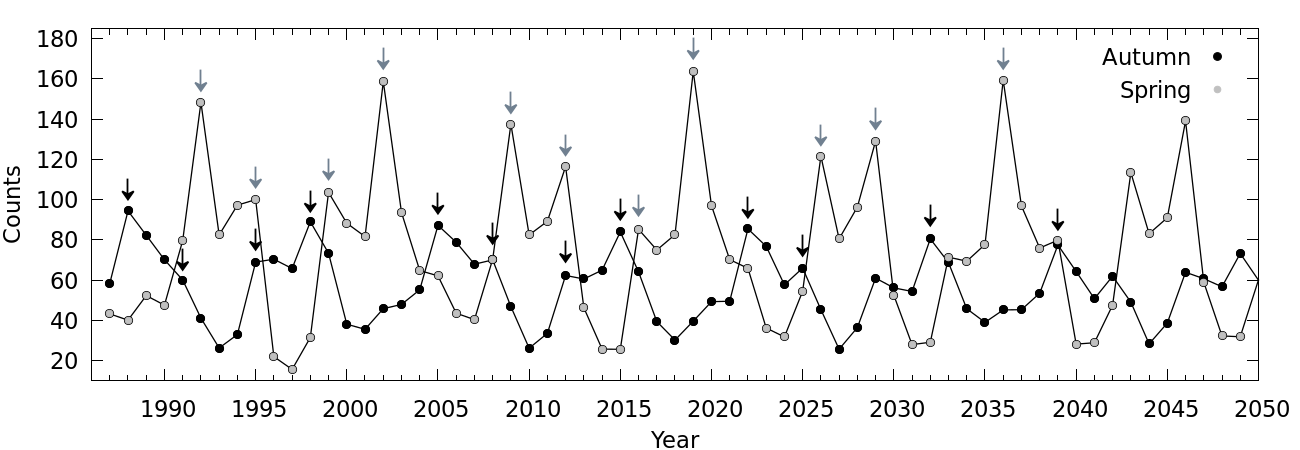}
    \caption{The number of particles simulated from comet 2P/Encke's clone A4 that are currently trapped in the 7:2 MMR with Jupiter and approach Earth's orbit in autumn (black) and spring (grey). Only particles with radii $>$1 cm and MOIDs with Earth's orbit below 0.02 AU were retained. Vertical arrows indicate the years of TSC returns predicted by \protect\cite{Asher1993b} in spring (grey) or autumn (black) between 1987 and 2039.}
    \label{fig:TSC_years}
\end{figure*}

Extending this analysis to the other 111 clones of comet Encke, we found that this dynamical situation for clone A4 is not unique. About 14\% of the clones  in our sample spent a few millennia in the 7:2 MMR during the past 30 ka\footnote{e.g., B12, B2-07, B4-13, B4-14, B4-16, B5-02, B5-03, B5-05, B5-15, B5-18 or B5-19.}, some of them being strongly trapped in the resonance during comparable periods of time as A4 (e.g., B5-18). However, we observe in Appendix E, Figure E2.3 that these clones fail in reproducing the peak times and the long duration of the Taurids. Releasing meteoroids within the 7:2 MMR during several millennia is therefore not a sufficient condition to reproduce the showers activity at the present epoch.

In contrast, meteoroids ejected from a body similar to Encke evolving in the 7:2 MMR may have contributed to the formation of the Taurid swarm complex. The TSC is expected to be rich in large meteoroids (small particles being removed faster from the resonance due to radiative forces) and to produce enhanced meteor activity at Earth every 3 to 7 years. Detailed modelling of the TSC and its impact on the Taurid meteor shower is presented in \cite{Asher1991} or \cite{Asher1993b}. While our focus is not to develop an exhaustive model of the TSC, we can identify components of our model that apply to the TSC. 

We performed a basic analysis of the A4 dataset to identify the presence of the TSC in our simulations. We examined the evolution of the $\sigma_{7:2}$ argument for each particle of radius greater than 1 cm (corresponding to a meteor of magnitude brighter than -2), in order to determine which simulated meteoroids are currently trapped in the 7:2 MMR with Jupiter. We found about 2500 resonant particles that approach Earth at the present epoch with a MOID $<$0.02 AU and a $\Delta T <$100 days. 

Figure \ref{fig:TSC_years} illustrates the number of resonant impactors predicted to encounter the Earth in autumn (black) and spring (grey) between 1987 and 2050. Despite the small number of particles predicted to impact Earth per year ($<$180), overall the model shows good agreement with the dates of TSC returns predicted by \cite{Asher1993b}, indicated by vertical arrows in the figure\footnote{\url{https://www.cantab.net/users/davidasher/taurid/swarmyears.html}, accessed in April 2022}. Our model displays a local maximum at each predicted swarm year (especially in spring), with the exception of 1991 that corresponded to a "near miss" resonance year \citep[e.g.][]{Asher1998,Beech2004,Johannink2006,Dubietis2007}. 

However, we observe that many of the individual swarm years do not stand out as extraordinary in Figure \ref{fig:TSC_years}, with similar particle numbers recorded in the years before and after the predicted TSC returns (indicated with vertical arrows in the figure). We interpret this artefact to be due to our resonant particle selection criterion, which allows a single particle to impact the Earth during several consecutive years without being removed. Unfortunately, the small number of particles retained when imposing more restrictive selection criteria ($<$10 for most years) prevents us from high fidelity modelling of the TSC. 

From our simple model, we predict an enhancement in the number of fireballs at TSC times in the southern branch in autumn and northern branch in spring. Most of the selected resonant fireballs in our simulations belong to the STA ($\simeq$48\%) and the ZPE ($\simeq$39\%), with a minor number associated with the BTA ($\simeq$10\%) and the NTA ($\simeq$3\%). This model is thus in good agreement with meteor observations that report enhanced TSC activity having radiants most closely associated with the STA \citep{Spurny2017}. 

In contrast to this good agreement for the years of expected TSC returns, we note that the timing within a given year of the modelled swarm returns are early by several days compared to the observed TSC apparitions (cf. Figure F2.2). This underscores that our model is not well-adapted to large meteoroids. Additional modelling from clone A4 (or using a slightly different ephemeris than A4) would be a good next step to further refine the past ephemeris through comparison with past TSC observations. 

In this regard, the detailed ephemeris of clone A4 is provided as supplementary material to this paper \citep{A4data}. The agreement between our simulations using A4 and the general TMC characteristics indicate that the specific orbital evolution of A4 is likely the closest available "history" of the recent dynamical evolution of 2P/Encke. 

We do not find in the A4 ephemeris any significant change in the body's orbit that could explain the discovery of Encke as late as in 1786. However, as the currently observable stream is fed by material  ejected only prior to ~2000-5000 BCE, our orbital constraints do not speak explicitly to this more recent epoch.   

For the same reason, we emphasize that the scenario where 2P/Encke alone following an orbit similar to A4 populates the current TMC cannot strictly rule out a fragmentation event around 3200 BCE which might have resulted in the separation of Encke and several NEAs. The accumulation of meteoroids ejected from several millennia of cometary outgassing from a large body (moving on an orbit similar to A4) with a significant fragmentation event around 3000 to 5000 BCE close to the 7:2 MMR would be consistent with the orbital dispersion of the TMC and the formation of the TSC. 

The results detailed in \cite{Egal2021} and Section \ref{sec:NEAs_ejection} show that the splitting of a large progenitor 5 to 6 thousand years ago can efficiently inject meteoroids into the 7:2 MMR, that would intersect the Earth's orbit at the time of the TSC (cf. Figure \ref{fig:NEAs_ms}). However, different ephemeris and/or fragmentation epochs than the ones considered in Section \ref{sec:NEAs_ejection} would be required to reproduce the radiants of the TSC, as these are only observed in the STA branch but are found to emanate from the NTA branch in our simulations. 

The nodal crossing locations of the meteoroids ejected from both the nominal (Figure \ref{fig:2P_nodes}) and A4 clone of comet Encke (cf. Figure F2.1) consistently show the wide dispersion of the stream. For each starting ephemeris, the time range of possible meteor activity exceeds 200$\degree$ in solar longitude, with the widest spread from old trails. The modelled nodal streamlets produced by the differential precession of the meteoroid trails intersect the Earth's orbit at multiple locations, suggestive that several additional minor meteor showers may be related to comet Encke. 

Table F2.1 in Appendix F summarizes the orbital elements of 15 meteor showers past authors have suggested were associated with the Taurid Complex, including the \#2 STA, \#17 NTA, \#173 BTA, \#172 ZPE, \#156 SMA, \#154 DEA, \#256 ORN, \#257 ORS, \#215 NPI, \#629 ATS, \#635 ATU, \#632 NET, \#634 TAT, \#726 DEG and the \#94 RGE \citep{Asher1991,Tomko2019}. The orbit of each stream is compared with the nodal streamlets produced by A4 and the nominal clone of Encke.  These were found to remain close to the parent body's orbit projected onto the ecliptic plane (cf. animation C1.1). Examination of Figure F2.1 indicates that comet Encke may indeed be the parent of several of the minor showers listed above. However, as shown in Sections \ref{sec:nominal_ejection} and \ref{sec:NEAs_ejection}, a careful comparison between observations and the modelled stream timing and radiants is necessary to confirm the link between the comet and these minor showers, a task beyond the scope of the current study. 

Consistent with \cite{Whipple1967, Wiegert2009}, the age, broadness, and low-inclination of our modelled streams suggests Encke is one of the most significant cometary contributors to the helion and anti-helion sporadic sources. 

The wide dispersion of the meteoroids close to the ecliptic plane also hints that material ejected by the comet may also impact other terrestrial planets. \cite{Christou2015} suggested that variations of the amount of calcium detected in Mercury's atmosphere are due to the encounter of the planet with Encke's stream. The authors concluded that the timing of the calcium peak observed in Mercury's atmosphere could be reproduced by impacts of mm-sized meteoroids ejected from 2P/Encke between 10 000 and 20 000 years ago.
 
Using model A4, about 12 000 simulated meteoroids approach Mercury with a MOID below 0.02 AU and with a $\Delta T < 10$ days every year at the present epoch. The distribution of Mercury's true anomaly angles $\nu$ during the encounter with the stream model is presented in Appendix H and compared with the calcium vaporization rate measured by the MESSENGER spacecraft. Consistent with \cite{Christou2015}'s model, we find that mm to cm-sized particles released by object A4 between 10 000 BCE and 20 000 BCE would indeed encounter the planet today between $\nu$=-9$\degree$ and $\nu$=60$\degree$, corresponding to the time of maximum calcium emission. Impacts with younger trails are located around $\nu$=60$\degree$, but only involve a small number of particles so would not influence the profile significantly. 

Similar to \cite{Christou2015}, our model shows a second enhancement in Taurid meteoroid delivery to Mercury between $\nu$=130-180$\degree$. The existence of enhanced calcium emission around $\nu$=150-170$\degree$ is still uncertain. However, model A4 shows good agreement with the impact distribution from Taurids on Mercury presented in \cite{Christou2015}. In addition, their stream age of 12 to 22 ka is consistent with the estimate of 7 to 21 ka obtained when calibrating our model via observed meteor activity on Earth (cf. Section \ref{section:A4_distribution}). 

\section{Conclusion}
 
The formation of the Taurid Meteoroid Complex (TMC) remains a thorny problem. While the genetic relationship between the Taurid meteor showers and comet Encke has been considered in many works \citep[e.g.,][]{Tomko2019}, different scenarios involving additional parent bodies have been widely explored. Such scenarios, generally based on the giant breakup hypothesis of \cite{Whipple1952} and \cite{Clube1984}, have been developed to explain the orbital dispersion of the TMC \citep{Steel1991}, the multiplicity and duration of the associated meteor showers and the formation of the Taurid resonant swarm \citep{Asher1991,Asher1993b}.

In this work, we detail the results of an extensive numerical modelling of the TMC, focusing on the meteoroid streams produced by comet Encke. Considering 113 possible ephemerides of the comet, we simulated the ejection of meteoroids from the nucleus since 30 000 BCE and analyzed their contribution to the current meteor activity. In total, our simulation set represents more than 6.7 million particles, that were integrated over 32 ka. Each stream model has been compared and calibrated on meteor observations of the four major Taurid showers (NTA, STA, BTA and ZPE) as measured by the CMOR radar, video networks (CAMS, DFN) and photographic records (IAU MDC database).

Despite the apparent stability of Encke’s evolution, we find that slight variations of the comet’s orbit significantly affect the time and radiant structure of the simulated meteor showers. Small differences in the comet’s integration, due to the initial orbit considered, the inclusion  of non-gravitational forces, the planetary solution or the integrator employed are found to have considerable impact on the TMC modelling, particularly timing and strength of predicted showers.

We find that the stream model produced with the nominal ephemeris of the comet, integrated without non-gravitational forces, successfully reproduces the radiant structure of the major Taurid showers. However, the model fails in explaining the peak time of the STA, BTA and ZPE and their duration. This example highlights the importance of examining both the time and the radiants when establishing a link between a meteor shower and its parent body.

To improve the agreement with meteor observations, we modelled meteoroids released by the fragmentation of a large cometary nucleus around 3200 BCE. Such a hypothetical event, was identified in \cite{Egal2021} as a possible cause of the separation of Encke and four NEAS associated with the Taurid Complex (2004 TG10, 2005 TF50, 2005 UR and 2015 TX24) from a similar orbit. 
This work explored the possibility of a collisional fragmentation, resulting in the punctual ejection of meteoroids at high velocity, and of a gentler separation of the parent fragments with ejection velocities of a few m/s. However, each scenario failed in reproducing both the radiant structure and the apparition time of each Taurid shower. 

An hybrid model, combining the regular outgassing of Encke since 30 000 BCE and a high-speed fragmentation event around 3200 BCE, was found to better reproduce the overall Taurid activity and orbital elements. However, this model failed in explaining the ZPE activity and the peak time of the BTA. It also produced some activity at the early STA, but with radiants located in the northern hemisphere (i.e., more compatible with the NTA). 

We suspect that including additional fragmentation events, as proposed by \cite{Steel1991} and \cite{Asher1991}, could improve the agreement between the simulations and radar/optical observations. The ejection of meteoroids at high velocity from a different location of the bodies' precession cycle could also reduce the discrepancy with the observations. However, since an unlimited range of scenarios becomes possible in this case, we restricted our analysis to the most plausible fragmentation event around 3200 BCE identified in \cite{Egal2021}. 

In addition to a fragmentation origin for the TMC, we examined the possibility that different dynamical histories for Encke might better reproduce present Taurid activity. In this regard, we explored 112 possible orbital histories of comet Encke, and probed the differences between the meteoroid streams produced. We identified one specific clone of the comet, called A4, which was able to largely reproduce the radiants, duration, magnitude and timing of the average NTA, STA, BTA and ZPE meteor showers. 
The model's sole failing is a poor match to the shape of the late STA ZHR$_v$ profile, and the ZPE main peak of activity around 77$\degree$ SL. However, the characteristics of the showers simulated from clone A4 are in much better agreement with Taurid observations than any of our prior simulations, including the nominal orbit of 2P/Encke (with or without non gravitational forces included).

With model A4, most meteoroids contributing to the present-day Taurids were ejected between 5000 BCE and 19 000 BCE. The youngest trails approaching the planet were released about 4 ka ago, but do not significantly contribute to contemporary Taurid activity. Our stream model suggests that trails 5 to 12 ka old are probably sufficient to account for the formation of the NTA and BTA, while ages of 7 to 18 ka and $>$14 ka are more plausible for the STA and ZPE meteor showers. Our simulations therefore suggest that the NTA/BTA branch contains a higher proportion of young material than the STA/ZPE branch, which is consistent with meteor observations. 

We find that Encke's clone A4 was trapped into the 7:2 MMR with Jupiter during most its recent evolution (between 2000 BCE and 28 000 BCE). This feature is not unique among our clones sample; about 14\% of the clones of 2P/Encke spent at least a few millennia in that resonance. However, the orbital history of clone A4 may explain the formation of the Taurid Swarm Complex (TSC), since most of the meteoroids contributing to current activity on Earth were initially released within the 7:2 MMR, making it much more likely that larger Taurids would remain trapped in the resonance. 

Because of the low-number of resonant particles approaching Earth at the current epoch in our simulations, model A4 does not efficiently reproduce all the characteristics of the TSC returns. We find that our model delivers resonant cm-sized particles mostly during the STA ($\simeq$48\%) and the ZPE ($\simeq$39\%), consistent with the fireballs rates reported in autumn. The modelled swarm years are also in general good agreement with the predictions of \cite{Asher1993b}, but our simulated TSC profile is shifted several days earlier compared to the observations \citep{Devillepoix2021}. Exploration of additional simulations, performed from A4 or an alternative ephemeris of this specific clone, are likely needed to reproduce all TSC characteristics. 

Despite these limitations, we find clone A4 of 2P/Encke is able to reproduce the broad characteristics of the TMC, including the radiants and timing of the four major Taurid showers, the expected age and strength of each branch, and the formation and return years of the TSC. Our results show that the general features of the TMC can be modelled with classic cometary ejection from an object similar to Encke between 5000 BCE and 19 000 BCE, that spent most of its recent evolution within the 7:2 MMR with Jupiter. 

A specific prediction of our simplified TSC model is that the autumn swarm return in 2022 should produce fireballs at a level comparable to the major swarm returns in 1998, 2005 and 2015. Observers are encouraged to monitor the Taurids during the last part of October and early-mid November 2022.  We also predict this will be the last strong TSC return until at least the mid-late 2030's. 

Because of the wide dispersion and low-inclination of the meteoroids simulated, we suspect comet Encke to be a significant contributor to the helion/anti-helion sporadic sources \citep{Wiegert2009}, and to deliver material to other planets of the solar system. In particular, the A4-produced meteoroid streams at Mercury support the findings of \cite{Christou2015} that variations in the amount of calcium detected in the planet's atmosphere may be due to impacts with Encke's dust stream. 

Our model is not incompatible with the hypothesis of large bodies fragmenting within the Taurid complex. In particular, the age of the meteoroids approaching Earth in our simulations do not exclude the possibility of a fragmentation event around 3200 BCE as proposed by \citet{Egal2021}. In this scenario, a fragmentation circa 3200 BCE led to the the orbital separation of Encke and NEAs 2004 TG10, 2005 TF50, 2005 UR and 2015 TX24. Modelling these additional events could help understanding more subtle characteristics of the Taurid complex, like specific trends in the meteoroids orbital elements or the formation of the TSC. 

To encourage future models of the TMC, we provide the ephemeris of Encke's clone A4 in \cite{A4data}. Despite the limitations of meteoroid ejected from this clone in reproducing all the characteristics of the TMC, the promising results obtained in matching the main features of the TMC is suggestive this represents a more likely past orbital history of 2P/Encke than other ephemerides. This underscores that meteor observations may ultimately be able to provide the missing piece of information required to reveal the origin of this peculiar comet. 

\section*{Acknowledgements}

We are highly thankful to Julio Castellano and the Cometas\_Obs amateur astronomers for providing measurements of 2P/Encke's dust production. We also thank the reviewer for his comments that helped improving this manuscript. Funding for this work was provided in part through NASA co-operative agreement 80NSSC21M0073. This work was funded in part by the Natural Sciences and Engineering Research Council of Canada Discovery Grants program (Grants no. RGPIN-2016-04433 \& RGPIN-2018-05659) and the Canada Research Chairs program. 

\section*{Data availability statement}
 	
The data underlying this article will be shared on reasonable request to the corresponding author.




\bibliographystyle{mnras}
\bibliography{references} 



\appendix

\bsp	
\label{lastpage}
\end{document}


\begin{center}
  \Large \textbf{SUPPLEMENTARY APPENDICES}\\[0.5cm]
  \end{center}

\setcounter{topnumber}{4}
\setcounter{totalnumber}{4}

\newcommand{\fignumprefix}{}
\renewcommand{\thefigure}{\fignumprefix\arabic{figure}}
\renewcommand{\theHfigure}{figure.\thefigure}
\newcommand{\setfignumprefix}[1]{%
	\renewcommand{\fignumprefix}{#1}
	\setcounter{figure}{0}
}

\newcommand{\tabnumprefix}{}
\renewcommand{\thetable}{\tabnumprefix\arabic{table}}
\renewcommand{\theHtable}{table.\thetable}
\newcommand{\settabnumprefix}[2]{%
	\renewcommand{\tabnumprefix}{#1}
	\setcounter{table}{0}
}

\newcommand{\missing}[1]{\color{red}[#1]\mbox{ }\color{black}}

\textbf{APPENDIX A: Orbital Evolution of comet 2P/Encke}\\[0.2cm]

\textbf{A1 Nominal orbit, with non gravitational forces (NGF) included. } 

\setfignumprefix{A}
\begin{figure}[!ht]
    \centering
    \includegraphics[width=.3\textwidth]{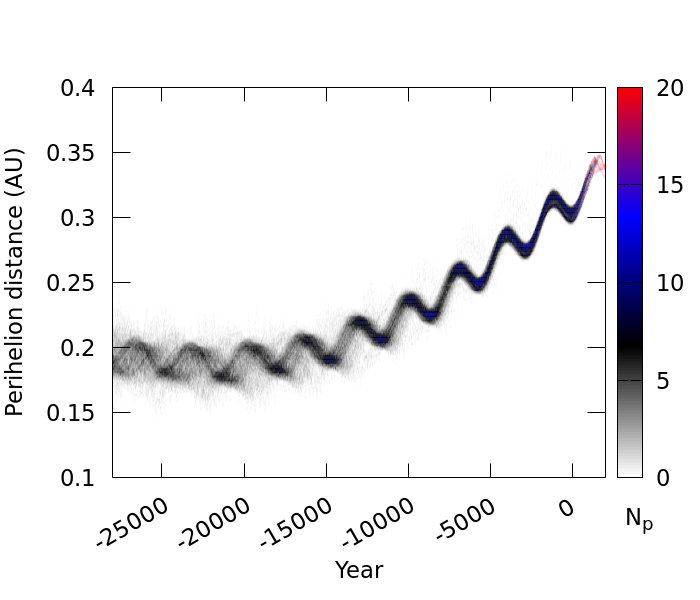}
    \includegraphics[width=.3\textwidth]{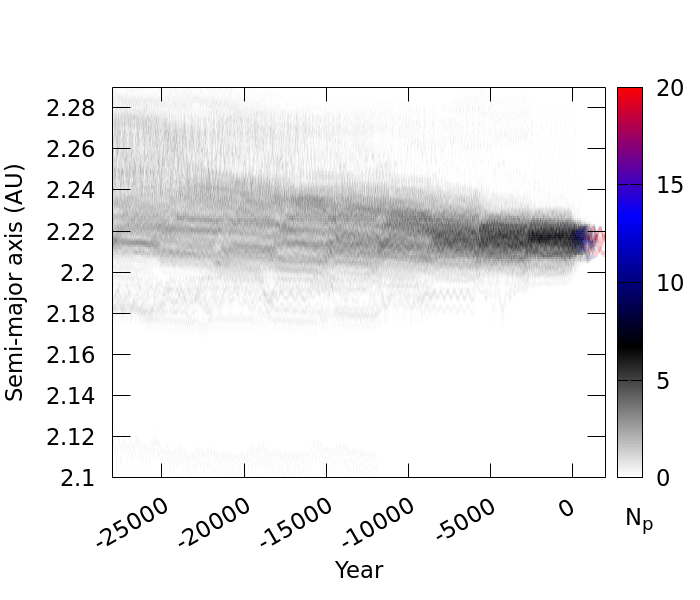}
    \includegraphics[width=.3\textwidth]{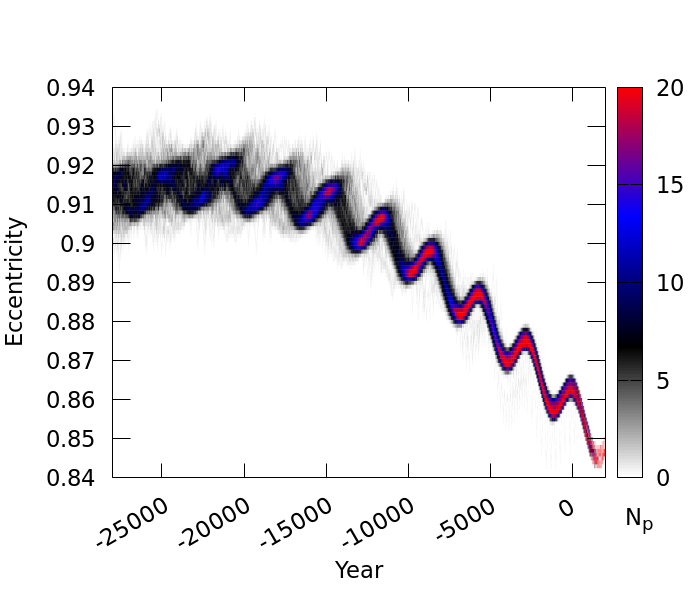}\\
    \includegraphics[width=.3\textwidth]{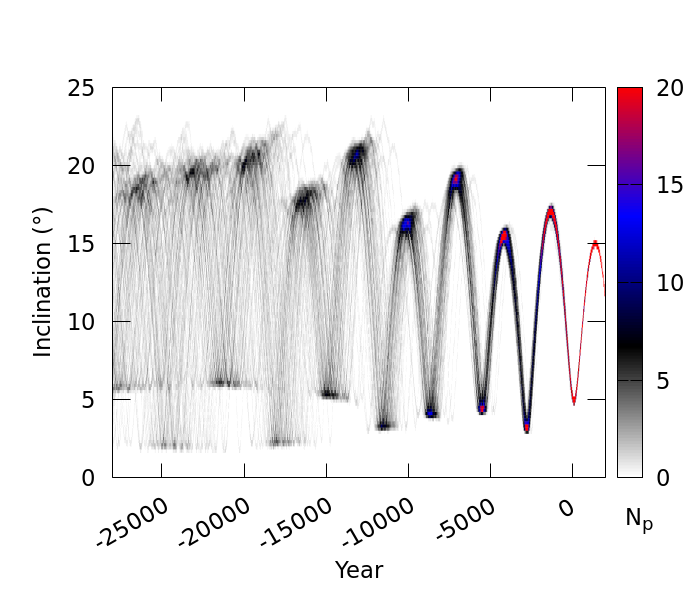}
    \includegraphics[width=.3\textwidth]{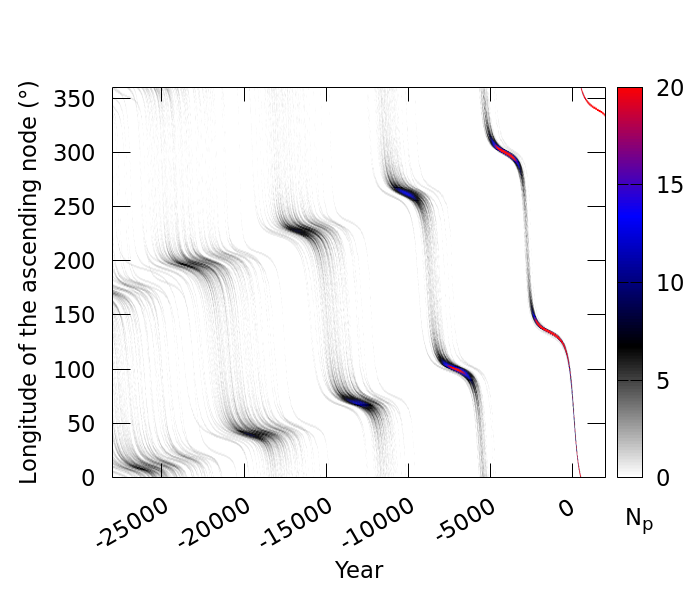}
    \includegraphics[width=.3\textwidth]{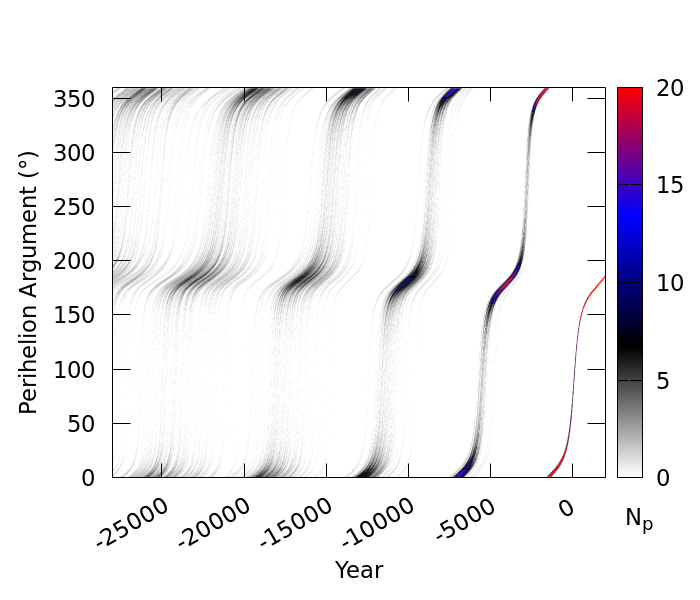}
    \caption{Time evolution of the orbital elements of a thousand clones of comet 2P/Encke, integrated from solution 1 in Table 1 using constant non-gravitational acceleration parameters. Colours represent the density of clones at each value of the orbital elements.}
    \label{2P:noNGF}
\end{figure}

\textbf{A2 Nominal orbit, without NGF} 
\begin{figure}[!ht]
    \centering
    \includegraphics[width=.3\textwidth]{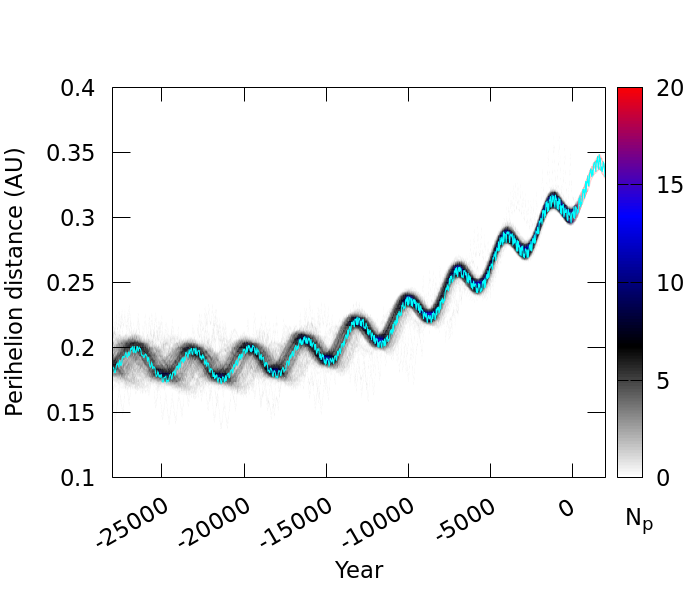}
    \includegraphics[width=.3\textwidth]{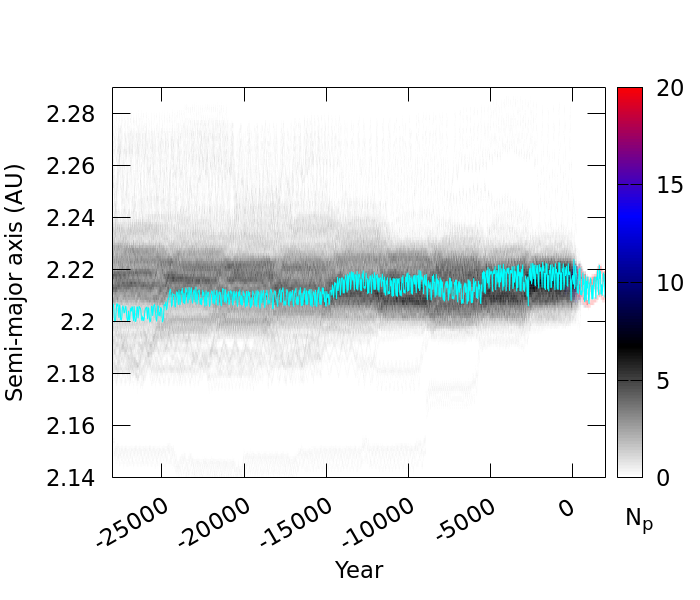}
    \includegraphics[width=.3\textwidth]{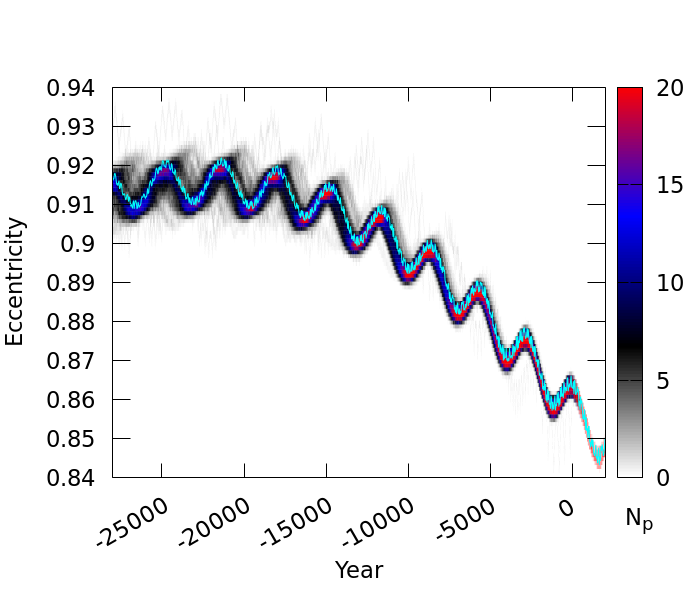}\\
    \includegraphics[width=.3\textwidth]{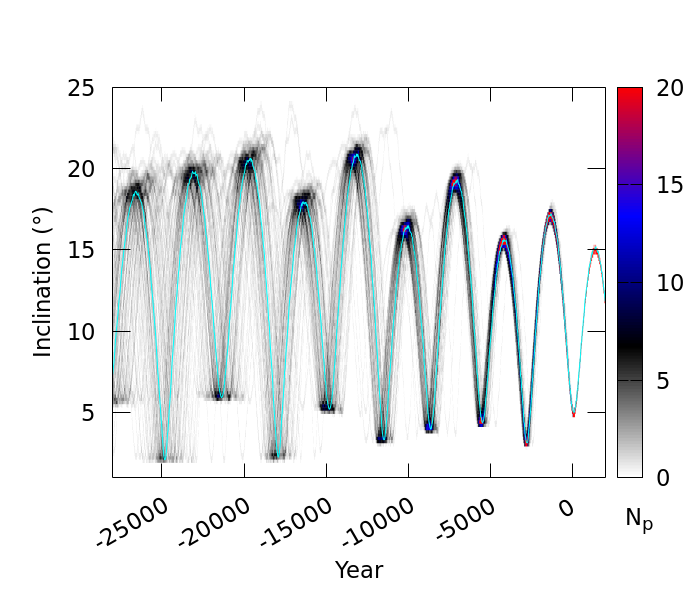}
    \includegraphics[width=.3\textwidth]{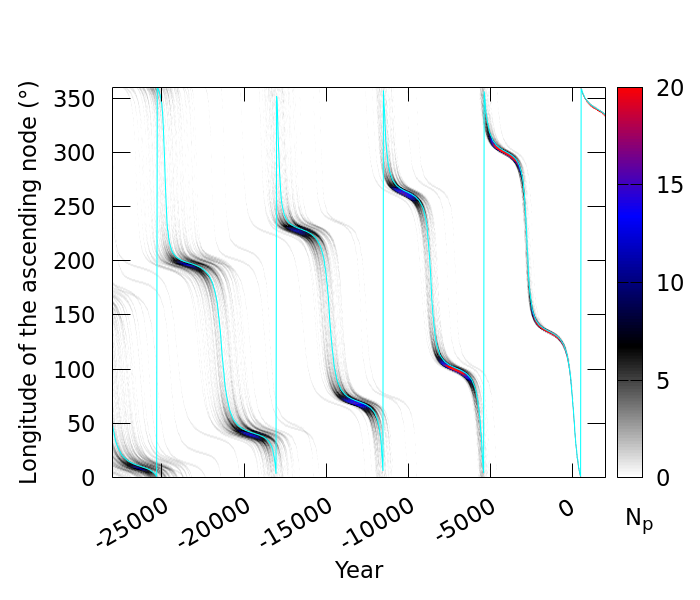}
    \includegraphics[width=.3\textwidth]{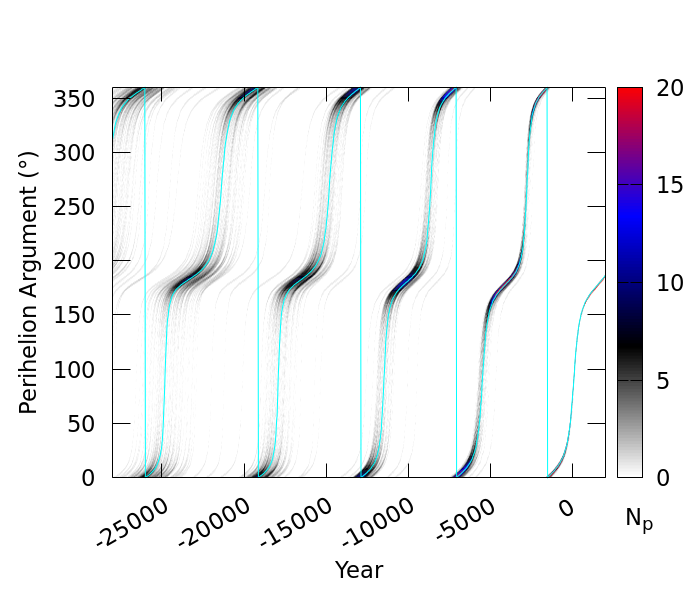}
    \caption{Time evolution of the orbital elements of a thousand clones of comet 2P/Encke, integrated from solution 2 in Table 1 ignoring the influence of  non-gravitational acceleration parameters. Colours represent the number of clones at each value of the orbital elements. The nominal ephemeris of 2P/Encke, used as initial condition of several meteoroid streams simulations performed in this work, is shown in cyan.}
    \label{2P:noNGF}
\end{figure}

\twocolumn
\textbf{APPENDIX B: Validity of Dust Heliocentric distance at ejection from 2P/Encke}\\

Here, we critically examine the conclusion of \cite{Epifani2001} that most of the mass lost by 2P/Encke around perihelion is concentrated in sub-mm particles, which are not detected by optical instruments. Based on this conclusion, the authors suggest that numerical models of the Taurid complex that assumed meteoroid ejection at perihelion needs to be updated. From the Af$\rho$ profile of Figure 1 (main text), we see that 78\% of 2P/Encke's modelled  dust production is confined to within an arc of 0.5 AU of perihelion, with the dust production peaking a few days before perihelion. 

If the heliocentric distance of the onset of dust ejection from the nucleus is a sensitive parameter in determining the outcome of dynamical meteoroid stream models \citep[see for example][]{Egal2019}, ejecting particles close to perihelion is an appropriate simplification when simulating the formation of meteoroid streams over long timescales. Since the heliocentric speed of the parent comet is higher at perihelion, meteoroids ejected around that location will disperse more quickly and fill in the appropriate orbital phase faster than the same number of particles released further away from the sun. Numerical simulations of the Taurids created by 2P/Encke by \cite{Tomko2019} indicated particles ejected at different speeds and locations formed similar meteoroid streams after 100 years, corresponding to about 30 orbital revolutions of the comet.

To check this result, we have performed two different sets of simulations. For each set, about 500 particles were ejected every 100 revolutions of 2P/Encke between 0 BCE and 30 000 BCE, representing a total of 45 000 simulated meteoroids. The particles were released from the comet as described in Section 3, using the orbital solution 2 of Table 1 and \cite{Crifo1997}'s ejection velocity model. The particles' radii were selected in the 1 to 10 mm range, corresponding to a magnitude of -2 to 5 for a density of 1000 kg.m$^{-3}$ (i.e., to meteors detectable with optical cameras). Our objective is to determine if the ejection circumstances of these meteoroids significantly impact their orbital distribution around Earth at the current epoch.

For the first set of simulations, named "P", all the particles were ejected at heliocentric distances $r_h$ below 0.5 AU (i.e., around perihelion). The second set of simulations, named "A", is composed of particles ejected along the comet's orbit below 3 AU where the sublimation of water ice is possible \citep{Combi2004}. The particles were integrated until 2021 CE, and their distribution around Earth's orbit examined. 

\begin{figure}
    \centering
    \includegraphics[width=.49\textwidth]{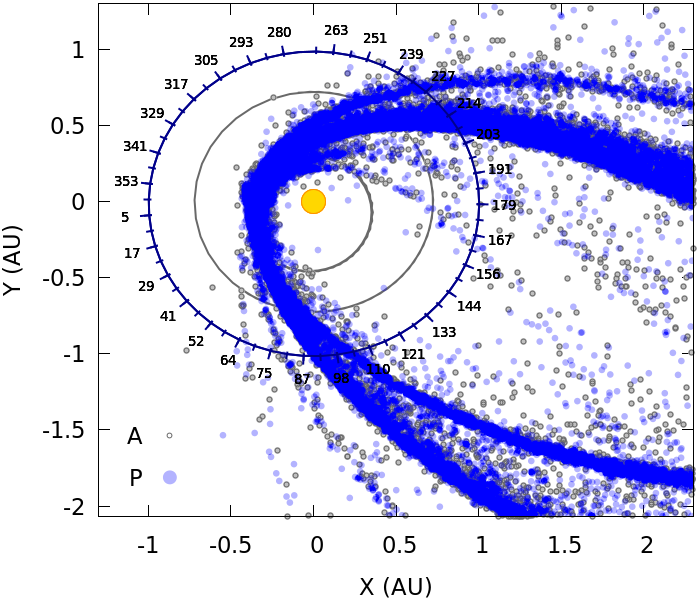}
    \includegraphics[width=.49\textwidth]{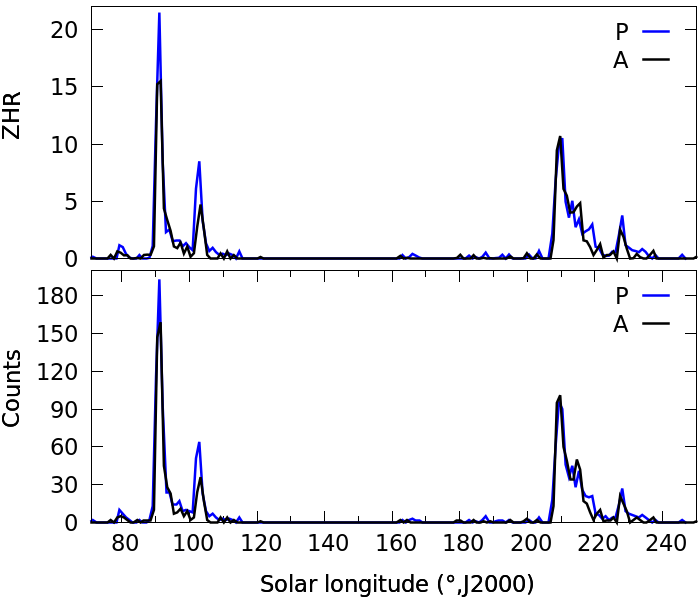}
    \caption{Top: nodal crossing locations in 2021 of the particles ejected around perihelion (in blue) or along 2P/Encke's orbit below 3 AU. Bottom: ZHR profile (weighted, top) and particles count (unweighted, bottom) obtained from each simulation set.}
    \label{fig:test_ejection}
\end{figure}

In the top panel of figure \ref{fig:test_ejection}, we present all the nodal crossing locations of the particles belonging to the sets "P" (in blue) or "A" (in grey) in 2021. We observe that the nodes structure for each simulation is similar, making it hard to distinguish them when the density of particles increases between 90-110$\degree$ and 205-235$\degree$ in solar longitude. In the bottom panel of Figure \ref{fig:test_ejection}, we present the number of particles approaching Earth's orbit in 2021 with a Minimum Orbital Distance (MOID) below 0.05 AU (bottom panel). In total, 1202 and 1458 particles were retained for the sets "A" and "P" respectively. In the bottom  panel of Figure \ref{fig:test_ejection}, we show the modelled ZHR profile obtained for each simulation set. This time, particles were weighted as a function of their physical properties and ejection circumstances, including their heliocentric distance at ejection (cf. Section 3.2.2).

We see that after a few millennia of integration, the simulated meteoroids lose the memory of their initial ejection location. We indeed see no significant difference in the stream's nodal location or activity profiles at the contemporary epoch between the two data sets. In contrast with \cite{Epifani2001}, we  conclude that models of the Taurid stream produced by particles ejected at perihelion only are valid, provided the stream motion is investigated over long timescales (i.e., more than a few centuries). Since this is the case for our present study, we are able to eject particles from 2P/Encke's orbit only below 0.5 AU, in order to reduce computing time without compromising the final results. 

\setfignumprefix{B}

\onecolumn

\newpage

\textbf{APPENDIX C: Taurid meteoroid stream produced from ejection of the nominal orbit of 2P/Encke}\\[0.2cm]

\textbf{C1 Orbital evolution of the meteoroid stream}\\

\setfignumprefix{C1.}
\begin{figure}[!ht]
    \centering
    \includegraphics[width=\textwidth]{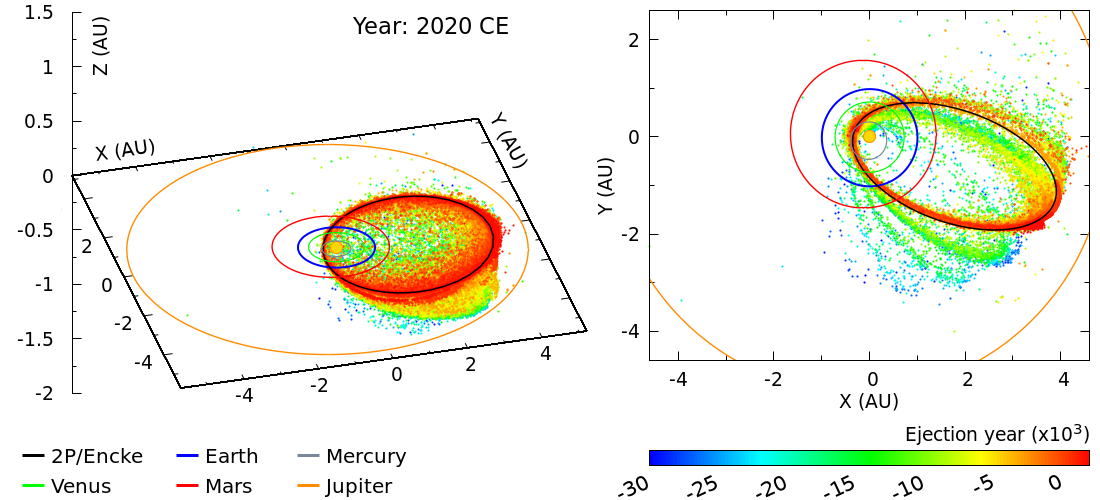}
    \caption{Orbital evolution of the meteoroid stream ejected from the nominal orbit of 2P/Encke (Run 1 dataset, solution 2) between 30 000 BCE and 2030. The meteoroids' 3D location (left panel) and nodal-crossing positions (right panel) are indicated with small dots, color-coded as function of their ejection epoch. The animated figure is provided online as a supplementary material to this paper, and can also be found at \url{https://youtu.be/Y_zCloube-M}.}
    \label{fig:animation_frame}
\end{figure}

\textbf{C2 Comparison of nominal model and observed average Taurid activity and radiants} 

\setfignumprefix{C2.}
\begin{figure}[!ht]
    \centering
    \includegraphics[width=.49\textwidth]{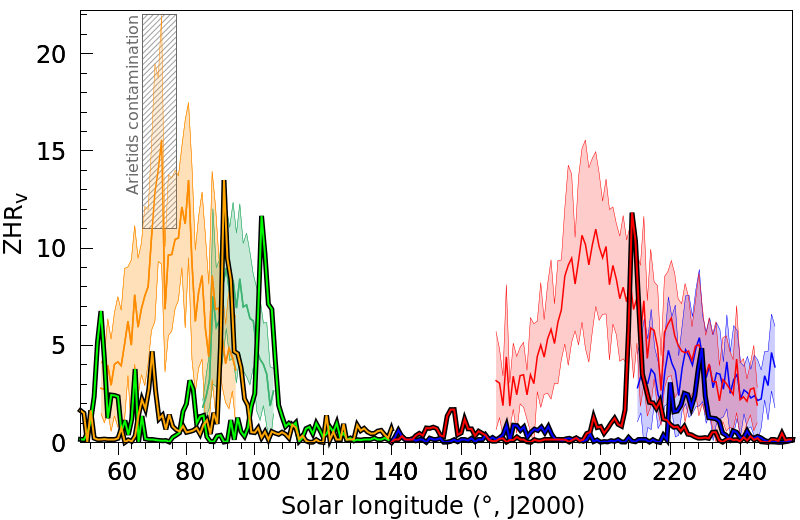}
    \includegraphics[width=.49\textwidth]{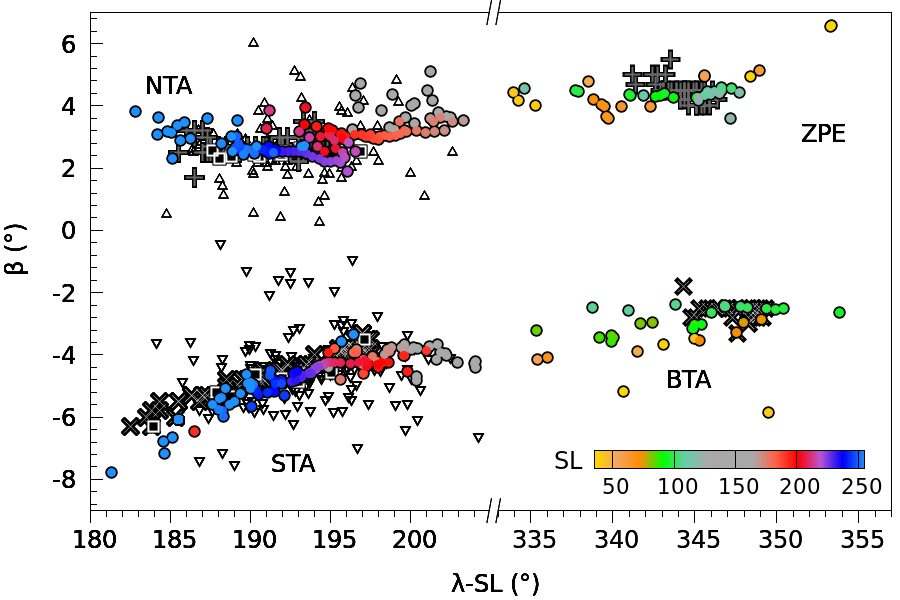}
    \caption{The average observed activity profile (colored and filled plots) compared to model predictions from nominal Encke (solid colored lines, left panel). Also shown are the average observed (grey and dark symbols) and model radiants (colored symbols, right panel) computed at each degree in solar longitude from the nominal Encke dataset (solution 2). }
    \label{fig:nominal_average}
\end{figure}

\clearpage
\newpage
\textbf{APPENDIX D: Taurid stream formation ejected from NEAs}\\[0.2cm]

\textbf{D1 Ejection from NEAs and Encke since 3200 BCE with relative speeds of a few m/s.} 

\setfignumprefix{D}
\begin{figure}[!ht]
    \centering
    \includegraphics[width=0.49\textwidth]{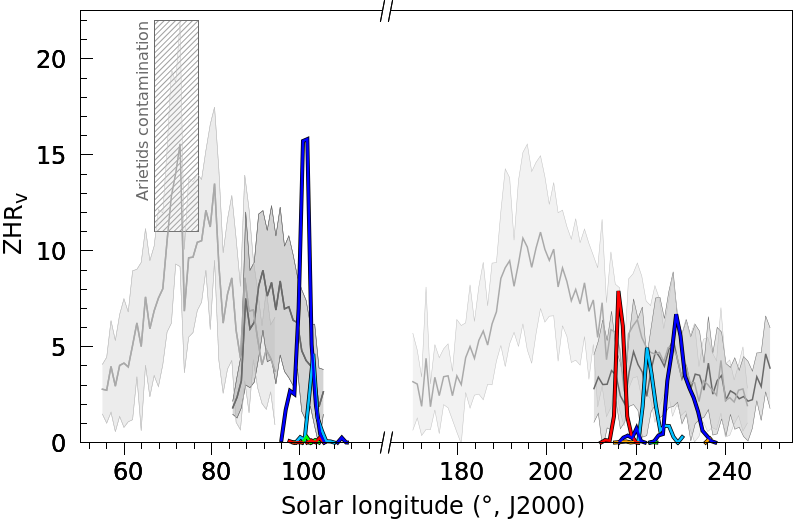}
    \includegraphics[width=0.49\textwidth]{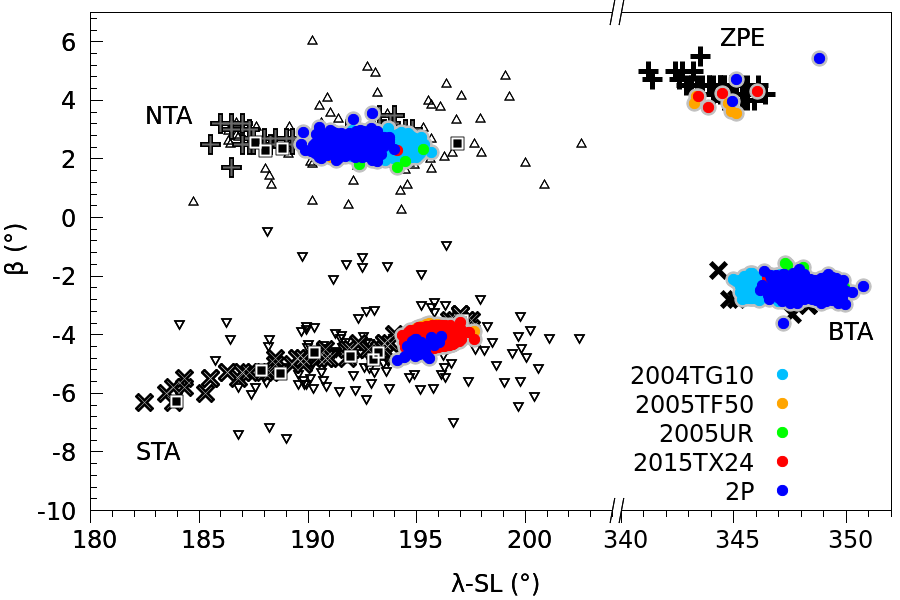}
    \caption{Simulated activity profiles (left) and sun-centered ecliptic radiants (right) of meteoroids approaching Earth's orbit in 2021. Particles were released from NEAs 2004 TG10 (cyan), 2005 TF50 (orange), 2005 UR (green), 2015 TX24 (red) and comet 2P/Encke (dark blue) at all apparitions beginning in 3200 BCE, with an ejection velocity of a few m/s. The average activity profile of the NTA, STA, BTA and ZPE determined from CMOR measurements is represented in grey.}
    \label{fig:NEAs_cont}
\end{figure}

\textbf{D2 Ejection at 3200 BCE} 

\begin{figure}[!ht]
    \centering
    \includegraphics[width=0.49\textwidth]{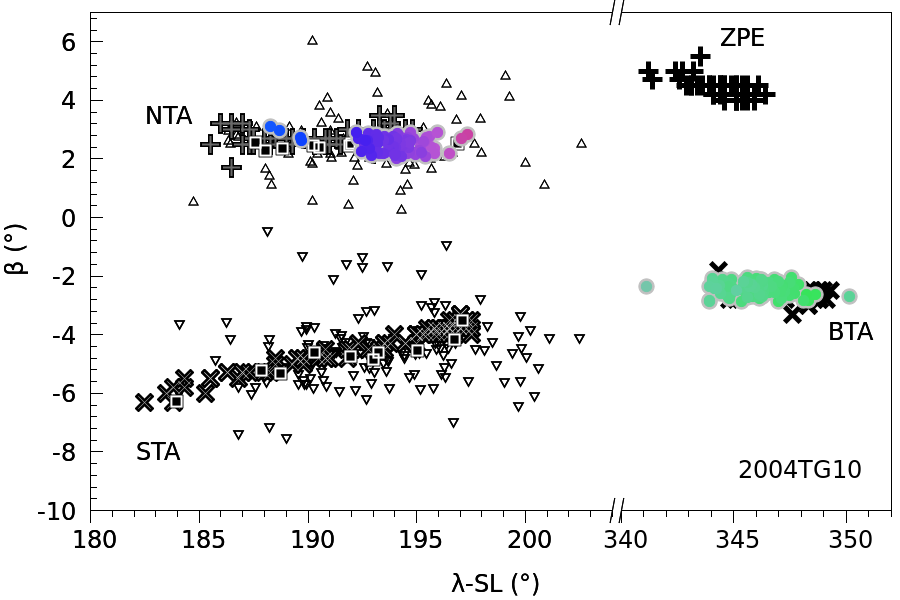}
    \includegraphics[width=0.49\textwidth]{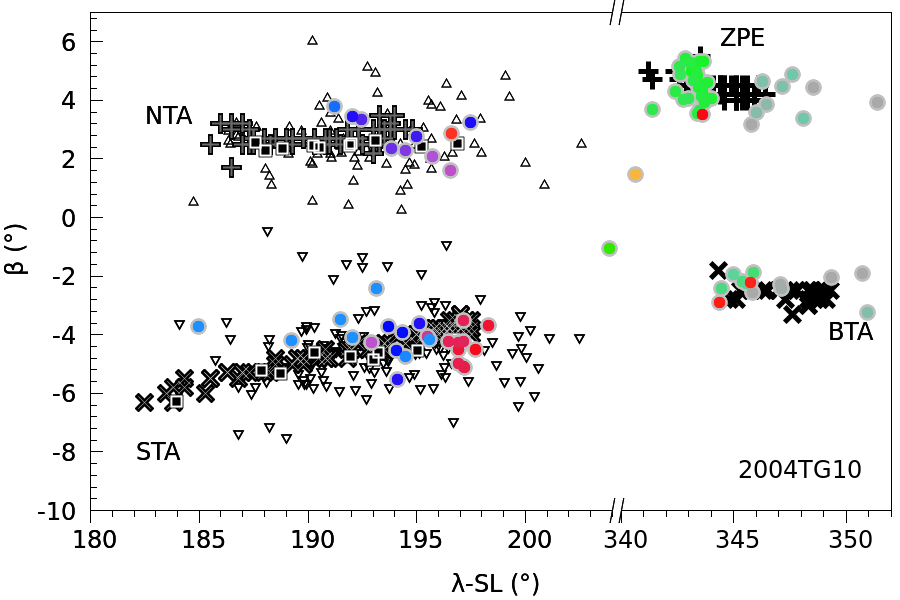}
    \includegraphics[width=0.49\textwidth]{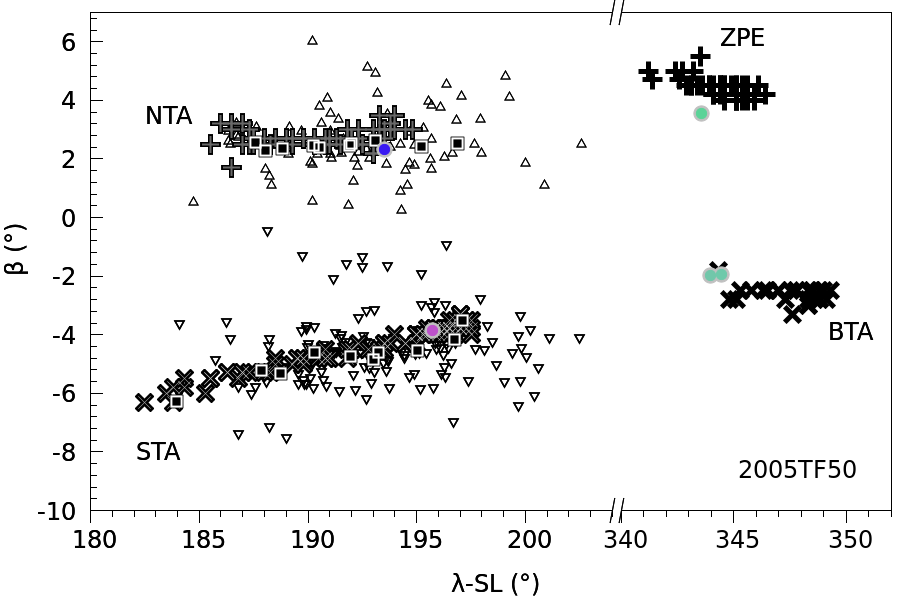}
    \includegraphics[width=0.49\textwidth]{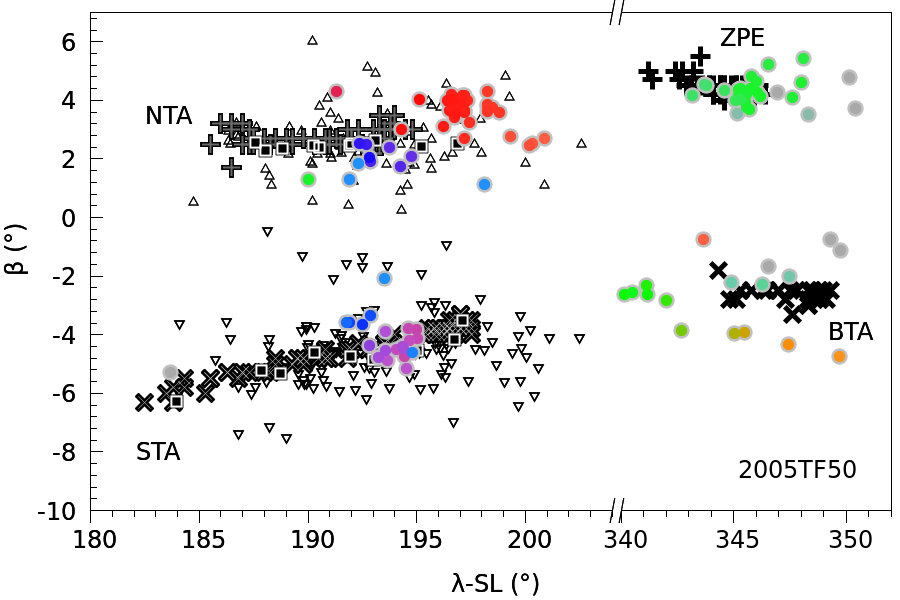}
    \includegraphics[width=0.6\textwidth]{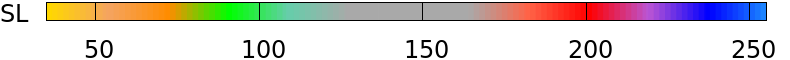}
    \caption{Sun-centered ecliptic longitude and latitude of meteoroids ejected from NEAs 2004 TG10 and 2005 TF50 that approach Earth with a MOID $<$ 0.01 AU in 2021. Meteoroids were ejected from the parent bodies in 3200 BCE only, with a velocity of a few m/s (left panel) or 1 km/s (right panel).}
    \label{fig:radSL1}
\end{figure}

\begin{figure}[!ht]
    \centering
    \includegraphics[width=0.49\textwidth]{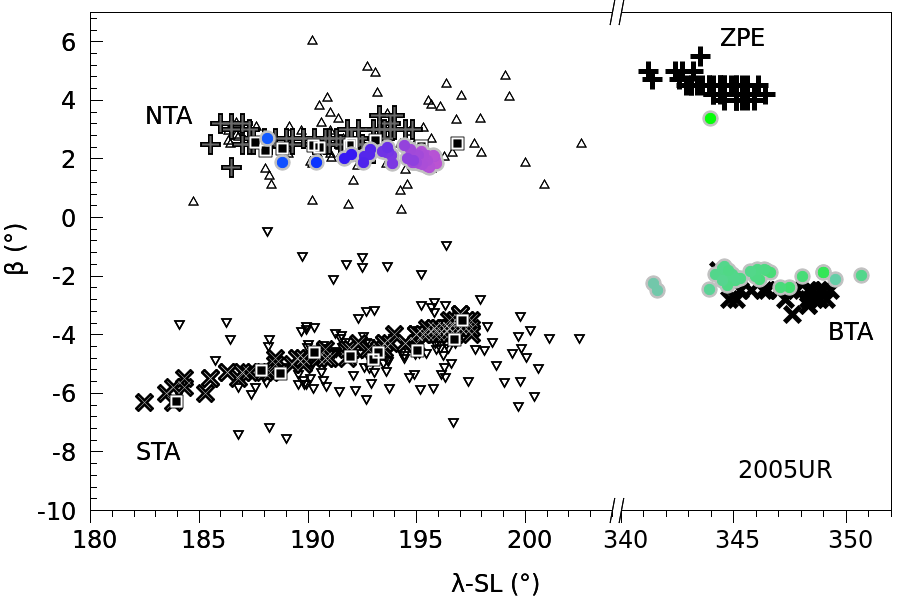}
    \includegraphics[width=0.49\textwidth]{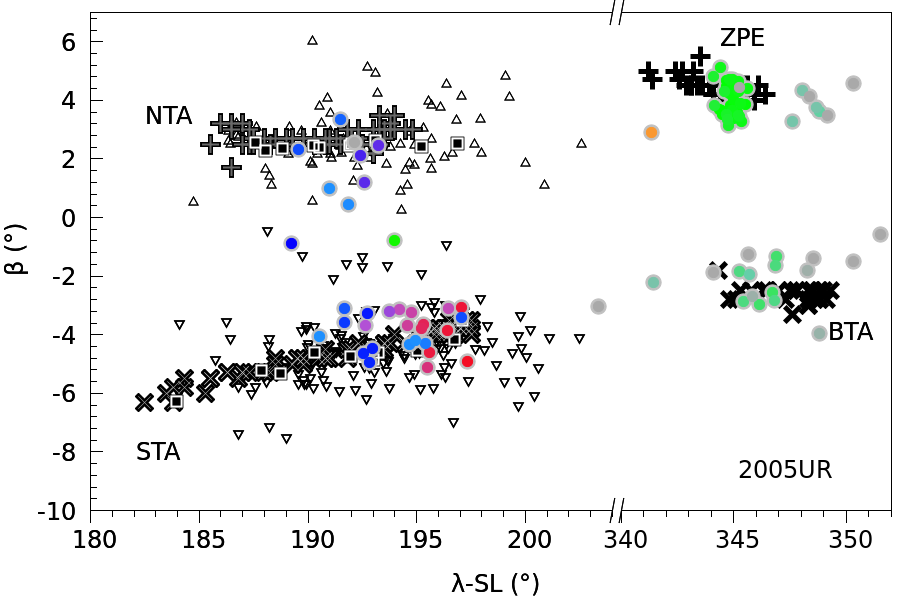}
    \includegraphics[width=0.49\textwidth]{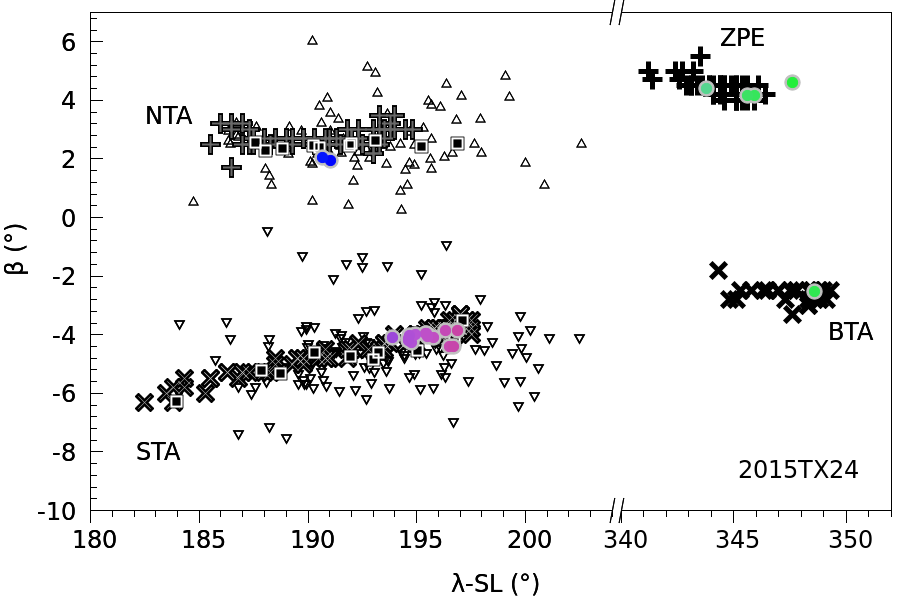}
    \includegraphics[width=0.49\textwidth]{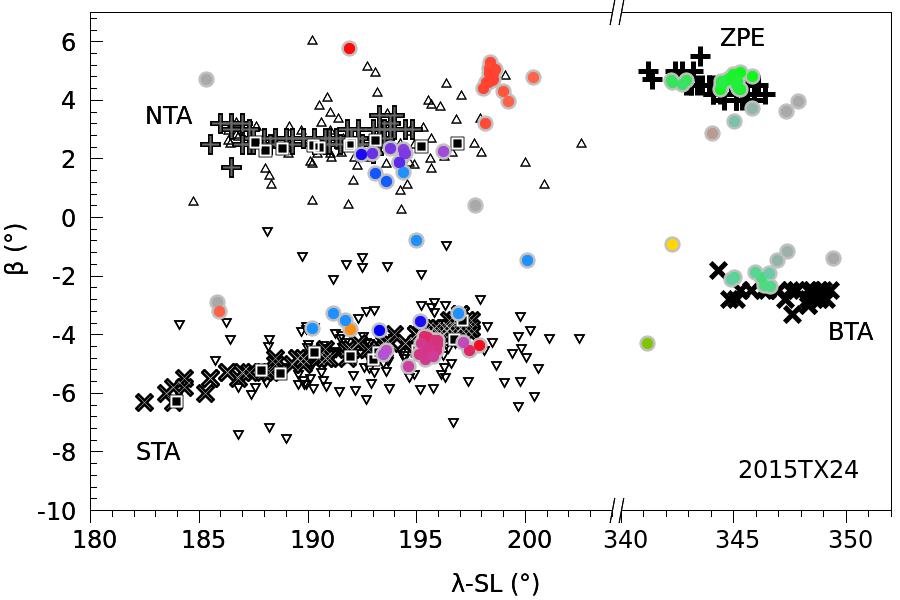}
    \includegraphics[width=0.49\textwidth]{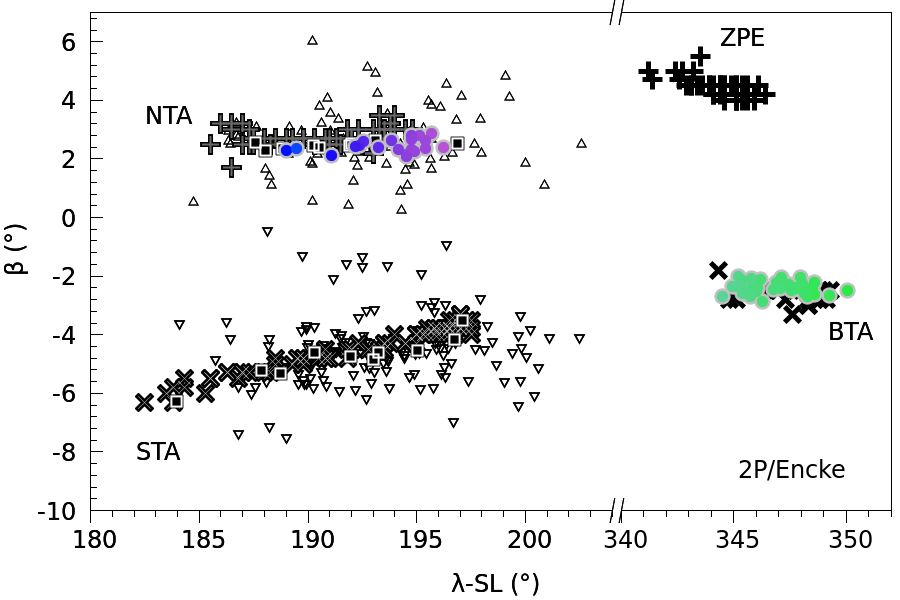}
    \includegraphics[width=0.49\textwidth]{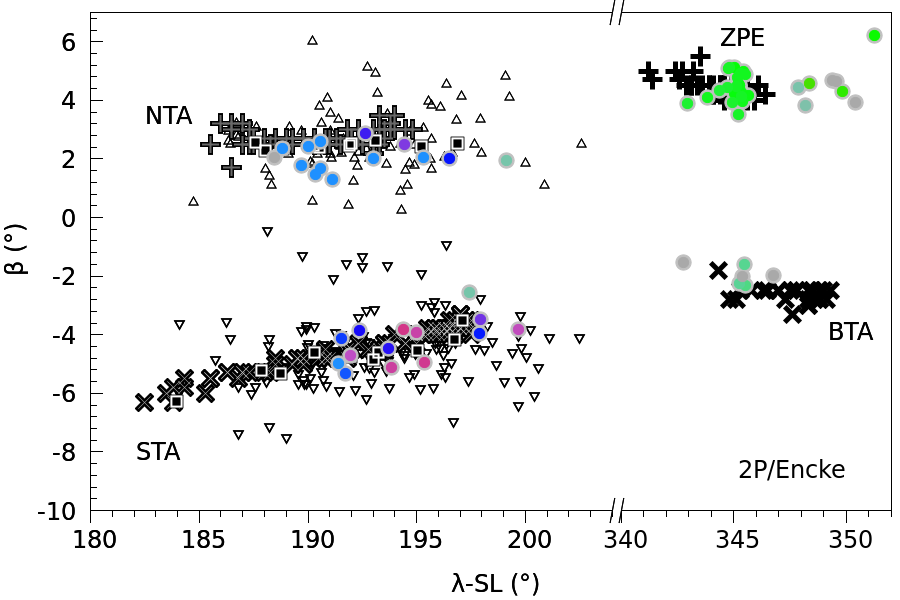}
    \includegraphics[width=0.6\textwidth]{SLbar.png}
    \caption{Sun-centered ecliptic longitude and latitude of meteoroids ejected from 2005 UR, 2015 TX24 and 2P/Encke that approach Earth with a MOID $<$ 0.01 AU in 2021. Meteoroids were ejected from the parent bodies in 3200 BCE, with a velocity of a few m/s (left panel) or 1 km/s (right panel).}
    \label{fig:radSL2}
\end{figure}

\begin{figure}
    \centering
    \includegraphics[width=0.32\textwidth]{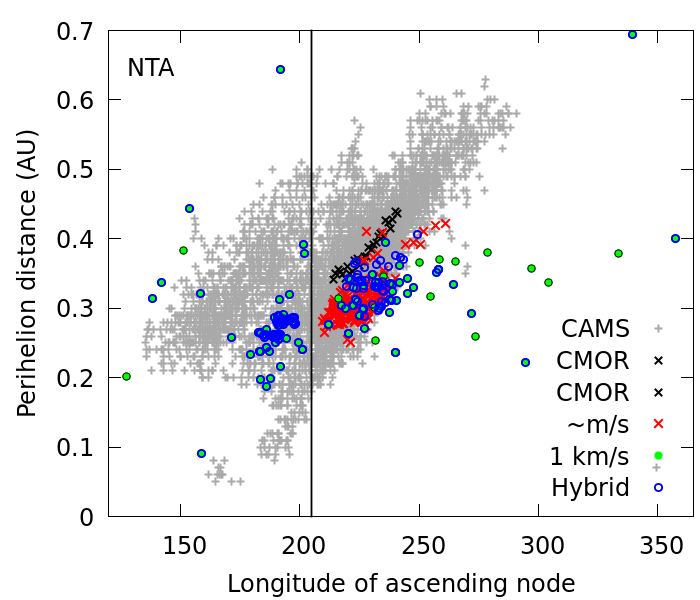}
    \includegraphics[width=0.32\textwidth]{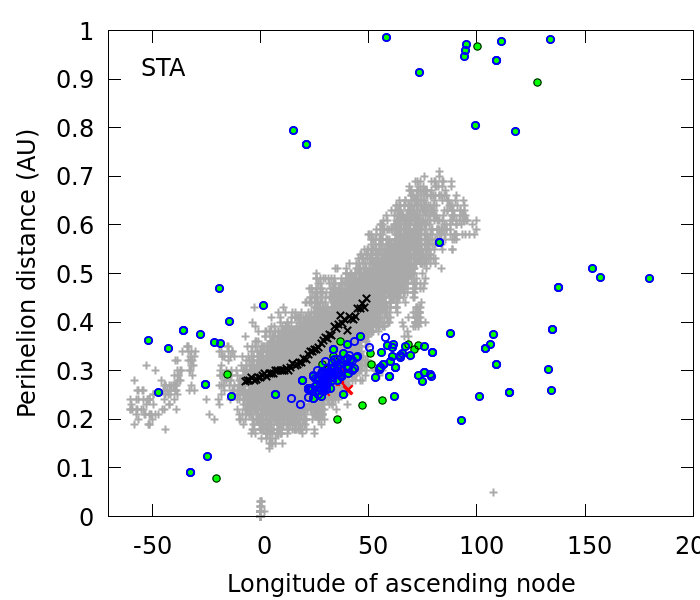}
    \includegraphics[width=0.32\textwidth]{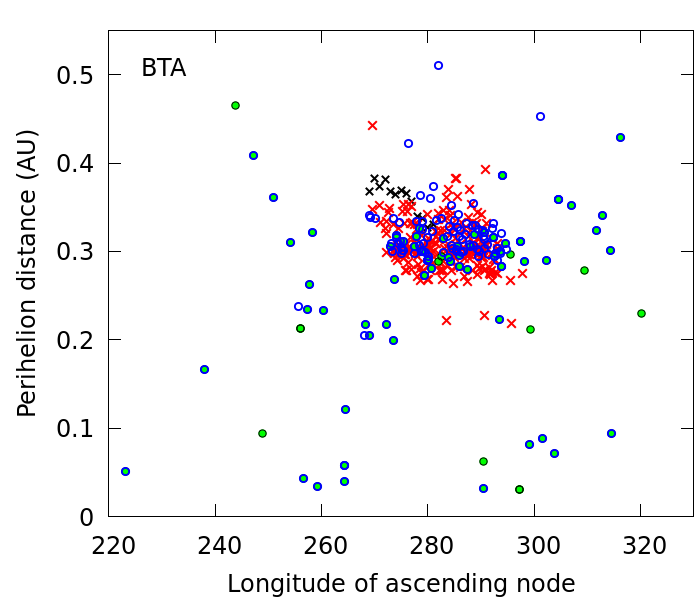}
    \includegraphics[width=0.32\textwidth]{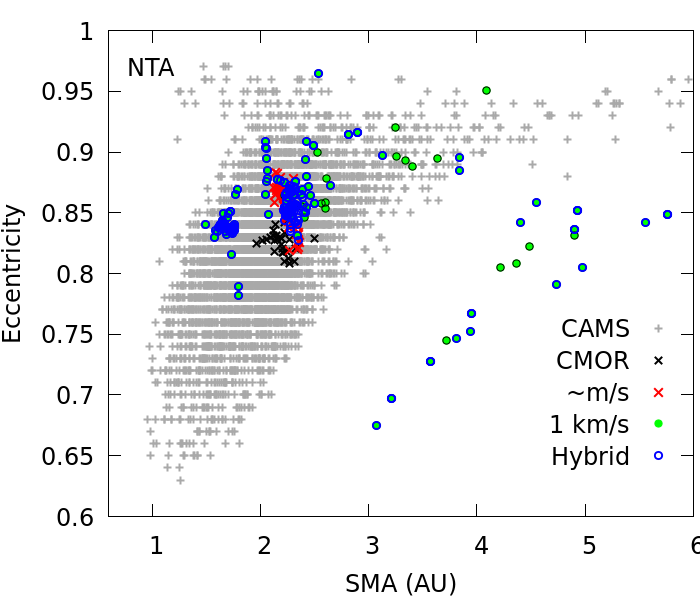}
    \includegraphics[width=0.32\textwidth]{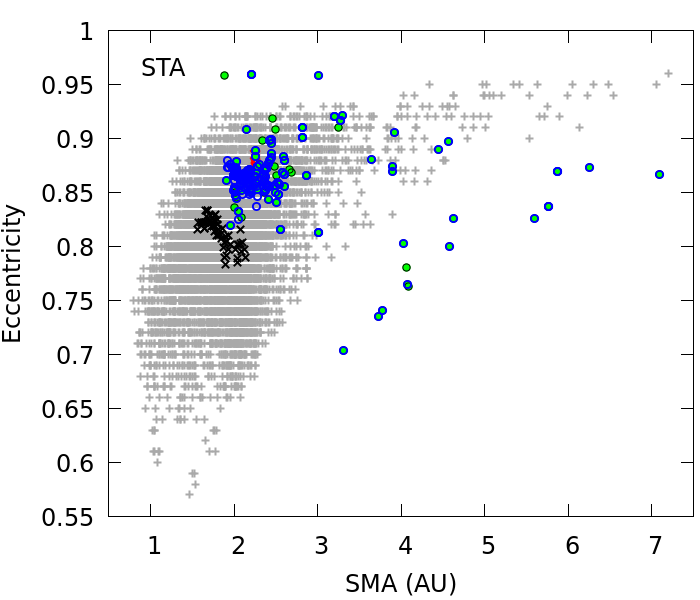}
    \includegraphics[width=0.32\textwidth]{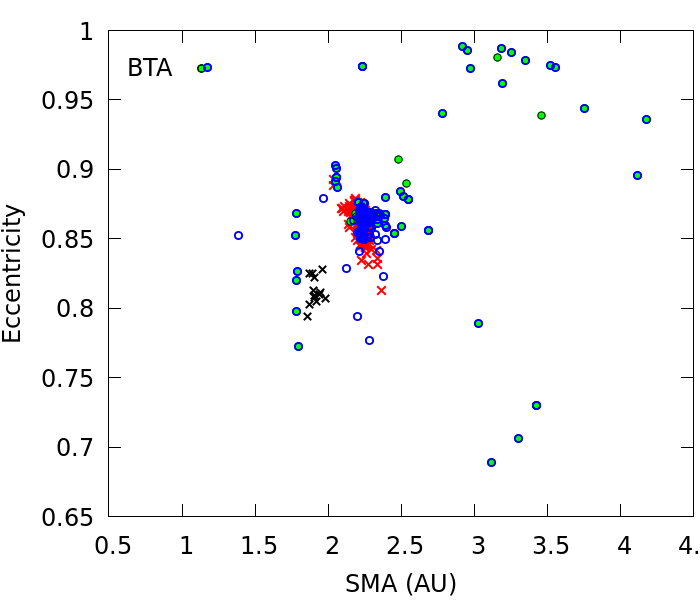}
    \includegraphics[width=0.32\textwidth]{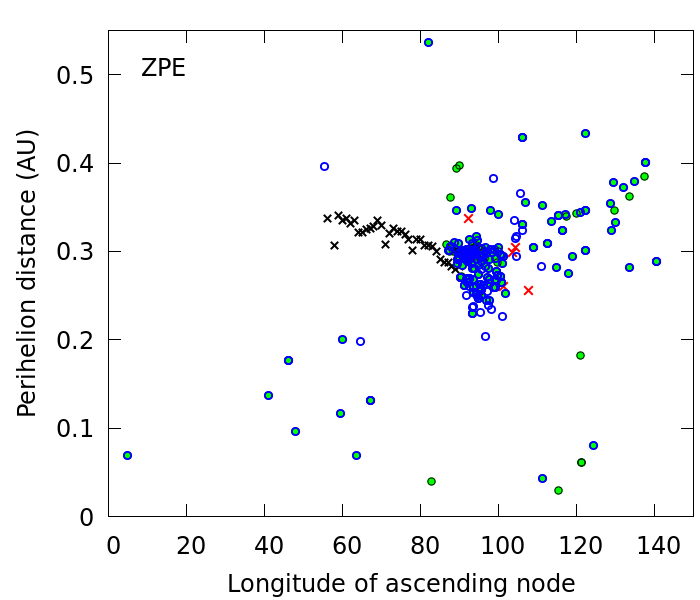}
    \includegraphics[width=0.32\textwidth]{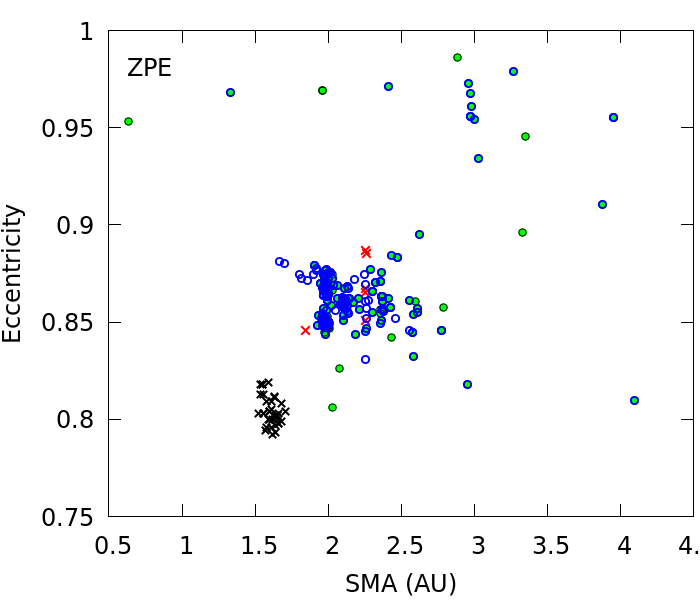}
    \caption{The perihelion distance, longitude of ascending node, semi-major axis and eccentricity of the NTA, STA, BTA and ZPE as measured by CAMS (grey crosses) and CMOR (wavelet analysis, black crosses). Colored crosses and dots represent the orbital elements of meteoroids approaching Earth in 2021 (MOID $<$ 0.01 AU), simulated following different scenarios: red and green points represent meteoroids ejected from 2P/Encke and NEAs 2004 TG10, 2005 TF50, 2005 UR and 2015 TX24 in 3200 BCE with velocities of a few m/s or 1 km/s respectively. Open blue circles refer to meteoroids obtained when combining the perihelion ejection of comet 2P/Encke since 30 000 BCE with the high-velocity break-up model around 3200 BCE (see text for details). }
    \label{fig:orbelt_NEAs}
\end{figure}

\newpage

\newpage
\clearpage
\textbf{APPENDIX E: Clones}\\ 

\textbf{E1. Initial states} \\

\settabnumprefix{E1.}	

\begin{table}[!ht]
  \resizebox{\textwidth}{!}{
   \begin{tabular}{lcccccccccccccc}
     ID & $Np_{/app}$ & $f_{app}$ & $Np_{tot}$ & JD & q & a & e & i & $\Omega$ & $\omega$ & m & A1 & A2 & A3 \\
        &  &  &  & & (AU) & (AU) & & ($\degree$) & ($\degree$) & ($\degree$) & ($\degree$) & (AU.d$^{-2}$) & (AU.d$^{-2}$) & (AU.d$^{-2}$) \\
     A92  & 100 & 100 & 9.7$\times10^2$ & 2457238.5 & 0.3360 & 2.2152 & 0.8483 & 11.78 & 334.57 & 186.55 & 185.41 & -4.74e$^{-10}$ &  5.02e$^{-12}$ &  2.81e$^{-8}$ \\     
     A254 & 100 & 100 & 9.7$\times10^2$ & 2457238.5 & 0.3360 & 2.2152 & 0.8483 & 11.78 & 334.57 & 186.55 & 185.41 &  5.03e$^{-10}$ & -3.95e$^{-12}$ &  1.06e$^{-7}$ \\     
     A509 & 100 & 100 & 9.7$\times10^2$ & 2457238.5 & 0.3360 & 2.2152 & 0.8483 & 11.78 & 334.57 & 186.55 & 185.41 & -4.45e$^{-10}$ & -1.78e$^{-13}$ &  1.55e$^{-8}$ \\     
     A742 & 100 & 100 & 9.7$\times10^2$ & 2457238.5 & 0.3360 & 2.2152 & 0.8483 & 11.78 & 334.57 & 186.55 & 185.41 & -3.77e$^{-10}$ &  7.18e$^{-13}$ &  3.58e$^{-8}$ \\     
     A852 & 100 & 100 & 9.7$\times10^2$ & 2457238.5 & 0.3360 & 2.2152 & 0.8483 & 11.78 & 334.57 & 186.55 & 185.41 & -6.65e$^{-11}$ &  4.42e$^{-12}$ &  7.11e$^{-8}$ \\     
     A869 & 100 & 100 & 9.7$\times10^2$ & 2457238.5 & 0.3360 & 2.2152 & 0.8483 & 11.78 & 334.57 & 186.55 & 185.41 & -4.26e$^{-10}$ &  4.33e$^{-12}$ &  1.13e$^{-7}$ \\     
     A565 & 100 & 100 & 9.7$\times10^2$ & 2457238.5 & 0.3360 & 2.2152 & 0.8483 & 11.78 & 334.57 & 186.55 & 185.41 & -5.01e$^{-10}$ & -1.87e$^{-13}$ &  6.40e$^{-8}$ \\     
     A931 & 100 & 100 & 9.7$\times10^2$ & 2457238.5 & 0.3360 & 2.2152 & 0.8483 & 11.78 & 334.57 & 186.55 & 185.41 & -4.27e$^{-10}$ &  3.92e$^{-12}$ & -9.35e$^{-9}$ \\     
     A944 & 100 & 100 & 9.7$\times10^2$ & 2457238.5 & 0.3360 & 2.2152 & 0.8483 & 11.78 & 334.57 & 186.55 & 185.41 & -4.00e$^{-10}$ & -4.50e$^{-14}$ & -5.46e$^{-8}$ \\     
     A995 & 100 & 100 & 9.7$\times10^2$ & 2457238.5 & 0.3360 & 2.2152 & 0.8483 & 11.78 & 334.57 & 186.55 & 185.41 &  9.69e$^{-11}$ &  2.10e$^{-12}$ & -6.58e$^{-8}$ \\     
     A16  & 100 & 100 & 9.7$\times10^2$ & 2457210.5 & 0.3360 & 2.2152 & 0.8483 & 11.78 & 334.57 & 186.55 & 177.04 &  3.32e$^{-12}$ &  1.13e$^{-12}$ &  3.58e$^{-9}$ \\    
     A60  & 100 & 100 & 9.7$\times10^2$ & 2457210.5 & 0.3360 & 2.2152 & 0.8483 & 11.78 & 334.57 & 186.55 & 177.04 &  3.00e$^{-11}$ & -6.12e$^{-13}$ &  3.94e$^{-9}$ \\    
     A69  & 100 & 100 & 9.7$\times10^2$ & 2457210.5 & 0.3360 & 2.2152 & 0.8483 & 11.78 & 334.57 & 186.55 & 177.04 & -4.52e$^{-11}$ & -3.31e$^{-12}$ &  3.78e$^{-9}$ \\    
     A76  & 100 & 100 & 9.7$\times10^2$ & 2457210.5 & 0.3360 & 2.2152 & 0.8483 & 11.78 & 334.57 & 186.55 & 177.04 &  1.70e$^{-11}$ & -6.05e$^{-14}$ &  3.77e$^{-9}$ \\    
     A91  & 100 & 100 & 9.7$\times10^2$ & 2457210.5 & 0.3360 & 2.2152 & 0.8483 & 11.78 & 334.57 & 186.55 & 177.04 &  1.13e$^{-11}$ & -1.43e$^{-12}$ &  3.87e$^{-9}$ \\   
     \hline
     A4  & 60 & 1 & 5.7$\times10^5$ & 2456618.24851 & 0.3360 & 2.2151 & 0.8483 & 11.78 & 334.57 & 186.54 & 0.00 & -1.41e$^{-11}$ & -3.78e$^{-11}$ &  7.70e$^{-11}$ \\
     A30 & 100 & 100 & 9.7$\times10^2$ & 2456618.24851 & 0.3360 & 2.2151 & 0.8483 & 11.78 & 334.57 & 186.54 & 0.00 & -1.0e$^{-8}$     &  0.0      & 0.0 \\  
     \hline  
   \end{tabular}}
   \caption{\label{tab:clones_auriane} Simulation parameters of meteoroid streams ejected from different clones of comet 2P/Encke. $Np_{/app}$ particles are ejected a each $f_{app}$ perihelion return of the comet since 30 000 BCE. Columns 4 to 15 indicate the time, orbital elements (perihelion distance $q$, semimajor axis $a$, eccentricity $e$, inclination $i$, longitude of ascending node $\Omega$, perihelion argument $\omega$, mean anomaly $m$) and nongravitational coefficients A$_1$, A$_2$ and A$_3$ used for the comet integration.}
\end{table}

  \begin{table}[!ht]
   \resizebox{\textwidth}{!}{
     \begin{tabular}{ccccccccccccccc}
        ID & $Np_{/app}$ & $f_{app}$ & $Np_{tot}$ & JD & q & a & e & i & $\Omega$ & $\omega$ & m & A1 & A2 & A3 \\
         &  &  &  & & (AU) & (AU) & & ($\degree$) & ($\degree$) & ($\degree$) & ($\degree$) & (AU.d$^{-2}$) & (AU.d$^{-2}$) & (AU.d$^{-2}$) \\
        B09 & 5 & 1 & 4.85$\times10^4$ & 2456618.24851 & 0.3360 & 2.2151 & 0.8483 & 11.78 & 334.57 & 186.54 & 0.00 &  0.0     &  0.0      & 0.0 \\
        B10 & 5 & 1 & 4.85$\times10^4$ & 2456618.24851 & 0.3360 & 2.2151 & 0.8483 & 11.78 & 334.57 & 186.54 & 0.00 &  0.0            &  0.0            & 0.0 \\
        B11 & 5 & 1 & 4.85$\times10^4$ & 2456618.24851 & 0.3360 & 2.2151 & 0.8483 & 11.78 & 334.57 & 186.54 & 0.00 & -1.90e$^{-11}$  & +1.73e$^{-13}$  & 3.50e$^{-9}$ \\
        B12 & 5 & 1 & 4.85$\times10^4$ & 2456618.24851 & 0.3360 & 2.2151 & 0.8483 & 11.78 & 334.57 & 186.54 & 0.00 & -1.90e$^{-10}$  & +1.73e$^{-12}$  & 3.50e$^{-9}$ \\
        B14 & 5 & 1 & 4.85$\times10^4$ & 2456618.24851 & 0.3360 & 2.2151 & 0.8483 & 11.78 & 334.57 & 186.54 & 0.00 &  0.0            &  -1.0e$^{-8}$   & 0.0 \\     
        B15 & 5 & 1 & 4.85$\times10^4$ & 2456618.24851 & 0.3360 & 2.2151 & 0.8483 & 11.78 & 334.57 & 186.54 & 0.00 &  0.0            &  0.0            & +1.0e$^{-8}$  \\      
        B16 & 5 & 1 & 4.85$\times10^4$ & 2456618.24851 & 0.3360 & 2.2151 & 0.8483 & 11.78 & 334.57 & 186.54 & 0.00 &  0.0            &  0.0            & -1.0e$^{-8}$  \\       
        B18 & 5 & 1 & 4.85$\times10^4$ & 2456618.24851 & 0.3360 & 2.2151 & 0.8483 & 11.78 & 334.57 & 186.54 & 0.00 &  0.0            &  -1.0e$^{-8}$   & +1.0e$^{-8}$  \\   
        B20 & 5 & 1 & 4.85$\times10^4$ & 2456618.24851 & 0.3360 & 2.2151 & 0.8483 & 11.78 & 334.57 & 186.54 & 0.00 &  0.0            &  -1.0e$^{-8}$   & -1.0e$^{-8}$  \\        
        B21 & 5 & 1 & 4.85$\times10^4$ & 2456618.24851 & 0.3360 & 2.2151 & 0.8483 & 11.78 & 334.57 & 186.54 & 0.00 &  0.0            &  +0.1e$^{-8}$   &  0.0 \\    
        B22 & 5 & 1 & 4.85$\times10^4$ & 2456618.24851 & 0.3360 & 2.2151 & 0.8483 & 11.78 & 334.57 & 186.54 & 0.00 &  0.0            & -0.1            &  0.0 \\
        B23 & 5 & 1 & 4.85$\times10^4$ & 2456618.24851 & 0.3360 & 2.2151 & 0.8483 & 11.78 & 334.57 & 186.54 & 0.00 &  0.0            &  0.0            & +0.1 \\
        B24 & 5 & 1 & 4.85$\times10^4$ & 2456618.24851 & 0.3360 & 2.2151 & 0.8483 & 11.78 & 334.57 & 186.54 & 0.00 &  0.0            &  0.0            & +0.1 \\
        B26 & 5 & 1 & 4.85$\times10^4$ & 2456618.24851 & 0.3360 & 2.2151 & 0.8483 & 11.78 & 334.57 & 186.54 & 0.00 &  0.0            & -0.1            & +0.1 \\
        B27 & 5 & 1 & 4.85$\times10^4$ & 2456618.24851 & 0.3360 & 2.2151 & 0.8483 & 11.78 & 334.57 & 186.54 & 0.00 &  0.0            & +0.1            & -0.1 \\
        B28 & 5 & 1 & 4.85$\times10^4$ & 2456618.24851 & 0.3360 & 2.2151 & 0.8483 & 11.78 & 334.57 & 186.54 & 0.00 &  0.0            & -0.1            & -0.1 \\
        B29 & 5 & 1 & 4.85$\times10^4$ & 2456618.24851 & 0.3360 & 2.2151 & 0.8483 & 11.78 & 334.57 & 186.54 & 0.00 & +1.0e$^{-8}$    &  0.0            &  0.0 \\   
        B30 & 5 & 1 & 4.85$\times10^4$ & 2456618.24851 & 0.3360 & 2.2151 & 0.8483 & 11.78 & 334.57 & 186.54 & 0.00 & -1.0e$^{-8}$    &  0.0            &  0.0 \\    
        B31 & 5 & 1 & 4.85$\times10^4$ & 2456618.24851 & 0.3360 & 2.2151 & 0.8483 & 11.78 & 334.57 & 186.54 & 0.00 & +1.0e$^{-9}$    &  0.0            &  0.0 \\  
        B32 & 5 & 1 & 4.85$\times10^4$ & 2456618.24851 & 0.3360 & 2.2151 & 0.8483 & 11.78 & 334.57 & 186.54 & 0.00 & -1.0e$^{-9}$    &  0.0            &  0.0 \\  
        \hline
      \end{tabular}}
        \caption{\label{tab:clones_paul1} Simulation parameters of meteoroid streams ejected from different clones of comet 2P/Encke, with extreme values of nongravitational coefficients A$_1$, A$_2$ and A$_3$. Same notation as in Table E1.1 }
    \end{table}

	  \begin{table}[!ht]
     \resizebox{\textwidth}{!}{
        \begin{tabular}{ccccccccccccccc}
        ID & $Np_{/app}$ & $f_{app}$ & $Np_{tot}$ & JD & q & a & e & i & $\Omega$ & $\omega$ & m & A1 & A2 & A3 \\
         &  &  &  & & (AU) & (AU) & & ($\degree$) & ($\degree$) & ($\degree$) & ($\degree$) & (AU.d$^{-2}$) & (AU.d$^{-2}$) & (AU.d$^{-2}$) \\
        B2-04  & 5 & 1 & 4.85$\times10^4$ & 2456618.24851 & 0.3360 & 2.2151 & 0.8483 & 11.78 & 334.57 & 186.54 & 0.00 &  2.26e$^{-11}$ & -6.71e$^{-11}$ &  6.34e$^{-11}$ \\
        B2-05  & 5 & 1 & 4.85$\times10^4$ & 2456618.24851 & 0.3360 & 2.2151 & 0.8483 & 11.78 & 334.57 & 186.54 & 0.00 & -6.47e$^{-11}$ &  6.38e$^{-11}$ &  2.97e$^{-11}$ \\
        B2-06  & 5 & 1 & 4.85$\times10^4$ & 2456618.24851 & 0.3360 & 2.2151 & 0.8483 & 11.78 & 334.57 & 186.54 & 0.00 &  9.34e$^{-11}$ & -8.96e$^{-11}$ & -8.29e$^{-11}$ \\
        B2-07  & 5 & 1 & 4.85$\times10^4$ & 2456618.24851 & 0.3360 & 2.2151 & 0.8483 & 11.78 & 334.57 & 186.54 & 0.00 &  1.53e$^{-11}$ & -5.65e$^{-11}$ &  7.27e$^{-11}$ \\
        B2-08  & 5 & 1 & 4.85$\times10^4$ & 2456618.24851 & 0.3360 & 2.2151 & 0.8483 & 11.78 & 334.57 & 186.54 & 0.00 &  2.03e$^{-11}$ &  8.62e$^{-11}$ & -4.64e$^{-11}$ \\
        B2-09  & 5 & 1 & 4.85$\times10^4$ & 2456618.24851 & 0.3360 & 2.2151 & 0.8483 & 11.78 & 334.57 & 186.54 & 0.00 &  7.87e$^{-11}$ & -8.10e$^{-11}$ &  7.45e$^{-11}$ \\
        B2-10  & 5 & 1 & 4.85$\times10^4$ & 2456618.24851 & 0.3360 & 2.2151 & 0.8483 & 11.78 & 334.57 & 186.54 & 0.00 &  1.46e$^{-11}$ &  2.30e$^{-11}$ & -1.83e$^{-11}$ \\
        B2-11  & 5 & 1 & 4.85$\times10^4$ & 2456618.24851 & 0.3360 & 2.2151 & 0.8483 & 11.78 & 334.57 & 186.54 & 0.00 &  7.54e$^{-11}$ &  1.18e$^{-11}$ &  6.12e$^{-11}$ \\
        B2-12  & 5 & 1 & 4.85$\times10^4$ & 2456618.24851 & 0.3360 & 2.2151 & 0.8483 & 11.78 & 334.57 & 186.54 & 0.00 &  3.03e$^{-12}$ &  8.30e$^{-11}$ & -4.08e$^{-11}$ \\
        
        B2-13  & 5 & 1 & 4.85$\times10^4$ & 2456618.24851 & 0.3360 & 2.2151 & 0.8483 & 11.78 & 334.57 & 186.54 & 0.00 & -3.86e$^{-11}$ &  5.94e$^{-11}$ &  1.70e$^{-11}$ \\
        B2-14  & 5 & 1 & 4.85$\times10^4$ & 2456618.24851 & 0.3360 & 2.2151 & 0.8483 & 11.78 & 334.57 & 186.54 & 0.00 & -5.41e$^{-11}$ &  2.28e$^{-11}$ & -1.16e$^{-11}$ \\
        B2-15  & 5 & 1 & 4.85$\times10^4$ & 2456618.24851 & 0.3360 & 2.2151 & 0.8483 & 11.78 & 334.57 & 186.54 & 0.00 & -3.75e$^{-11}$ &  8.75e$^{-13}$ &  4.18e$^{-11}$ \\
        B2-16  & 5 & 1 & 4.85$\times10^4$ & 2456618.24851 & 0.3360 & 2.2151 & 0.8483 & 11.78 & 334.57 & 186.54 & 0.00 & -3.62e$^{-11}$ & -4.49e$^{-11}$ &  8.99e$^{-11}$ \\
        B2-17  & 5 & 1 & 4.85$\times10^4$ & 2456618.24851 & 0.3360 & 2.2151 & 0.8483 & 11.78 & 334.57 & 186.54 & 0.00 &  1.15e$^{-11}$ & -9.96e$^{-11}$ &  2.47e$^{-11}$ \\
        \hline
    \end{tabular}}
    \caption{\label{tab:clones_paul2a} Simulation parameters of meteoroid streams ejected from different clones of comet 2P/Encke, considering more realistic nongravitational forces. Same notation as in Table E1.1 }
  \end{table}

  \begin{table}[!ht]
     \resizebox{\textwidth}{!}{
        \begin{tabular}{ccccccccccccccc}
        ID & $Np_{/app}$ & $f_{app}$ & $Np_{tot}$ & JD & q & a & e & i & $\Omega$ & $\omega$ & m & A1 & A2 & A3 \\
         &  &  &  & & (AU) & (AU) & & ($\degree$) & ($\degree$) & ($\degree$) & ($\degree$) & (AU.d$^{-2}$) & (AU.d$^{-2}$) & (AU.d$^{-2}$) \\
        B2-18  & 5 & 1 & 4.85$\times10^4$ & 2456618.24851 & 0.3360 & 2.2151 & 0.8483 & 11.78 & 334.57 & 186.54 & 0.00 &  9.87e$^{-11}$ & -4.14e$^{-11}$ & -9.38e$^{-11}$ \\
        B2-19  & 5 & 1 & 4.85$\times10^4$ & 2456618.24851 & 0.3360 & 2.2151 & 0.8483 & 11.78 & 334.57 & 186.54 & 0.00 &  5.74e$^{-11}$ & -6.91e$^{-11}$ &  8.57e$^{-11}$ \\
        B2-20  & 5 & 1 & 4.85$\times10^4$ & 2456618.24851 & 0.3360 & 2.2151 & 0.8483 & 11.78 & 334.57 & 186.54 & 0.00 &  6.92e$^{-11}$ &  3.02e$^{-11}$ & -9.22e$^{-11}$ \\
        B2-21  & 5 & 1 & 4.85$\times10^4$ & 2456618.24851 & 0.3360 & 2.2151 & 0.8483 & 11.78 & 334.57 & 186.54 & 0.00 & -3.19e$^{-11}$ & -2.47e$^{-11}$ & -1.07e$^{-11}$ \\
        B2-22  & 5 & 1 & 4.85$\times10^4$ & 2456618.24851 & 0.3360 & 2.2151 & 0.8483 & 11.78 & 334.57 & 186.54 & 0.00 & -4.81e$^{-11}$ &  8.49e$^{-11}$ & -7.62e$^{-11}$ \\
        B2-23  & 5 & 1 & 4.85$\times10^4$ & 2456618.24851 & 0.3360 & 2.2151 & 0.8483 & 11.78 & 334.57 & 186.54 & 0.00 &  3.80e$^{-11}$ &  6.07e$^{-11}$ & -8.75e$^{-11}$ \\
        B2-24  & 5 & 1 & 4.85$\times10^4$ & 2456618.24851 & 0.3360 & 2.2151 & 0.8483 & 11.78 & 334.57 & 186.54 & 0.00 & -4.24e$^{-11}$ &  2.46e$^{-11}$ & -9.86e$^{-11}$ \\ 
            B3-04  & 5 & 1 & 4.85$\times10^4$ & 2456618.24851 & 0.3360 & 2.2151 & 0.8483 & 11.78 & 334.57 & 186.54 & 0.00 &  1.41e$^{-11}$ &  4.46e$^{-11}$ &  2.10e$^{-11}$ \\
        B3-05  & 5 & 1 & 4.85$\times10^4$ & 2456618.24851 & 0.3360 & 2.2151 & 0.8483 & 11.78 & 334.57 & 186.54 & 0.00 &  2.62e$^{-11}$ &  5.20e$^{-11}$ &  6.75e$^{-13}$ \\
        B3-06  & 5 & 1 & 4.85$\times10^4$ & 2456618.24851 & 0.3360 & 2.2151 & 0.8483 & 11.78 & 334.57 & 186.54 & 0.00 &  7.65e$^{-11}$ &  7.43e$^{-11}$ & -2.86e$^{-11}$ \\
        B3-07  & 5 & 1 & 4.85$\times10^4$ & 2456618.24851 & 0.3360 & 2.2151 & 0.8483 & 11.78 & 334.57 & 186.54 & 0.00 & -2.87e$^{-11}$ & -7.79e$^{-11}$ &  8.71e$^{-11}$ \\
      B4-00  & 5 & 1 & 4.85$\times10^4$ & 2456618.24851 & 0.3360 & 2.2151 & 0.8483 & 11.78 & 334.57 & 186.54 & 0.00 &  9.19e$^{-12}$ &  6.32e$^{-12}$ &  5.55e$^{-12}$ \\
      B4-01  & 5 & 1 & 4.85$\times10^4$ & 2456618.24851 & 0.3360 & 2.2151 & 0.8483 & 11.78 & 334.57 & 186.54 & 0.00 & -8.33e$^{-11}$ &  9.88e$^{-11}$ & -1.15e$^{-11}$ \\
      B4-02  & 5 & 1 & 4.85$\times10^4$ & 2456618.24851 & 0.3360 & 2.2151 & 0.8483 & 11.78 & 334.57 & 186.54 & 0.00 &  8.72e$^{-11}$ &  5.30e$^{-11}$ &  3.53e$^{-11}$ \\
      B4-03  & 5 & 1 & 4.85$\times10^4$ & 2456618.24851 & 0.3360 & 2.2151 & 0.8483 & 11.78 & 334.57 & 186.54 & 0.00 &  6.55e$^{-11}$ &  4.49e$^{-12}$ & -6.11e$^{-11}$ \\
      B4-04  & 5 & 1 & 4.85$\times10^4$ & 2456618.24851 & 0.3360 & 2.2151 & 0.8483 & 11.78 & 334.57 & 186.54 & 0.00 & -1.41e$^{-11}$ & -3.78e$^{-11}$ &  7.70e$^{-11}$ \\
      B4-05  & 5 & 1 & 4.85$\times10^4$ & 2456618.24851 & 0.3360 & 2.2151 & 0.8483 & 11.78 & 334.57 & 186.54 & 0.00 &  2.02e$^{-12}$ & -6.13e$^{-11}$ & -3.86e$^{-11}$ \\
      B4-06  & 5 & 1 & 4.85$\times10^4$ & 2456618.24851 & 0.3360 & 2.2151 & 0.8483 & 11.78 & 334.57 & 186.54 & 0.00 & -3.49e$^{-11}$ & -6.33e$^{-11}$ & -1.66e$^{-11}$ \\
      B4-07  & 5 & 1 & 4.85$\times10^4$ & 2456618.24851 & 0.3360 & 2.2151 & 0.8483 & 11.78 & 334.57 & 186.54 & 0.00 & -7.24e$^{-11}$ &  6.61e$^{-11}$ & -2.10e$^{-11}$ \\
      B4-08  & 5 & 1 & 4.85$\times10^4$ & 2456618.24851 & 0.3360 & 2.2151 & 0.8483 & 11.78 & 334.57 & 186.54 & 0.00 &  7.54e$^{-11}$ &  5.21e$^{-11}$ &  6.73e$^{-11}$ \\
       B4-09  & 5 & 1 & 4.85$\times10^4$ & 2456618.24851 & 0.3360 & 2.2151 & 0.8483 & 11.78 & 334.57 & 186.54 & 0.00 & -1.65e$^{-11}$ &  7.84e$^{-11}$ &  9.72e$^{-11}$ \\
      B4-10  & 5 & 1 & 4.85$\times10^4$ & 2456618.24851 & 0.3360 & 2.2151 & 0.8483 & 11.78 & 334.57 & 186.54 & 0.00 & -1.55e$^{-11}$ & -7.28e$^{-11}$ & -5.75e$^{-11}$ \\
      B4-11  & 5 & 1 & 4.85$\times10^4$ & 2456618.24851 & 0.3360 & 2.2151 & 0.8483 & 11.78 & 334.57 & 186.54 & 0.00 & -4.97e$^{-11}$ & -1.98e$^{-11}$ & -1.27e$^{-11}$ \\
      B4-12  & 5 & 1 & 4.85$\times10^4$ & 2456618.24851 & 0.3360 & 2.2151 & 0.8483 & 11.78 & 334.57 & 186.54 & 0.00 & -6.39e$^{-11}$ &  9.99e$^{-11}$ &  7.97e$^{-11}$ \\
      B4-13  & 5 & 1 & 4.85$\times10^4$ & 2456618.24851 & 0.3360 & 2.2151 & 0.8483 & 11.78 & 334.57 & 186.54 & 0.00 & -8.04e$^{-12}$ & -1.48e$^{-11}$ &  9.81e$^{-12}$ \\
      B4-14  & 5 & 1 & 4.85$\times10^4$ & 2456618.24851 & 0.3360 & 2.2151 & 0.8483 & 11.78 & 334.57 & 186.54 & 0.00 & -9.78e$^{-11}$ & -1.48e$^{-11}$ & -2.77e$^{-11}$ \\
      B4-15  & 5 & 1 & 4.85$\times10^4$ & 2456618.24851 & 0.3360 & 2.2151 & 0.8483 & 11.78 & 334.57 & 186.54 & 0.00 &  8.53e$^{-11}$ &  6.53e$^{-11}$ & -4.08e$^{-11}$ \\
      B4-16  & 5 & 1 & 4.85$\times10^4$ & 2456618.24851 & 0.3360 & 2.2151 & 0.8483 & 11.78 & 334.57 & 186.54 & 0.00 & -6.25e$^{-11}$ & -1.68e$^{-11}$ &  9.30e$^{-11}$ \\
      B4-17  & 5 & 1 & 4.85$\times10^4$ & 2456618.24851 & 0.3360 & 2.2151 & 0.8483 & 11.78 & 334.57 & 186.54 & 0.00 &  5.18e$^{-11}$ & -4.93e$^{-11}$ &  2.18e$^{-11}$ \\
      B4-18  & 5 & 1 & 4.85$\times10^4$ & 2456618.24851 & 0.3360 & 2.2151 & 0.8483 & 11.78 & 334.57 & 186.54 & 0.00 &  3.16e$^{-11}$ & -2.21e$^{-11}$ &  4.97e$^{-11}$ \\
      B4-19  & 5 & 1 & 4.85$\times10^4$ & 2456618.24851 & 0.3360 & 2.2151 & 0.8483 & 11.78 & 334.57 & 186.54 & 0.00 & -4.02e$^{-11}$ &  9.01e$^{-11}$ & -7.06e$^{-11}$ \\
      B4-20  & 5 & 1 & 4.85$\times10^4$ & 2456618.24851 & 0.3360 & 2.2151 & 0.8483 & 11.78 & 334.57 & 186.54 & 0.00 &  1.30e$^{-11}$ &  9.47e$^{-11}$ & -8.96e$^{-11}$ \\
      B4-21  & 5 & 1 & 4.85$\times10^4$ & 2456618.24851 & 0.3360 & 2.2151 & 0.8483 & 11.78 & 334.57 & 186.54 & 0.00 &  3.05e$^{-11}$ & -9.81e$^{-11}$ &  9.66e$^{-11}$ \\
      B4-22  & 5 & 1 & 4.85$\times10^4$ & 2456618.24851 & 0.3360 & 2.2151 & 0.8483 & 11.78 & 334.57 & 186.54 & 0.00 & -1.45e$^{-11}$ &  7.75e$^{-11}$ &  8.62e$^{-11}$ \\
      B4-23  & 5 & 1 & 4.85$\times10^4$ & 2456618.24851 & 0.3360 & 2.2151 & 0.8483 & 11.78 & 334.57 & 186.54 & 0.00 & -5.32e$^{-11}$ & -6.01e$^{-11}$ &  4.10e$^{-11}$ \\
      B4-24  & 5 & 1 & 4.85$\times10^4$ & 2456618.24851 & 0.3360 & 2.2151 & 0.8483 & 11.78 & 334.57 & 186.54 & 0.00 &  2.24e$^{-11}$ &  9.52e$^{-11}$ & -2.36e$^{-11}$ \\
      B5-02  & 5 & 1 & 4.85$\times10^4$ & 2456618.24851 & 0.3360 & 2.2151 & 0.8483 & 11.78 & 334.57 & 186.54 & 0.00 & -6.35e$^{-11}$ & -4.64e$^{-11}$ &  8.36e$^{-11}$ \\
      B5-03  & 5 & 1 & 4.85$\times10^4$ & 2456618.24851 & 0.3360 & 2.2151 & 0.8483 & 11.78 & 334.57 & 186.54 & 0.00 &  1.81e$^{-11}$ & -4.49e$^{-11}$ &  9.50e$^{-11}$ \\
      B5-04  & 5 & 1 & 4.85$\times10^4$ & 2456618.24851 & 0.3360 & 2.2151 & 0.8483 & 11.78 & 334.57 & 186.54 & 0.00 &  1.66e$^{-11}$ & -4.65e$^{-11}$ & -2.12e$^{-11}$ \\
      B5-05  & 5 & 1 & 4.85$\times10^4$ & 2456618.24851 & 0.3360 & 2.2151 & 0.8483 & 11.78 & 334.57 & 186.54 & 0.00 &  3.54e$^{-11}$ &  3.03e$^{-11}$ &  4.08e$^{-11}$ \\
      B5-06  & 5 & 1 & 4.85$\times10^4$ & 2456618.24851 & 0.3360 & 2.2151 & 0.8483 & 11.78 & 334.57 & 186.54 & 0.00 &  6.52e$^{-11}$ &  8.96e$^{-11}$ &  1.38e$^{-11}$ \\
      B5-07  & 5 & 1 & 4.85$\times10^4$ & 2456618.24851 & 0.3360 & 2.2151 & 0.8483 & 11.78 & 334.57 & 186.54 & 0.00 & -1.74e$^{-11}$ & -7.53e$^{-11}$ &  2.02e$^{-11}$ \\
      B5-08  & 5 & 1 & 4.85$\times10^4$ & 2456618.24851 & 0.3360 & 2.2151 & 0.8483 & 11.78 & 334.57 & 186.54 & 0.00 & -9.65e$^{-11}$ & -5.89e$^{-11}$ & -7.83e$^{-11}$ \\
      B5-09  & 5 & 1 & 4.85$\times10^4$ & 2456618.24851 & 0.3360 & 2.2151 & 0.8483 & 11.78 & 334.57 & 186.54 & 0.00 & -7.54e$^{-11}$ & -8.91e$^{-11}$ & -2.28e$^{-11}$ \\
      B5-10  & 5 & 1 & 4.85$\times10^4$ & 2456618.24851 & 0.3360 & 2.2151 & 0.8483 & 11.78 & 334.57 & 186.54 & 0.00 &  9.41e$^{-11}$ &  4.93e$^{-11}$ & -5.83e$^{-11}$ \\
      B5-11  & 5 & 1 & 4.85$\times10^4$ & 2456618.24851 & 0.3360 & 2.2151 & 0.8483 & 11.78 & 334.57 & 186.54 & 0.00 & -5.01e$^{-11}$ &  9.57e$^{-11}$ & -7.47e$^{-11}$ \\
      B5-12  & 5 & 1 & 4.85$\times10^4$ & 2456618.24851 & 0.3360 & 2.2151 & 0.8483 & 11.78 & 334.57 & 186.54 & 0.00 &  6.05e$^{-11}$ & -5.81e$^{-11}$ & -9.11e$^{-11}$ \\
      B5-13  & 5 & 1 & 4.85$\times10^4$ & 2456618.24851 & 0.3360 & 2.2151 & 0.8483 & 11.78 & 334.57 & 186.54 & 0.00 &  3.86e$^{-11}$ & -5.56e$^{-11}$ &  2.38e$^{-11}$ \\
      B5-14  & 5 & 1 & 4.85$\times10^4$ & 2456618.24851 & 0.3360 & 2.2151 & 0.8483 & 11.78 & 334.57 & 186.54 & 0.00 & -4.22e$^{-11}$ &  9.72e$^{-11}$ &  3.46e$^{-11}$ \\
      B5-15  & 5 & 1 & 4.85$\times10^4$ & 2456618.24851 & 0.3360 & 2.2151 & 0.8483 & 11.78 & 334.57 & 186.54 & 0.00 &  9.30e$^{-11}$ & -9.31e$^{-11}$ &  2.27e$^{-11}$ \\
      B5-16  & 5 & 1 & 4.85$\times10^4$ & 2456618.24851 & 0.3360 & 2.2151 & 0.8483 & 11.78 & 334.57 & 186.54 & 0.00 &  7.62e$^{-11}$ &  4.53e$^{-12}$ &  8.05e$^{-11}$ \\
      B5-17  & 5 & 1 & 4.85$\times10^4$ & 2456618.24851 & 0.3360 & 2.2151 & 0.8483 & 11.78 & 334.57 & 186.54 & 0.00 &  9.05e$^{-11}$ &  6.72e$^{-11}$ &  6.62e$^{-11}$ \\
      B5-18  & 5 & 1 & 4.85$\times10^4$ & 2456618.24851 & 0.3360 & 2.2151 & 0.8483 & 11.78 & 334.57 & 186.54 & 0.00 &  2.13e$^{-12}$ & -4.69e$^{-11}$ & -3.48e$^{-11}$ \\
      B5-19  & 5 & 1 & 4.85$\times10^4$ & 2456618.24851 & 0.3360 & 2.2151 & 0.8483 & 11.78 & 334.57 & 186.54 & 0.00 & -1.88e$^{-11}$ & -2.83e$^{-11}$ & -6.62e$^{-11}$ \\
      B5-20  & 5 & 1 & 4.85$\times10^4$ & 2456618.24851 & 0.3360 & 2.2151 & 0.8483 & 11.78 & 334.57 & 186.54 & 0.00 & -5.17e$^{-11}$ &  5.32e$^{-11}$ &  4.89e$^{-12}$ \\
      B5-21  & 5 & 1 & 4.85$\times10^4$ & 2456618.24851 & 0.3360 & 2.2151 & 0.8483 & 11.78 & 334.57 & 186.54 & 0.00 & -1.86e$^{-11}$ & -4.57e$^{-11}$ & -2.22e$^{-11}$ \\
      B5-22  & 5 & 1 & 4.85$\times10^4$ & 2456618.24851 & 0.3360 & 2.2151 & 0.8483 & 11.78 & 334.57 & 186.54 & 0.00 &  2.04e$^{-11}$ &  9.71e$^{-11}$ & -2.34e$^{-11}$ \\
      B5-23  & 5 & 1 & 4.85$\times10^4$ & 2456618.24851 & 0.3360 & 2.2151 & 0.8483 & 11.78 & 334.57 & 186.54 & 0.00 &  2.99e$^{-11}$ & -7.73e$^{-11}$ & -2.02e$^{-11}$ \\
      B5-24  & 5 & 1 & 4.85$\times10^4$ & 2456618.24851 & 0.3360 & 2.2151 & 0.8483 & 11.78 & 334.57 & 186.54 & 0.00 & -5.53e$^{-11}$ &  2.69e$^{-11}$ &  5.49e$^{-11}$ \\  
      \hline
    \end{tabular}}
    \caption{\label{tab:clones_paul2b} Simulation parameters of meteoroid streams ejected from different clones of comet 2P/Encke, considering more realistic nongravitational forces. Same notation as in Table E1.1 }
  \end{table}

\newpage
\clearpage
\textbf{E2. Past Orbital Evolution of 2P/Encke and Clones}\\

\setfignumprefix{E2.}	
\begin{figure}[!ht]
  \includegraphics[width=.33\textwidth]{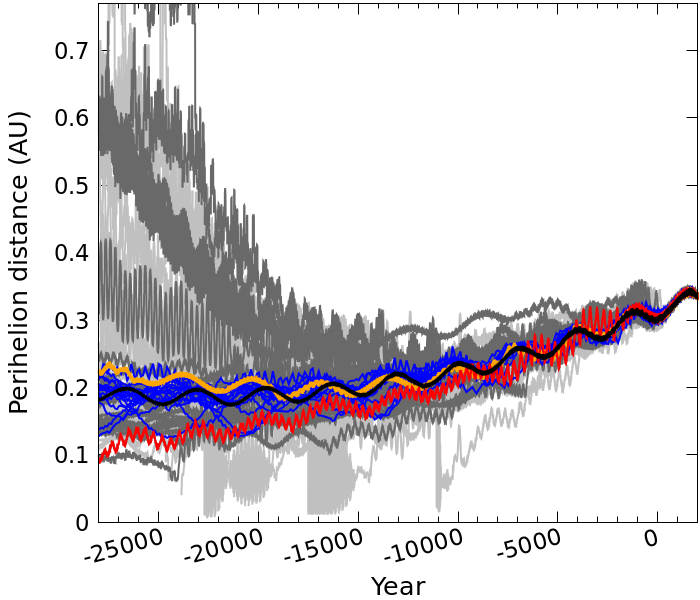}
  \includegraphics[width=.33\textwidth]{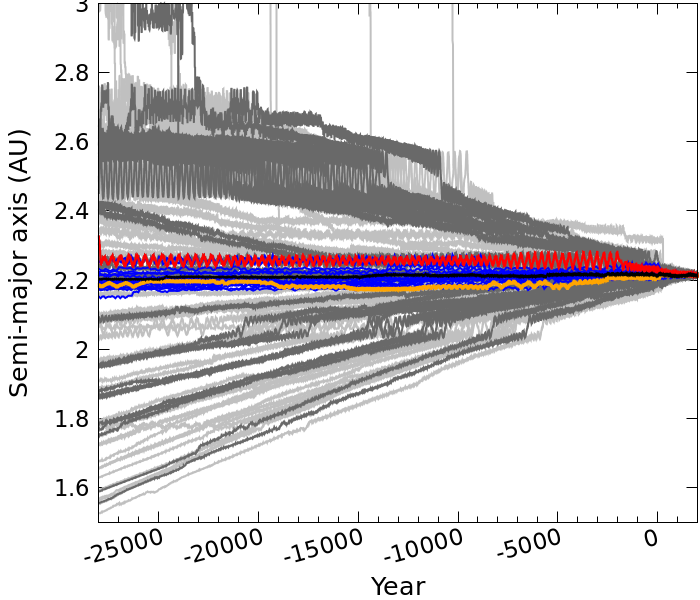}
  \includegraphics[width=.33\textwidth]{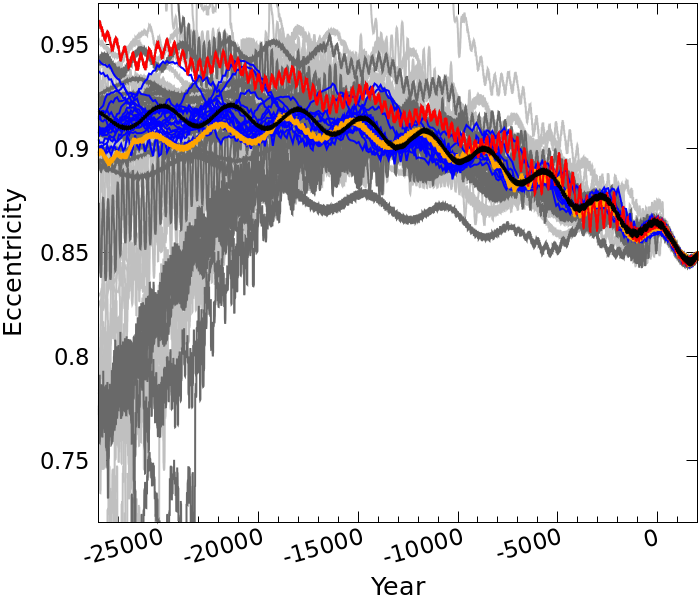}\\
  \includegraphics[width=.33\textwidth]{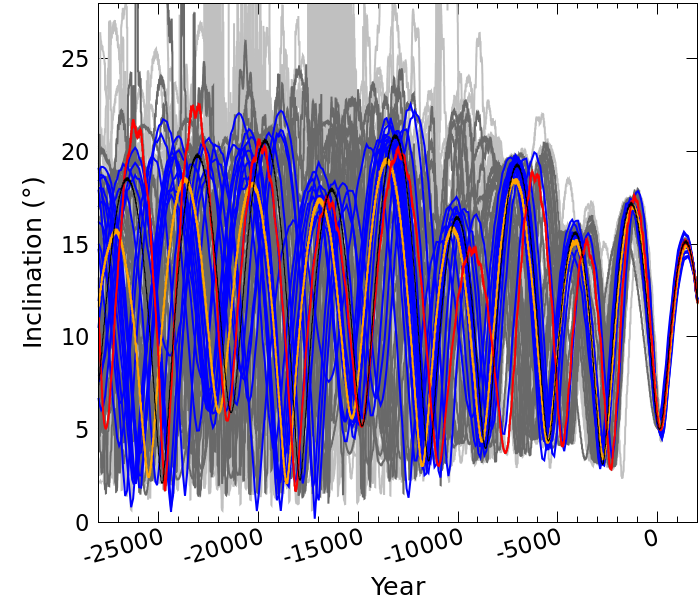}
  \includegraphics[width=.33\textwidth]{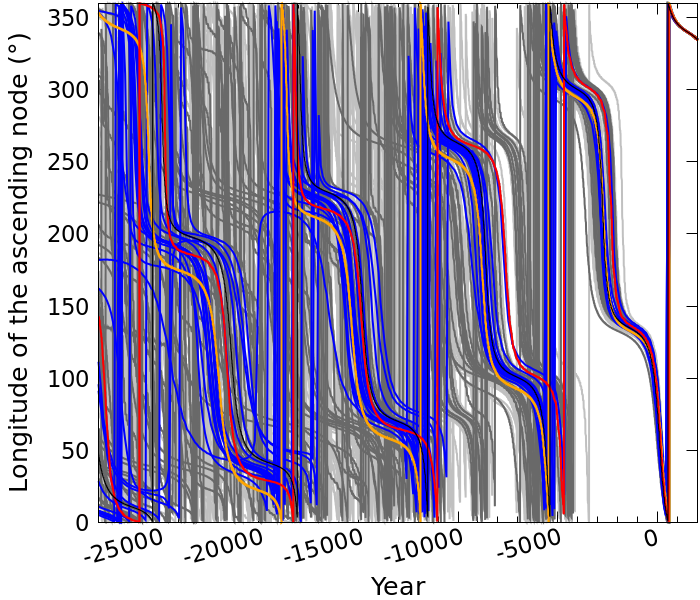}
  \includegraphics[width=.33\textwidth]{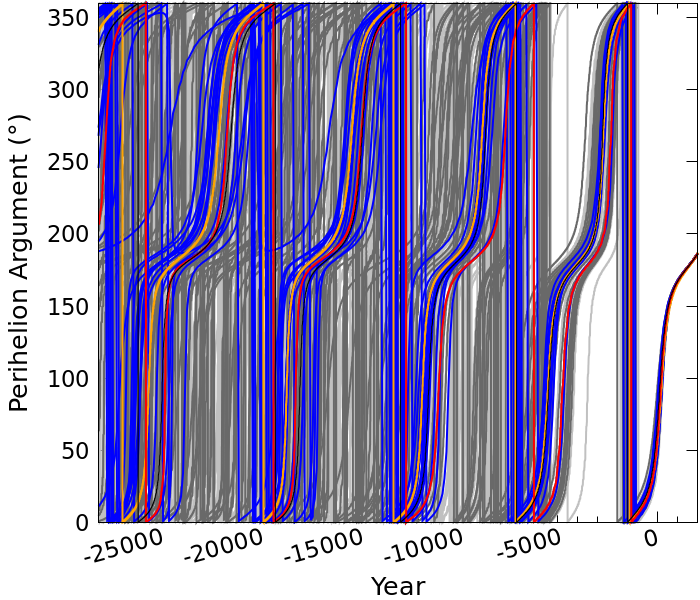}
  \caption{Orbital evolution of the clones generated from 2P/Encke's nominal orbit over 30 Ka. The black line represents the nominal evolution of comet 2P/Encke, while the light and dark grey lines illustrate the evolution of the clones presented in tables \ref{tab:clones_paul1}, \ref{tab:clones_paul2a} and \ref{tab:clones_paul2b}. The evolution of clones presented in table \ref{tab:clones_auriane} are presented in dark blue, with the peculiar clone 30 and clone 4 illustrated by the orange and red curve respectively. }
\end{figure}

\textbf{E3 Observed and Modelled Shower Activity Profiles for Different Clones of 2P/Encke}\\

\begin{figure*}[!ht]
\centering
 \includegraphics[width=.98\textwidth]{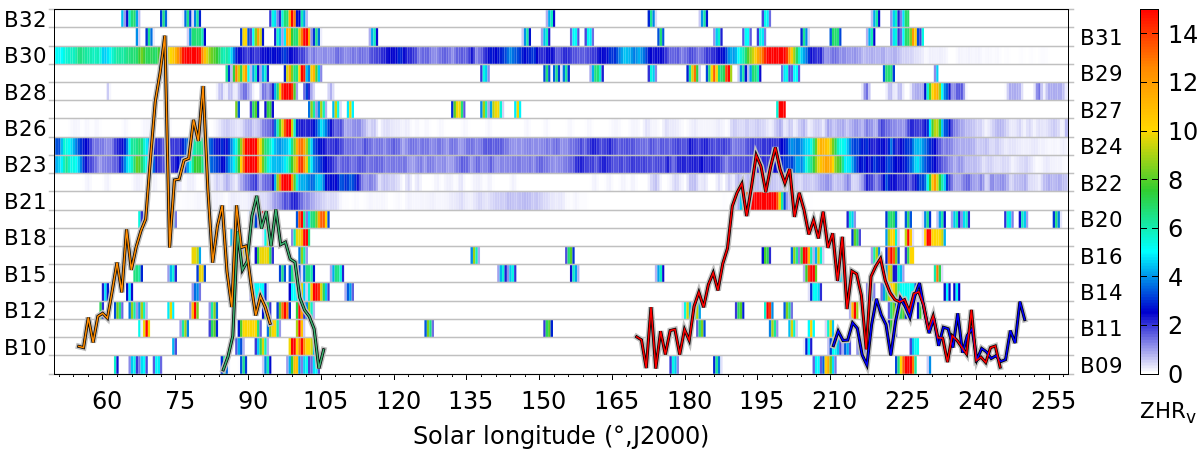}
 \caption{\label{fig:clonesB1_profiles} Meteor activity produced in 2021 by meteoroid streams modelled from different ephemerides of comet 2P/Encke (sample B). The ZHR$_v$ profile of each clone is represented by a horizontal band color-coded as a function of the modelled meteor rates. Maximum activity regions are represented by yellow to red areas, while moderate rates are illustrated in blue. The average activity profile of the NTA, STA, BTA and ZPE as measured by CMOR is presented for comparison with blue, red, green and orange lines respectively. }
\end{figure*}

\begin{figure*}[!ht]
\centering
 \includegraphics[width=1\textwidth]{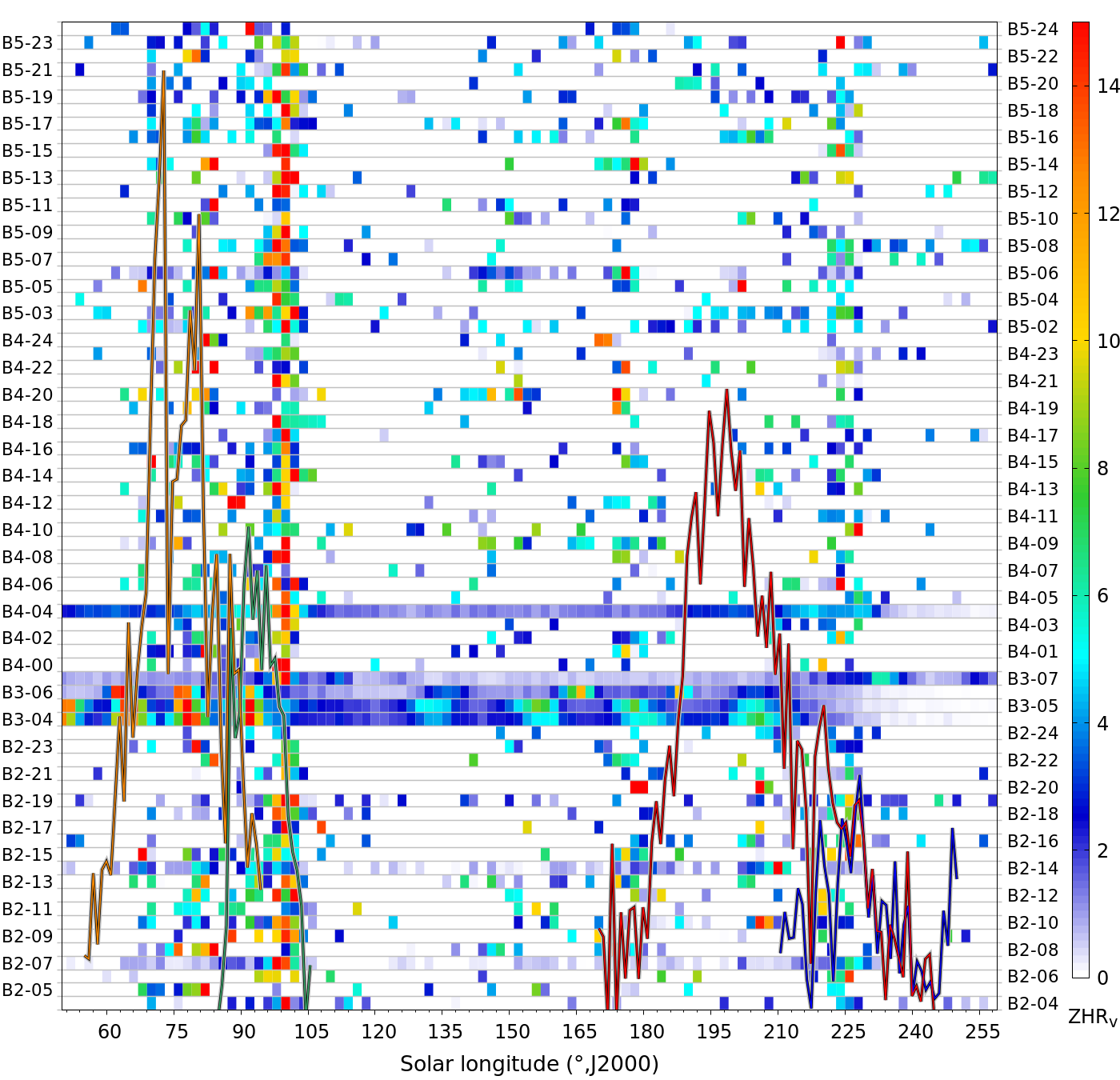}
 \caption{\label{fig:clonesB2_profiles}   Meteor activity produced in 2021 by meteoroid streams modelled from different ephemerides of comet 2P/Encke (sample B). The ZHR$_v$ profile of each clone is represented by an horizontal band color-coded as a function of the expected meteor rates. Maximum activity regions are represented by yellow to red areas, while moderate rates are illustrated in blue. The average activity profile of the NTA, STA, BTA and ZPE as measured by CMOR is presented for comparison with blue, red, green and orange lines respectively.}
\end{figure*}

\clearpage
\newpage

\begin{figure}
\textbf{E4 Radiants from Different Clones of 2P/Encke}\\[-0.3cm] 

\centering
	\includegraphics[trim={0 0.0 3.9cm 0.75cm},clip,width=.36\textwidth]{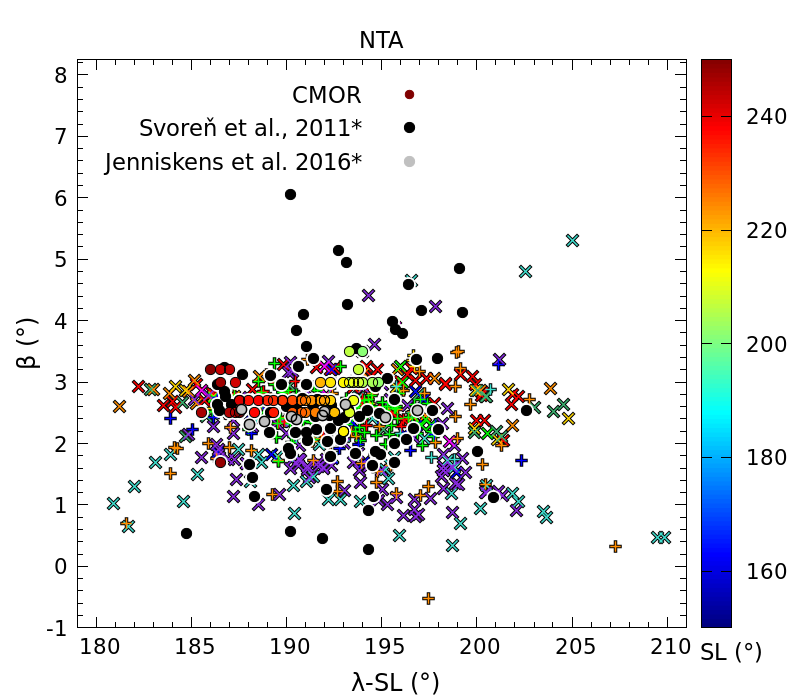}
	\includegraphics[trim={0 0 3.9cm 0.75cm},clip,width=.36\textwidth]{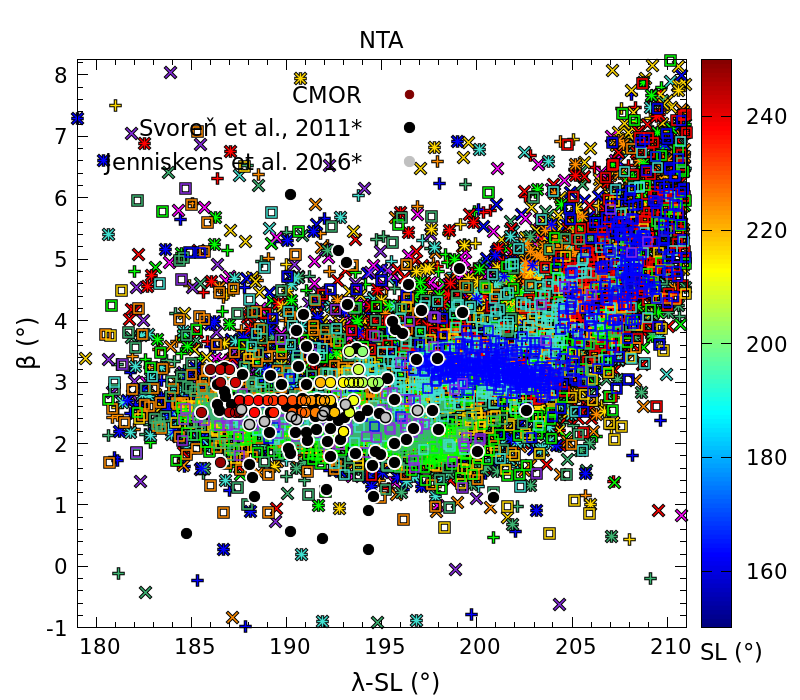}\\[-0.12cm]
	\includegraphics[trim={0 0 3.9cm 0.75cm},clip,width=.36\textwidth]{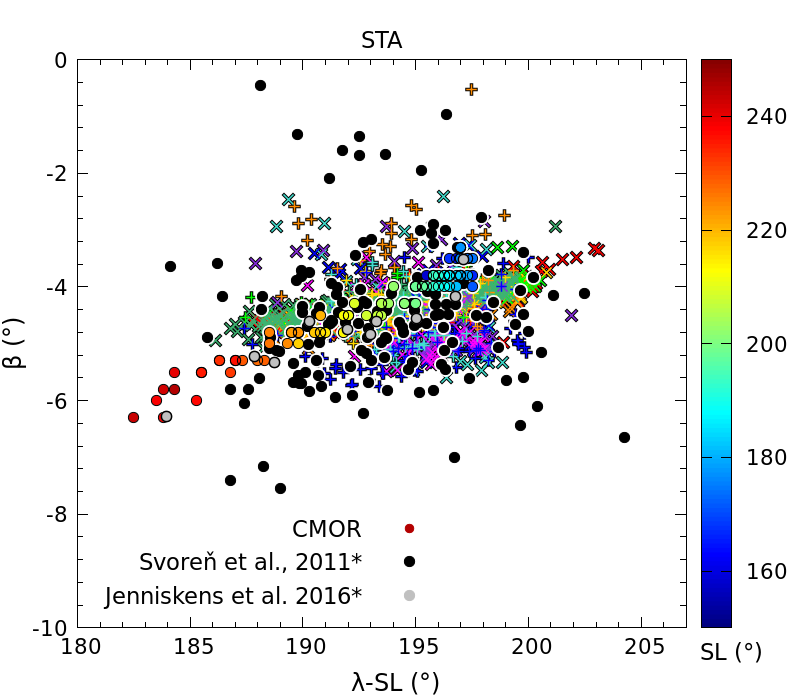}
	\includegraphics[trim={0 0 3.9cm 0.75cm},clip,width=.36\textwidth]{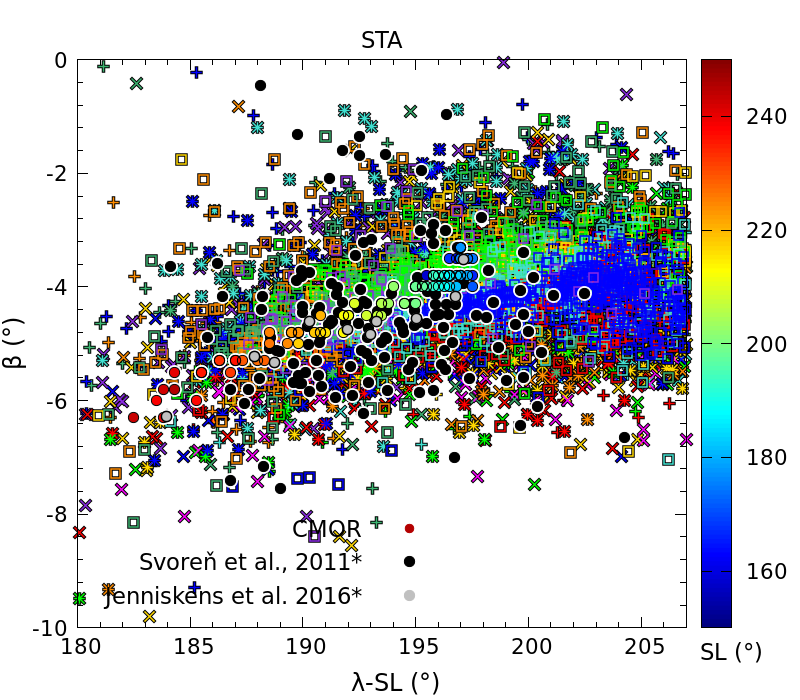}\\[-0.12cm]
	\includegraphics[trim={0 0 3.9cm 0.75cm},clip,width=.36\textwidth]{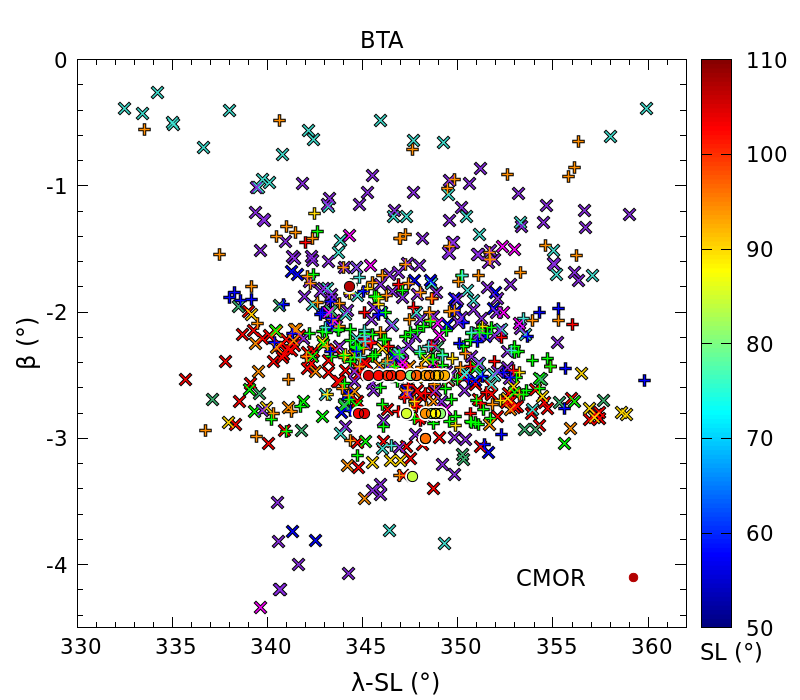}
	\includegraphics[trim={0 0 3.9cm 0.75cm},clip,width=.36\textwidth]{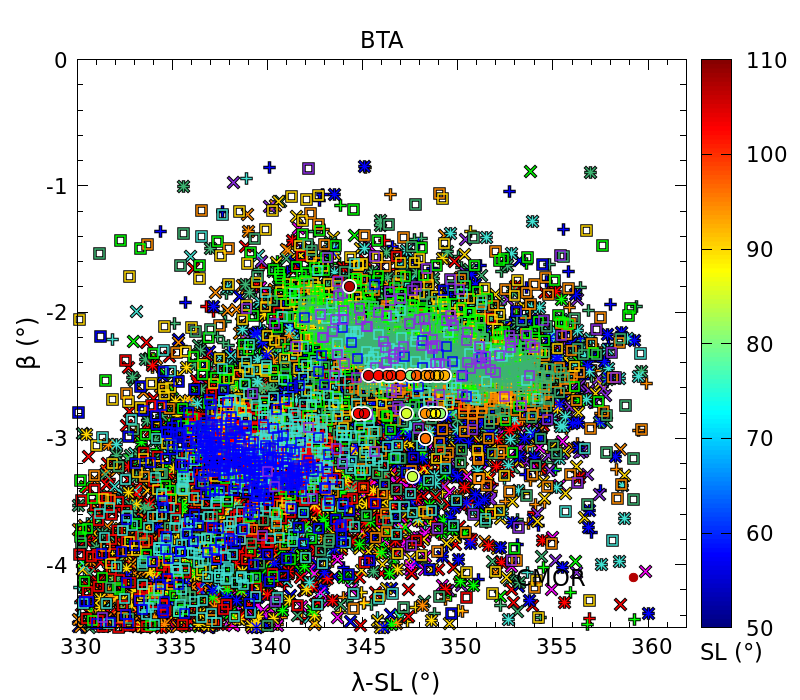}\\[-0.12cm]
	\includegraphics[trim={0 0 3.9cm 0.75cm},clip,width=.36\textwidth]{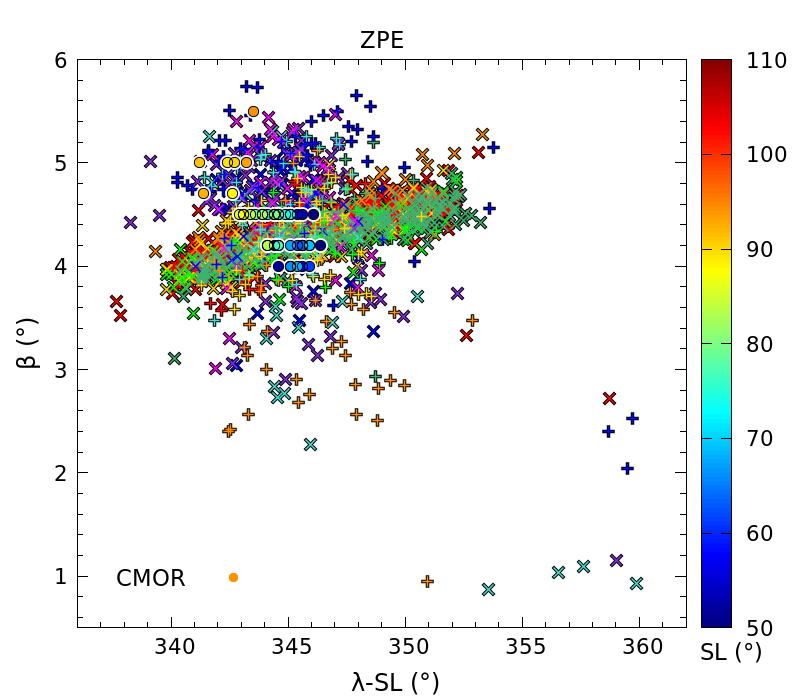}
	\includegraphics[trim={0 0 3.9cm 0.75cm},clip,width=.36\textwidth]{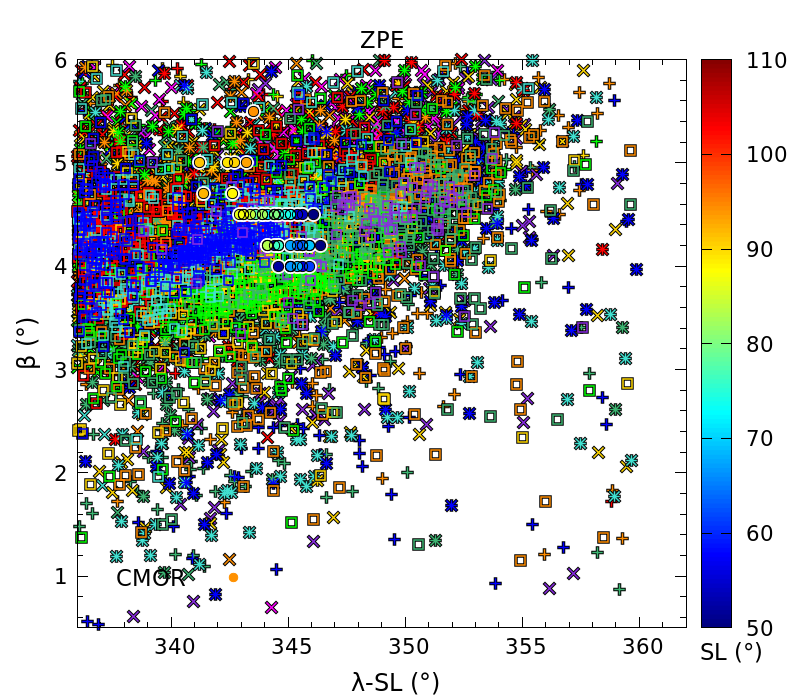}
	\caption{Sun-centered ecliptic longitude and latitude of the clones presented in table \ref{tab:clones_auriane} (left panel) or \ref{tab:clones_paul1}, \ref{tab:clones_paul2a} and \ref{tab:clones_paul2b} (right panel) that approach Earth in 2021 with a MOID below 5$\times10^{-3}$ AU. Meteoroids from different clones of the comet are represented by symbols of different colors. The simulations are compared with the radiants of the showers observed by CMOR (circles color-coded in function of the solar longitude), and presented in \protect\cite{Svoren2011} and \protect\cite{Jenniskens2016}.}
\end{figure}

\newpage
\clearpage
\textbf{APPENDIX F: Comparing clone A4 with the nominal solution of 2P/Encke}\\ 

\textbf{F1 Orbital elements} \\

\setfignumprefix{F1.}
\begin{figure}[!ht]
    \centering
    \includegraphics[width=\textwidth]{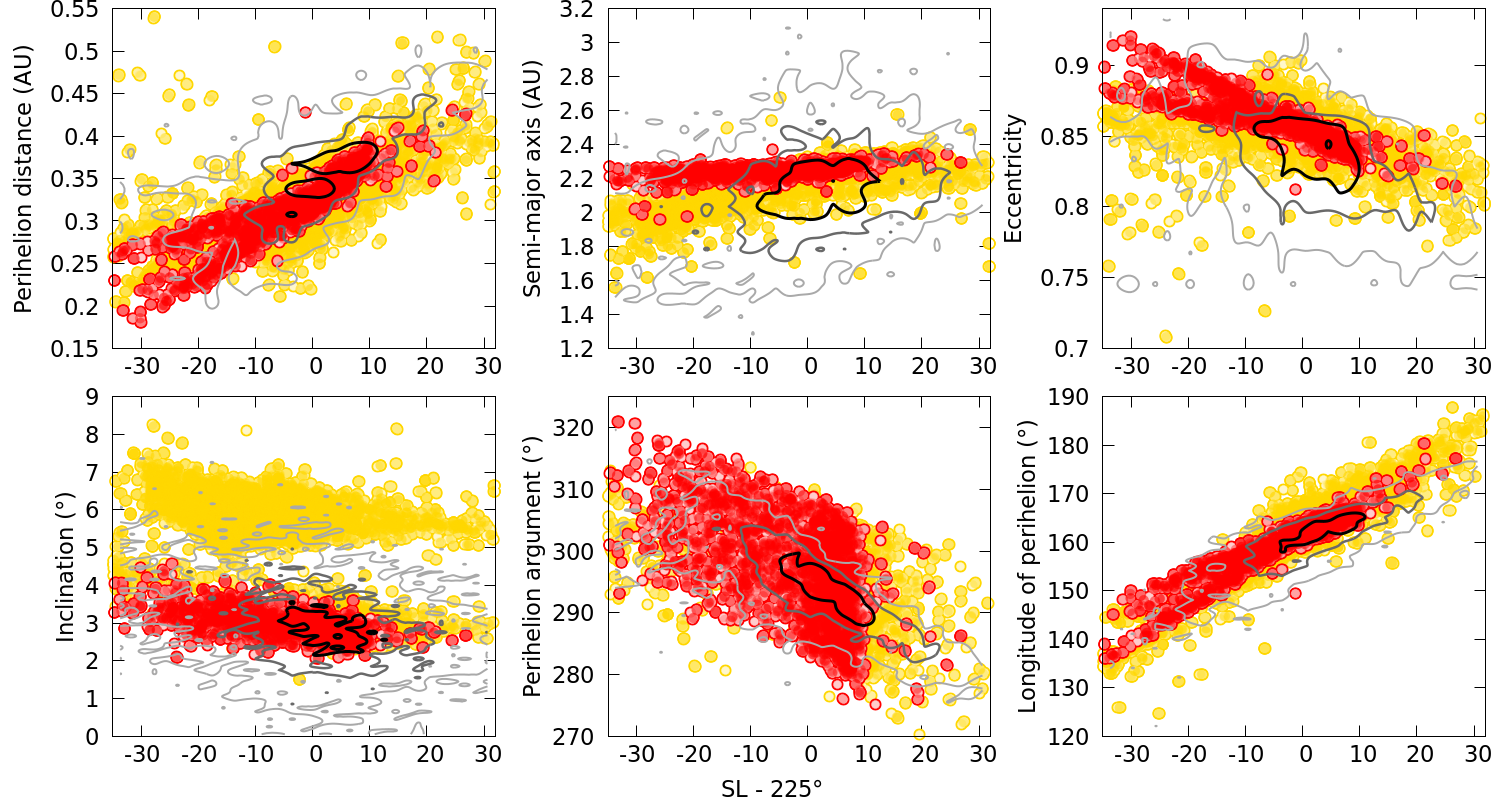}
    \caption{Perihelion distance, semi-major axis, eccentricity, inclination, perihelion argument and longitude of perihelion of NTA meteors simulated from 2P/Encke's nominal ephemeris (yellow) and clone A4 (red) approaching Earth with a MOID below 0.01 AU between 2002 and 2021. Contour lines represent the density of meteoroids recorded in the CAMS database, from low (light grey) to high (black) as presented in \protect\cite{Egal2022}, Figure C2.1.1.}
\end{figure}

\begin{figure}[!ht]
    \centering
    \includegraphics[width=\textwidth]{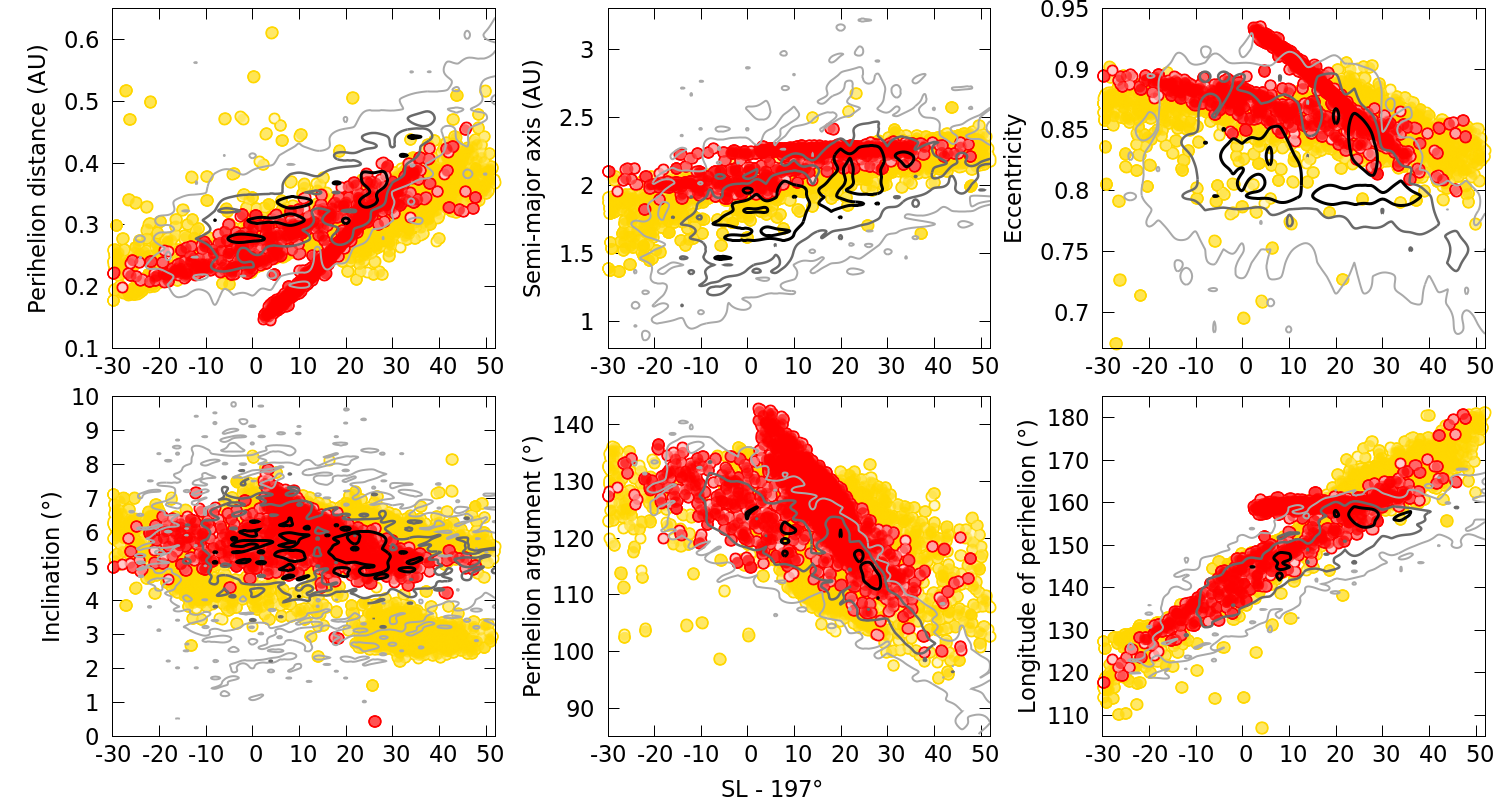}
    \caption{Perihelion distance, semi-major axis, eccentricity, inclination, perihelion argument and longitude of perihelion of STA meteors simulated from 2P/Encke's nominal ephemeris (yellow) and clone A4 (red) approaching Earth with a MOID below 0.01 AU between 2002 and 2021. Contour lines represent the density of meteoroids recorded in the CAMS database, from low (light grey) to high (black) as presented in \protect\cite{Egal2022}, Figure C2.1.2. }
\end{figure}

\begin{figure}[!ht]
    \centering
    \includegraphics[width=\textwidth]{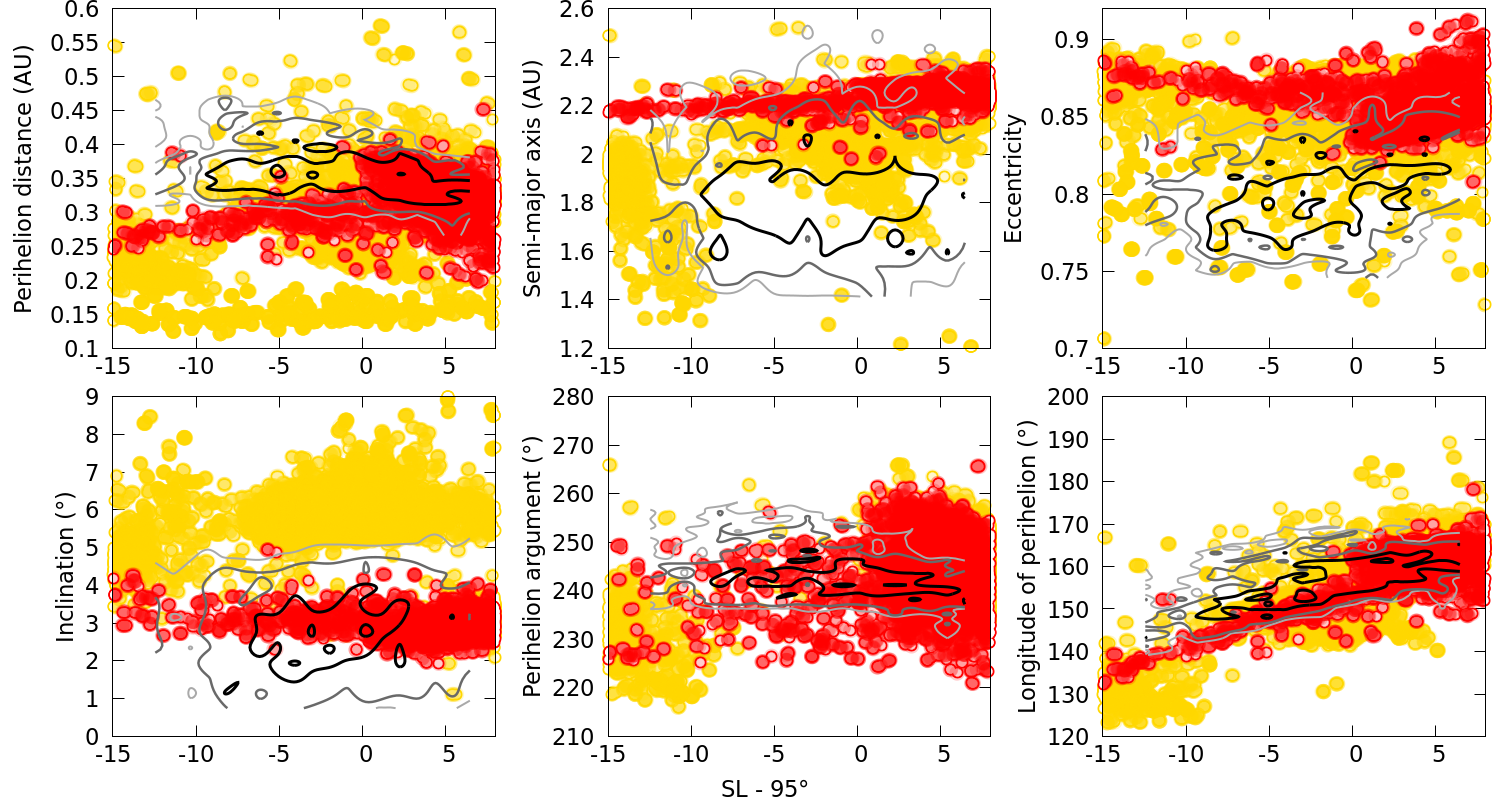}
    \caption{Perihelion distance, semi-major axis, eccentricity, inclination, perihelion argument and longitude of perihelion of BTA meteors simulated from 2P/Encke's nominal ephemeris (yellow) and clone A4 (red) approaching Earth with a MOID below 0.01 AU between 2002 and 2021. Contour lines represent the density of meteoroids recorded by CMOR, from low (light grey) to high (black) as presented in \protect\cite{Egal2022}, Figure C2.2.3. }
\end{figure}

\begin{figure}[!ht]
    \centering
    \includegraphics[width=\textwidth]{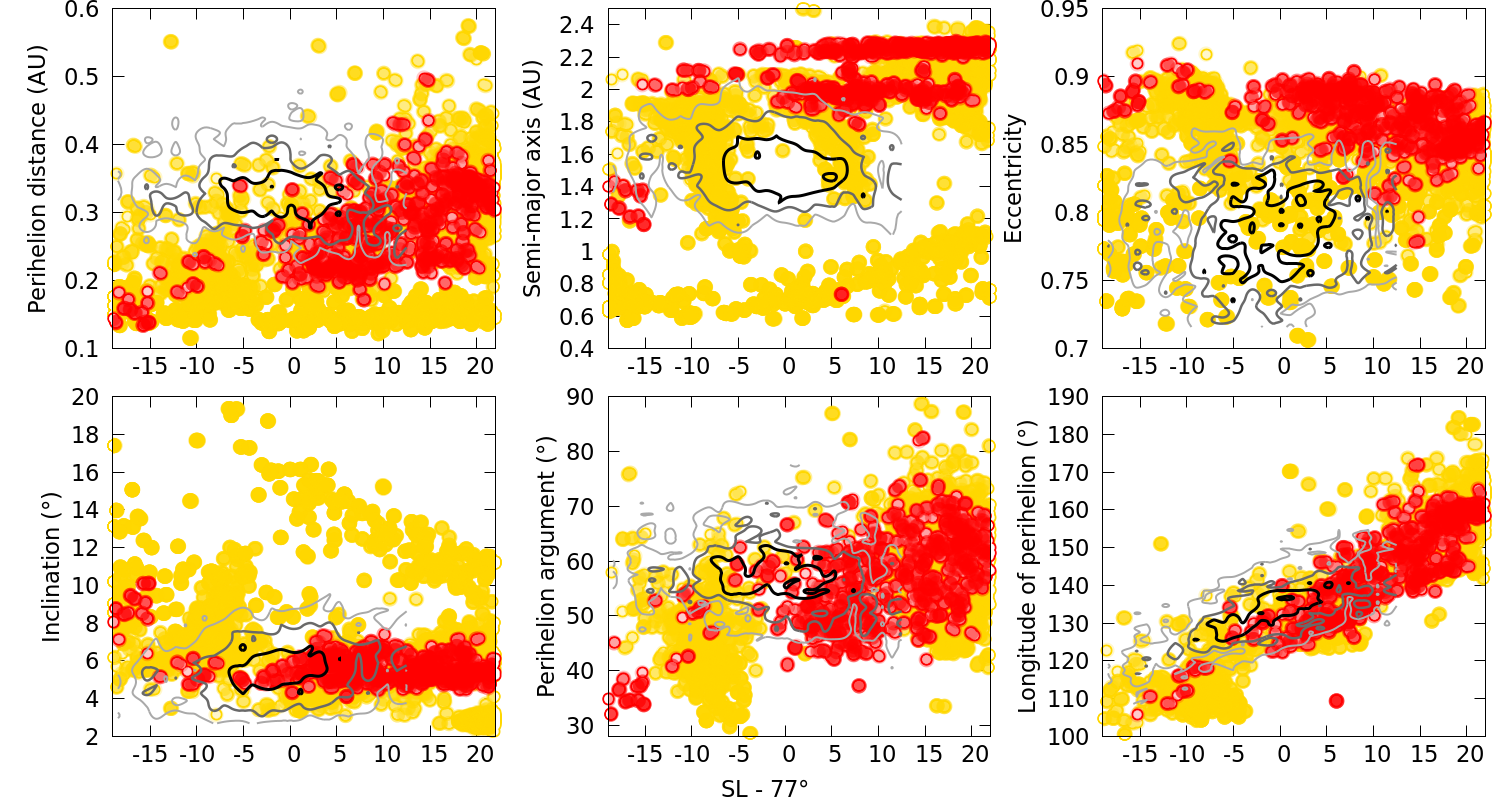}
    \caption{Perihelion distance, semi-major axis, eccentricity, inclination, perihelion argument and longitude of perihelion of ZPE meteors simulated from 2P/Encke's nominal ephemeris (yellow) and clone A4 (red) approaching Earth with a MOID below 0.01 AU between 2002 and 2021. Contour lines represent the density of meteoroids recorded by CMOR, from low (light grey) to high (black) as presented in \protect\cite{Egal2022}, Figure C2.2.3.}
\end{figure}

\clearpage
\newpage
\clearpage
\textbf{F2 Clone A4 Meteoroid Stream Nodal crossing locations} \\

\setfignumprefix{F2.}
\begin{figure}[!ht]
    \centering
    \includegraphics[trim={0.9cm 0 6.0cm 0}, clip,height=10.8cm]{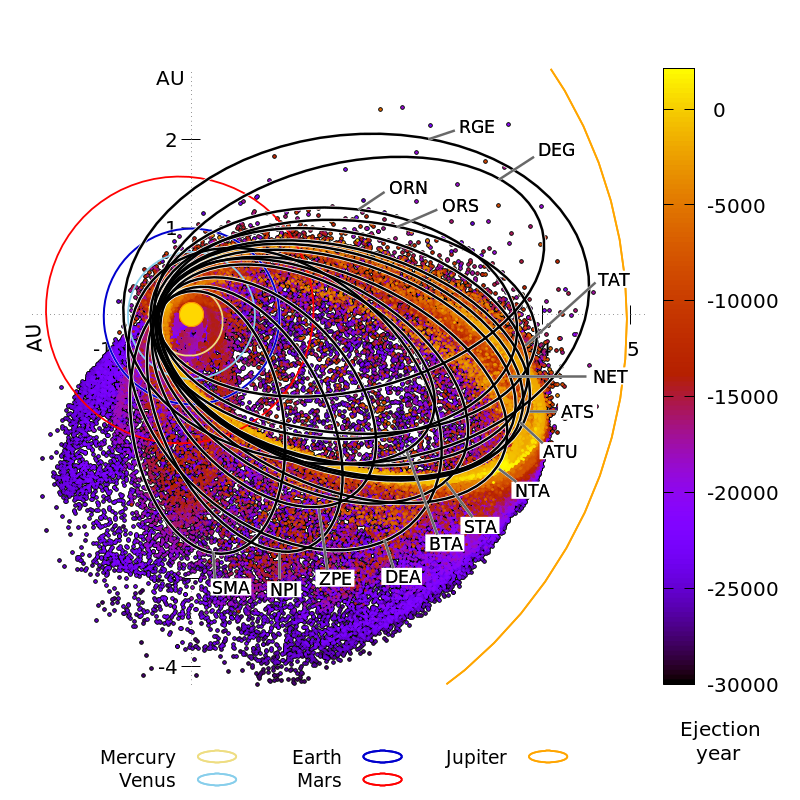}
    \includegraphics[trim={0.9cm 0 0.8cm 0},clip,height=10.8cm]{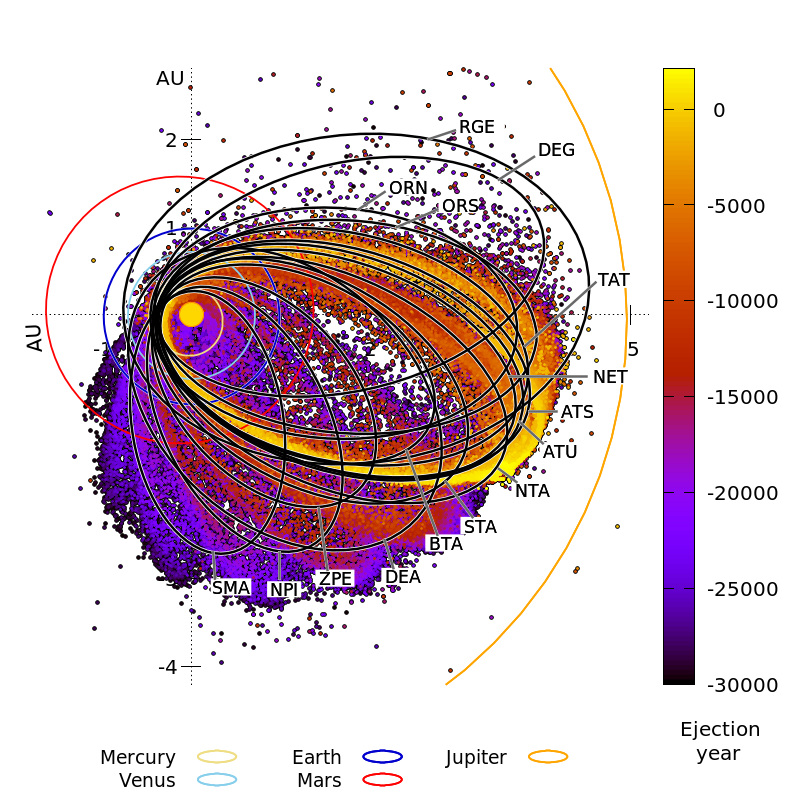}
    \caption{Nodal-crossing location of the particles ejected from clone A4 (left panel) or the nominal ephemeris of comet 2P/Encke (right panel). Particles are color-coded as a function of the ejection year. The orbits of several minor meteor showers associated with the Taurid complex (cf. Table \ref{tab:minor_showers}), projected on the ecliptic plane, are represented in black.}
    \label{fig:TC_showers}
\end{figure}

\settabnumprefix{F2.}	

\begin{table}[!h]
    \centering
    \begin{tabular}{llcccccl}
    Code & No & a (AU) & e & i ($\degree$) & $\omega$ ($\degree$) & $\Omega$ ($\degree$) & Reference \\
    \hline
SMA &  156 & 1.51  & 0.820 & 5.1 & 227.1 & 231.7 &  \cite{Brown2008}   \\ 
DEA &  154 & 2.026 & 0.708 & 2.8 & 48.1  & 90.0  &  \cite{Sekanina1976}    \\
ORN &  256 & 2.09  & 0.778 & 2.2 & 255.9 & 280.8 &  \cite{Jenniskens2016}   \\ 
ORS &  257 & 2.16  & 0.828 & 5.3 & 64.3  & 111.3 &  \cite{Jenniskens2016}   \\ 
NPI &  215 & 1.63  & 0.82  & 4.9 & 172.0 & 305.1 &  \cite{Jenniskens2016}   \\ 
ATS &  629 & 2.19  & 0.82  & 2.7 & 290.5 & 233.4 &  \cite{Jenniskens2016}    \\
ATU &  635 & 2.16  & 0.82  & 2.6 & 291.2 & 231.2 &  \cite{Jenniskens2016}    \\
NET &  632 & 2.11  & 0.83  & 2.8 & 294.3 & 227.1 &  \cite{Jenniskens2016}    \\
TAT &  634 & 2.17  & 0.80  & 2.5 & 284.7 & 243.3 &  \cite{Jenniskens2016}   \\ 
DEG &  726 & 2.26  & 0.82  & 2.3 & 285.6 & 268.0 &  \cite{Jenniskens2016}   \\  
RGE &  94  & 2.66  & 0.71  & 3.5 & 243.4 & 302.1 &  \cite{Lindblad1971} \\ 
\hline
    \end{tabular}
    \caption{Orbital elements of several minor meteor showers associated with the Taurid complex}
    \label{tab:minor_showers}
\end{table}
\newpage
\clearpage

\begin{figure}[!ht]
    \centering
    \includegraphics[width=.6\textwidth]{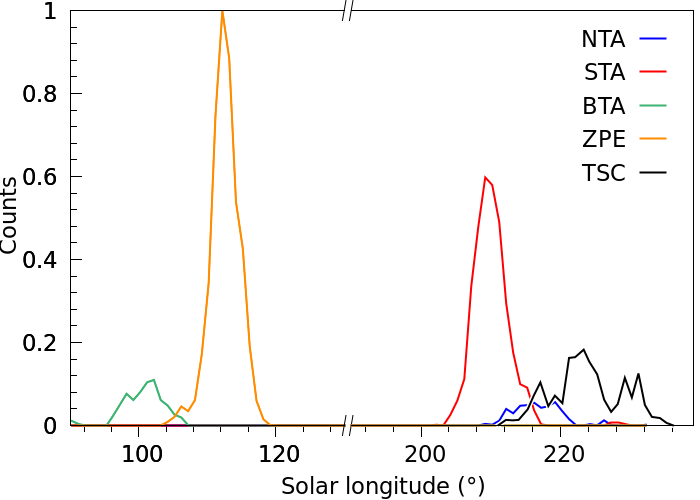}
    \caption{Normalized activity profiles produced by resonant meteoroids ejected from 2P/Encke's clone A4. Only particles of radius $>$1 cm trapped into the 7:2 MMR with Jupiter that approach Earth between 1985 and 2050 are retained for the computation. The modelled swarm in autumn precedes by a few days the 2015 TSC return (black line) as reported by  \protect\cite{Devillepoix2021}.}
    \label{fig:TSC_SL}
\end{figure}

\textbf{APPENDIX H: Comet A4 Stream Impacts with Mercury}\\ 

\setfignumprefix{H.}
\begin{figure}[!ht]
    \centering
    \includegraphics[width=.7\textwidth]{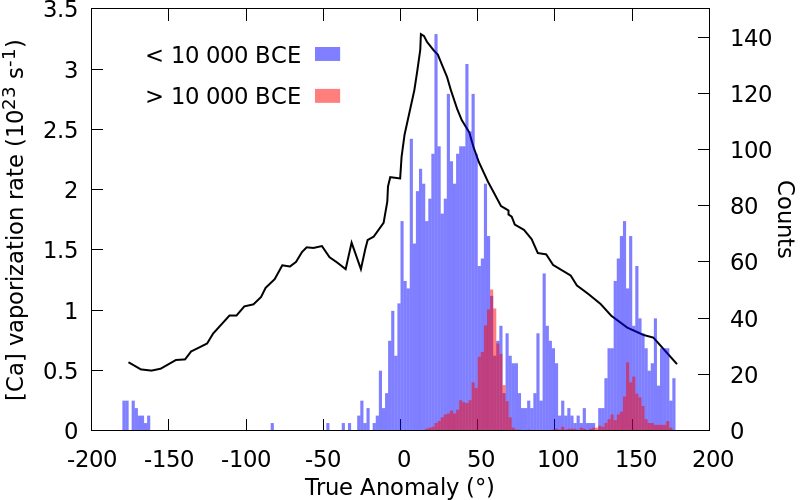}
    \caption{Distribution of Mercury's true anomaly when impacting the simulated A4 stream (particles count). Only particles approaching the planet with a MOID below 0.02 AU and a $\Delta T$ of 10 days were retained for the computation. The contribution of particles emitted between 20 000 BCE and 10 000 BCE or after 10 000 BCE are represented in blue and red respectively. The distributions are normalized to the maximum Ca vaporization rate in order to ease the comparison of the peak emission time. The black line corresponds to the average Ca emission rate measured by the MASCS instrument on-board the MESSENGER spacecraft, published in \protect\cite{Christou2015}. The first bump in the observed Ca rate around true anomaly -50$\degree$, successfully reproduced with a disk dust model, is not expected to be caused by the impact with a meteoroid stream \protect\citep{Christou2015}.}
    \label{fig:ca_mercury}
\end{figure}

\newpage
\clearpage
\bibliographystyle{mnras}
\bibliography{references} 